\documentstyle[12pt]{article}
\begin{document}

\begin{center}
{\large\bf QUANTUM GROUP AND QUANTUM SYMMETRY}

\bigskip

Zhe Chang

\medskip

{\em
International Centre for Theoretical Physics, Trieste, Italy}

{\em INFN, Sezione di Trieste, Trieste, Italy}

and

{\em Institute of High Energy Physics, Academia Sinica, Beijing, China}
\end{center}

\bigskip\bigskip

\tableofcontents

\newpage

\bigskip

\centerline{\bf Abstract}

\medskip This is a self-contained review on the theory of quantum group
and its applications to modern physics. A brief introduction is given
to the Yang-Baxter equation in integrable quantum field theory and
lattice statistical physics. The quantum group is primarily
introduced as a systematic method for solving the Yang-Baxter equation.
Quantum group theory is presented within the framework of quantum double
through quantizing Lie bi-algebra. Both the highest weight and the cyclic
representations are investigated for the quantum group and emphasis is
laid on the new features of representations for $q$ being a root of unity.
Quantum symmetries are explored in selected topics of modern physics. For
a Hamiltonian system the quantum symmetry is an enlarged symmetry that
maintains invariance of equations of motion and allows a deformation of
the Hamiltonian and symplectic form. The configuration space of the
integrable lattice model is analyzed in terms of the representation
theory of quantum group. By means of constructing the Young operators
of quantum group, the Schr\"{o}dinger equation of the model is transformed
to be  a set of coupled linear equations that can be solved by the standard
method.  Quantum symmetry of the minimal model and the WZNW model in conformal
field theory is a hidden symmetry expressed in terms of screened vertex
operators, and has a deep interplay with the Virasoro algebra. In quantum
group approach a complete description for vibrating and rotating diatomic
molecules is given. The exact selection rules and wave functions are
obtained. The Taylor expansion of the analytic formulas of the approach
reproduces the famous Dunham expansion.

\newpage
\section{Introduction}
It is well-known that symmetry plays an important role in modern physics.
Gauge symmetry leads to the Standard Model in high energy physics;
crystallographic space symmetry is fundamental to solid state physics,
conformal symmetry is crucial to string theory and critical phenomenon.
In a sense, the progress of modern physics is accompanied by study
of symmetry.

The mathematical counterpart of all the above motioned symmetries
and other popular ones in physics is group. Recently there has
been great interest in the study of quantum
group and quantum symmetry. Quantum group in contrast to its literal
meaning is not a group, even not a semi-group. However quantum group
is the deformation of the universal enveloping algebra of a
finite-dimensional semi-simple Lie algebra introduced by Drinfel'd
\cite{Drinfeld86} and Jimbo \cite{Jimbo85} in their study of the Yang-Baxter
equation (YBE). In 1967, Yang \cite{Yang}
discovered that if a certain consistency condition is satisfied,  the quantum
mechanical many body problem on a line with the potential $\displaystyle
c\sum_{i<j}\delta(x_i-x_j)$
can be solved exactly by making use  of Bethe Ansatz.
The consistency condition in Yang's paper and the commutivity
condition for the transfer matrix of the eight-vertex model in
statistical physics found by Baxter \cite{Baxter72a} are referred as
the YBE \cite{Faddeev79,Faddeev82}.

At this stage we would like to draw readers attention that the
word ``quantum'' in quantum group is from the YBE:  solutions of
the classical YBE are closely related with classical or semi-simple group,
while solutions of the quantum YBE related with quantum group.
The ``quantum'' thus  really differs from the canonical quantization and
possesses different meanings for different systems.
Conventionally, quantum group is the Hopf algebra which is neither commutative
nor cocommutative. A Hopf algebra is endowed with the algebra homomorphisms:
comultiplication $\Delta$ and counit $\epsilon$, and the algebra
anti-homomorphism: antipode $S$.

As long as $q$ is not a root of unity, the representation theory of
quantum group runs parallel to the classical theory \cite{Rosso88}.
The representations
are labeled by the same highest weight vectors. Most of the standard
expressions  valid for classical algebras have $q$-analogs which often amount
to
replacing ordinary number by $q$-numbers. However, if $q$ is a root of
unity, representation theory changes drastically
\cite{Concini89}--\cite{Pasquier90}. This is one of
the most intriguing situations which are absent in the ``classical'' case.
The highest weight representations are still well defined, however, they are
no longer irreducible in general. Many of representations appearing in
the decomposition of tensor products of irreducible representations
are reducible but not fully reducible \cite{Pasquier90}. If $q$ is a root
of unity,
quantum group contains a large ``standard'' central subalgebra, and new
representations appear, the so-called cyclic representations
\cite{Concini89}.

Quantum group (quantum symmetry) is now a popular topic in different fields of
modern physics due to its richer structure than that of Lie group.
Quantum group is used as the natural structure to characterize and classify
rational conformal field theory \cite{Alvarez89}--\cite{Moore89}.
Integrable lattice models are constructed and solved by quantum group
theory. In Hamiltonian systems, quantum group is an enlarged symmetry
that maintains invariance of equations of motion. In quantum group
approach, a consistent description of vibrating and rotating molecular
spectra is given. Quantum group is also crucial in knot theory
\cite{Resh87a}--\cite{Lee92}, quantum optics \cite{Chai90}--\cite{CChang}
and gauge field theory \cite{Chau93}.

Quantum group theory has been  developed in different directions. In the
quantum space approach \cite{Manin88,Manin87}, the initial object is a
quadratic algebra which is considered being
as the polynomial algebra on a quantum linear space. Quantum group appears
like a group of automorphisms of the quantum linear space.
The crucial ingredient of the matrix pseudogroup approach
\cite{Woronowicz87a}-- \cite{Woronowicz88}
states that any commutative $C^*$ algebra with unit element
is isomorphic to an algebra of all continuous functions on some compact
topological manifolds.
Other important approaches can be found in References
\cite{Reshitikhin89,Faddeev,Wess-Zumino}.
We do not  discuss all of them and relations between
them, merely restrict ourselves to  investigating  quantum group as
quantized
universal enveloping algebra, because most of the applications of quantum
group in modern physics are expressed in this approach.

This presentation is by no means an exhaustive overview of this rapid
developing
subject.  Likewise the references we give have no  pretense to
completeness. Further materials can be  found, for example, in References
\cite{Jimbo}--\cite{Ma93} and references therein.

This article is organized as follows. In Sec.$2$, we start to briefly
review the subject of the YBE, indeed the attempts of solving the YBE
motivated Drinfel'd and Jimbo to introduce the theory of quantum group.
Sec.$3$ begins with stating the basic facts of Hopf algebras,
the language in which the quantum group theory is written. After defining the
suitable form of quantization for Lie bi-algebra, quantum group
is presented as quantum double. In Sec.$4$, the representation theory of
quantum group is discussed in detail, for both generic and
non-generic $q$. Both the highest weight and cyclic representations are
investigated.  Sec.$5$---Sec.$8$ are devoted to discussing
applications of quantum group, i.e., quantum symmetry in modern physics.
In Sec.$5$, by means of symplectic geometry, it is shown that quantum group
can be realized in a Hamiltonian system -- symmetric top system. The
Hamiltonian system, in which quantum group is realized, obeys the same
equation of motion as the standard Hamiltonian system, but these two
Hamiltonian systems have different constants of motion. In Sec.$6$, the
integrable lattice model are mapped into the XXZ spin chain through the
very anisotropic limit. Quantum symmetry in the model is shown. By
constructing a basis of a primitive left ideal from the Young operators
of quantum group, we transform the Schr\"{o}dinger equation of the model
to be a set of coupled linear equations which can be solved by the standard
method.  In Sec.$7$, the quantum symmetry of the minimal model and the WZNW
model in their Coulomb gas version, is formulated in terms of a type of
screened vertex operators, which define the representation space of a
quantum group. The conformal properties of these operators exhibit a deep
interplay between the quantum group and the Virasoro algebra. In Sec.$8$
 we give a complete treatment of vibrating and
rotating molecules in quantum group theoretic approach. The energy spectra of
the model possessing
quantum symmetry exhibit the properties of the infrared,
vibrational and rotational Raman spectra as well as vibrational and
rotational structures of electronic transitions of molecules.  The exact
selection rules and wave functions are obtained. The Taylor expansion of
the analytic formulas of the approach reproduces
Dunham expansion.

\section{Yang-Baxter equation}

\subsection{Integrable quantum field theory}
For over two decades the YBE has been studied as
the master equation in lattice statistical physics and integrable quantum
field theory. The YBE first manifested itself in the work of Yang
\cite{Yang}. He considered a one dimensional quantum mechanical many body
problem with Hamiltonian

\begin{equation}
H=-\sum_{i=1}^N\frac{\partial^2}{\partial x_i^2}+2c\sum_{i<j}
  \delta(x_i-x_j)~,~~~c>0~.
\end{equation}
Use is making of Bethe Ansatz, to  write the wave function of the system
in the following form

\begin{equation}
\psi=\sum_PA_P(Q)\exp(ik_Px_Q)~,
\end{equation}
where $k_Px_Q\equiv\displaystyle\sum_{i=1}^Nk_{p_i}x_{q_i}$, $k_i$ is a
set of unequal numbers, $x_i$ is the position of the $i$-th particle and
satisfies
$0<x_{q_1}<x_{q_2}\cdots<x_{q_N}<L$,
$P=\left(p_1,p_2, \cdots ,p_N\right)$
and
$Q=\left(q_1,q_2, \cdots ,q_N \right)$  are two permutations of $N$ bodies.\\
Then we get the eigenenergy of the system as

\begin{equation}
E=\sum_{i=1}^Nk_i^2~.
\end{equation}
The solution of the problem thus reduces to obtain $k_i~(i=1,~2,~\cdots,~N)$
and $A_P(Q)$. Let us introduce the operator $T_R$ by

\begin{equation}
T_RA_P(Q)\equiv A_P(QR)~.
\end{equation}
For convenience, we denote $T_{ij}=T_{(ij)}, ~T_i=T_{(i~i+1)}$. \\
For the continuity of $\psi$,

\begin{equation}
A_{P}(Q)+A_{P'}(Q)=A_{P}(Q')+A_{P'}(Q')
\end{equation}
and the discontinuity of its derivative,

\begin{equation}
i(k_{p_{i+1}}-k_{p_i})\left(A_{P}(Q)
         -A_{P'}(Q')\right)=c\left(A_{P}(Q)+A_{P'}(Q)\right)~,
\end{equation}
to be satisfied as required by the $\delta$-type interaction between
particles, it is sufficient to demand

\begin{equation}
\begin{array}{rcl}
A_{P}(Q)&=&\displaystyle\frac{i(k_{p_{i+1}}-k_{p_i}) T_i+c}
{i(k_{p_{i+1}}-k_{p_i})-c}A_{P'}(Q)\\[4mm]
&\equiv&Y_i(k_{p_{i+1}}-k_{p_i})A_{P'}(Q)~,
\end{array}
\end{equation}
where we have used the notation
$Y_i(u)\equiv\displaystyle\frac{iuT_i+c}{iu-c}$,
$P'$ and $Q'$ are two permutations of $N$ bodies and related with $P$ and $Q$
through $P'=P(i~i+1)$ and $Q'=Q(i~i+1)$. \\
Using the operator $Y_i(u)$ repeatedly, we can express $A_P(Q)$ in terms of
$A_E(Q)$, where $E$ is the unit element of the permutation group $S_N$.
For example, when $P=(321)$ we have

\begin{equation}
\begin{array}{rcl}
A_{(321)}(Q)&=&Y_1(k_2-k_3)Y_2(k_1-k_3)Y_1(k_1-k_2)A_{(123)}(Q)\\
            &=&Y_2(k_1-k_2)Y_1(k_1-k_3)Y_2(k_2-k_3)A_{(123)}(Q)~.
\end{array}
\end{equation}
{}From the above equation, we see that $Y_1$ and $Y_2$ must satisfy the
following consistency condition

\begin{equation}
Y_1(k_2-k_3)Y_2(k_1-k_3)Y_1(k_1-k_2)=Y_2(k_1-k_2)Y_1(k_1-k_3)Y_2(k_2-k_3)~.
\end{equation}
In general case, the consistency condition possesses the form

\begin{equation}\label{0.yb1}
\begin{array}{l}
Y_i(u)Y_j(v)=Y_j(v)Y_i(u)~,~~~~|i-j| \geq 2~,\\
Y_i(u)Y_i(-u)=1~,\\
Y_i(u)Y_{i+1}(u+v)Y_i(v)=Y_{i+1}(v)Y_i(u+v)Y_{i+1}(u)~.
\end{array}
\end{equation}
Imposing the periodic boundary condition of $\psi$, \\  \\
$\displaystyle\sum_PA_P(q_1,q_2,\cdots,q_N)
\exp \left(i\left(k_{p_2}x_{q_2}+k_{p_3}x_{q_3}+\cdots
+k_{p_N}x_{q_N}\right)\right)$
\begin{equation}\label{0.07}
=\sum_PA_P(q_2,q_3,\cdots,q_N,q_1)\exp
\left(i\left(k_{p_1}x_{q_2}+k_{p_2}x_{q_3}+\cdots
+k_{p_{N-1}}x_{q_N}+k_{p_N}L\right)\right)~,
\end{equation}
we obtain\\ \\
$Y_1(k_1-k_i)Y_2(k_2-k_i)\cdots Y_{i-1}(k_{i-1}-k_i)A_E(Q)$
\begin{equation}
=\exp\left(ik_iL\right)T_1T_2\cdots T_{N-1}Y_{N-1}(k_i-k_N)Y_{N-2}
(k_i-k_{N-1})\cdots Y_{i}(k_i-k_{i+1})A_E(Q)~.
\end{equation}
Defining

\begin{equation}\label{0.yb3}
\begin{array}{l}
R_{i(i+1)}(u)\equiv T_iY_i(u)=\displaystyle\frac{u-icT_i}{u+ic}~,\\[4mm]
R_{ij}(u) \equiv \displaystyle\frac{u-icT_{ij}}{u+ic} ~,
\end{array}
\end{equation}
we have, similar with Eq.(\ref{0.yb1}),

\begin{equation}\label{0.YB4}
\begin{array}{l}
R_{ij}(u)R_{kl}(v)=R_{kl}(v)R_{ij}(u)~,\\
R_{ij}(u)R_{ij}(-u)=1~,
\end{array}
\end{equation}
and

\begin{equation}\label{0.yb5}
R_{ij}(u)R_{ik}(u+v)R_{jk}(v)=R_{jk}(v)R_{ik}(u+v)R_{ij}(u)~.
\end{equation}
Equation  (\ref{0.yb5}) is just the YBE in integrable quantum field
theory.  Yang provided  first solution of the YBE, Eq.(\ref{0.yb3}).

In terms of $R_i$, the periodic boundary condition (\ref{0.07})
can be cast into the form

\begin{equation}
{\cal M}_iA_E(Q)=e^{ik_iL}A_E(Q),
\end{equation}
where
$${\cal M}_i\equiv R_{(i+1)i}(k_{i+1}-k_i)\cdots
R_{Ni}(k_{N}-k_i)R_{1i}(k_{1}-k_i)\cdots
R_{(i-1)i}(k_{i-1}-k_i)~.$$
It is straightforward to show that the operators ${\cal M}_i$ commute with
each other.  As a result, $A_E(Q)$ is a common eigenstate
of the operators ${\cal M}_i$ with eigenvalue $e^{ik_iL}$. The problem
thus reduces to an eigenvalue problem for $k_i$. This eigenvalue problem
can be solved by a second use of Bethe Ansatz \cite{Yang}.

\subsection{Lattice statistical physics}
Independently, the YBE has arisen in Baxter's papers \cite{Baxter72a},
\cite{Baxter73a}--\cite{Baxter82} as the commutivity
condition for the transfer matrices of the so called eight-vertex model
in lattice statistical physics.

Consider a two-dimensional square lattice \cite{Baxter82}
(Fig.1). Degrees of freedom of
vertex model are associated with  links of the lattice and
interact at the vertices (Fig.2).
For a horizontal row of the lattice with the adjacent vertex edges,
let $\alpha=\{\alpha_i,\cdots,\alpha_n\}$ be the state variables on the
lower row of the vertical edges, $\alpha'=\{\alpha'_i,\cdots,\alpha'_n\}$
be the
state variables on the upper row,  and $\beta=\{\beta_i,\cdots,\beta_n\}$
be the state variables on the horizontal edges (Fig.3).

\setlength{\unitlength}{5pt}
\thicklines
\begin{picture}(30,40)
\multiput(25,10)(0,5){5}{\line(1,0){25}}
\multiput(27.5,7.5)(5,0){5}{\line(0,1){25}}
\put(27.5,5.5){1}
\put(32.5,5.5){2}
\put(47.5,5.5){n}
\put(23,10){1}
\put(23,15){2}
\put(23,30){m}

\end{picture}

\hspace{10em}Fig.1. Square lattice.

\setlength{\unitlength}{5pt}
\thicklines
\begin{picture}(20,30)
\put(30,10){\line(1,0){10}}
\put(35,5){\line(0,1){10}}
\put(28,10){i}
\put(35,3){j}
\put(41,10){k}
\put(35,16){l}
\end{picture}

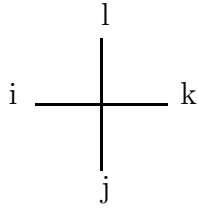
\hspace{7em}Fig.2. Vertex $w(i,j,k,l)$, the Boltzmann

\hspace{9.5em} weight of the vertex model.

\thicklines
\setlength{\unitlength}{10pt}
\begin{picture}(50,15)(20,3)
\put(25,10){\line(1,0){30}}
\multiput(27.5,7)(5,0){6}{\line(0,1){6}}
\put(27.5,5){$\alpha_1$}
\put(32.5,5){$\alpha_2$}
\put(47.5,5){$\alpha_{n-1}$}
\put(52.5,5){$\alpha_{n}$}
\put(27.5,14){$\alpha'_1$}
\put(32.5,14){$\alpha'_2$}
\put(47.5,14){$\alpha'_{n-1}$}
\put(52.5,14){$\alpha'_{n}$}
\put(25.5,10.5){$\beta_1$}
\put(30.5,10.5){$\beta_2$}
\put(35.5,10.5){$\beta_3$}
\put(50.5,10.5){$\beta_{n}$}
\put(55.5,10.5){$\beta_{n+1}=\beta_1$}
\label{fig1}
\end{picture}

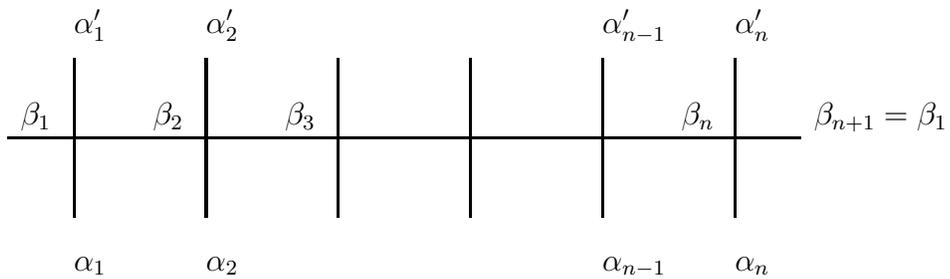
\hspace{4.5em}Fig.3. Row-to-Row transfer matrix ${\cal V}^{(n)}$ for the Vertex
model.

It is convenient to adopt a Hamiltonian picture, in which ``time'' flows
upward
on the lattice, and the various configurations of vertical links are considered
as independent possible states of the system at a given time. Time evolution
is carried out by the row-to-row transfer matrix ${\cal V}^{(n)}(u)$,
whose matrix elements ${\cal V}^{(n)}_{\alpha,\alpha'}(u)$  is defined by

\begin{equation}\label{0.2.11}
{\cal V}^{(n)}_{\alpha,\alpha'}(u)=\sum_{\beta_1\cdots\beta_n}
                       w(\beta_1,\alpha_1,\beta_2,\alpha_1'|u)
                                   w(\beta_2,\alpha_2,\beta_3,\alpha_2'|u)
                \cdots w(\beta_n,\alpha_n,\beta_1,\alpha_n'|u)~,
\end{equation}
where $w(\beta_i,\alpha_i,\beta_i',\alpha_i'|u)$ is Boltzmann weight
of the vertex.
In terms of the transfer matrix ${\cal V}^{(n)}(u)$, the partition function
$Z_N$ and the free energy per site $f$ are given by

\begin{equation}
\begin{array}{l}
Z_N=\displaystyle\sum_{\alpha^{(1)}\alpha^{(2)}\cdots \alpha^{(m)}}
{\cal V}^{(n)}_{\alpha^{(1)},\alpha^{(2)}}(u) {\cal V}^{(n)}_{\alpha^{(2)},
\alpha^{(3)}}(u)\cdots {\cal V}^{(n)}_{\alpha^{(m-1)},\alpha^{(m)}}(u)
{\cal V}^{(n)}_{\alpha^{(m)},\alpha^{(1)}}(u)={\rm Tr}\left({\cal
V}^{(n)}(u)\right)^m~,\\
f=-k_BT\displaystyle\lim_{N\to\infty}N^{-1}\log Z_N~,
\end{array}
\end{equation}
where $N=mn$ is the number of lattice sites and summation is taken over all
configurations of arrows. \\
Let ${\cal V}^{(n)}(u+v)$ be another transfer matrix where the Boltzmann
weight $w(\beta_i,\alpha_i,\beta_{i+1},\alpha_i'|u)$ is
replaced by $w(\beta_i,\alpha_i,\beta_{i+1},\alpha_i'|u+v)$. From
Eq.(\ref{0.2.11}),
we have

\begin{equation}\label{0.2.13}   \begin{array}{rcl}
\left({\cal V}^{(n)}(u){\cal V}^{(n)}(u+v)\right)_{\alpha,\alpha'}&=&
\displaystyle\sum_{\gamma}{\cal V}^{(n)}(u)_{\alpha,\gamma}
{\cal V}^{(n)}(u+v)_{\gamma,\alpha'}\\
                      &=&\displaystyle\sum_{\beta\bar{\beta}}\prod_{i=1}^{n}
X(\beta_i,\bar{\beta}_i|\beta_{i+1}, \bar{\beta}_{i+1}|\alpha_i,\alpha'_i)~,
\end{array}
\end{equation}
where

\begin{equation}\label{0.2.14}
X(\beta_i,\bar{\beta}_i|\beta_{i+1}, \bar{\beta}_{i+1}|\alpha_i,\alpha_i')
          =\sum_{\gamma_i} w(\beta_i,\alpha_i,\beta_{i+1},\gamma_i|u)
                        w(\bar{\beta}_i,\gamma_i,
                        \bar{\beta}_{i+1},\alpha_i'|u+v)~.
\end{equation}
Equation (\ref{0.2.13}) can be written  as a compact form

\begin{equation}\label{0.2.15}
\left({\cal V}^{(n)}(u){\cal V}^{(n)}(u+v)\right)_{\alpha,\alpha'}={\rm
Tr}X(\alpha_1,\alpha'_1)X(\alpha_2,\alpha'_2)
                    \cdots X(\alpha_n,\alpha'_n)~,
\end{equation}
where $X(\alpha,\alpha')$ is the matrix with element
$X(\beta_i,\bar{\beta}_i|\beta_{i+1},\bar{\beta}_{i+1}|\alpha_i,\alpha_i')$
in row $(\beta_i,\bar{\beta}_i)$ and in column
$(\beta_{i+1},\bar{\beta}_{i+1})$.

Similarly, we define $X'$ with
$w(\beta_i,\alpha_i,\beta_{i+1},\alpha_i'|u)$
and $w(\beta_i,\alpha_i,\beta_{i+1},\alpha_i'|u+v)$ interchanged in
Eq.(\ref{0.2.14}), which lead to

\begin{equation}\label{0.2.16}
({\cal V}^{(n)}(u){\cal V}^{(n)}(u+v))_{\alpha,\alpha'}={\rm
Tr}X'(\alpha_1,\alpha_1')X'(\alpha_2,\alpha_2')
                    \cdots X'(\alpha_n,\alpha_n')~.
\end{equation}
{}From Eqs.(\ref{0.2.15}) and (\ref{0.2.16}), we see that ${\cal V}^{(n)}(u)$
and ${\cal V}^{(n)}(u+v)$ commute if there exists a matrix $W$ satisfying

\begin{equation}\label{0.0010}
X(\alpha_i,\alpha_i')=WX'(\alpha_i,\alpha_i')W^{-1}~.
\end{equation}
The matrix $W$ has rows labeled by $(\beta_i,\bar{\beta}_i)$, and columns
labeled by $(\beta_{i+1},\bar{\beta}_{i+1})$ with elements
$w(\bar{\beta}_i,\beta_i,\beta_{i+1},\bar{\beta}_{i+1}|v)$.
Equation (\ref{0.0010}) can then be cast into the following
form\\ \\ $\displaystyle\sum_{\gamma_i\beta_i''\bar{\beta}_i''}
w(\beta_i,\alpha_i,\beta_i'',\gamma_i|u)
w(\bar{\beta}_i,\gamma_i,\bar{\beta}_i'',\alpha_i'|u+v)
w(\bar{\beta}''_i,\beta''_i,\beta'_i,\bar{\beta}_i'|v)$
\begin{equation}\label{0.2.18}
=\sum_{\gamma_i\beta''_i\bar{\beta}''_i}
w(\bar{\beta}_i,\beta_i,\beta''_i,\bar{\beta}''_i|v)
w(\beta''_i,\alpha_i,\beta'_i,\gamma_i|u+v)w(\bar{\beta}''_i,\gamma_i,
\bar{\beta}'_i,\alpha'_i|u)~.
\end{equation}
This is the YBE in lattice statistical physics.\\

\subsection{Yang-Baxter equation}
When we identify

\begin{equation}
\begin{array}{l}
w(\beta_i,\alpha_i,\beta''_i,\gamma_i|u)=R^{\beta_i\beta''_i}_{\alpha_i\gamma_i}
(u)~,\\
w(\bar{\beta}_i,\gamma_i,\bar{\beta}''_i,\alpha'_i|u+v)=
R^{\bar{\beta}_i\bar{\beta}''_i}_{\gamma_i\alpha'_i}(u+v)~,\\
w(\bar{\beta}''_i,\beta''_i,\beta'_i,\bar{\beta}'_i|v)=
R^{\bar{\beta}''_i\beta'_i}_{\beta''_i\bar{\beta}'_i}(v)~,
\end{array}
\end{equation}
Eq.(\ref{0.2.18}) becomes Eq.(\ref{0.yb5}). Furthermore, the quantities
$R$ and $w(\beta,\alpha,\beta',\alpha')$ can be obviously
interpreted as an operator ${\cal R}$ in the tensor product space
$V^{\otimes 2}$.
In the space $V^{\otimes 3}$ we introduce three
operators ${\cal R}_{12}$, ${\cal R}_{13}$, ${\cal R}_{23}$
corresponding to the three canonical embeddings of $V^{\otimes 2}$ into
$V^{\otimes 3}$ (for example, ${\cal R}_{12}={\cal R}\otimes I$,
${\cal R}_{23}=I\otimes {\cal R}$). Then Eqs.(\ref{0.yb5}), (\ref{0.2.18})
can be rewritten as

\begin{equation}\label{0.cd1}
{\cal R}_{12}(u){\cal R}_{13}(u+v){\cal R}_{23}(v)={\cal R}_{23}(v)
   {\cal R}_{13}(u+v){\cal R}_{12}(u)~.
\end{equation}
The variables $u$ is called as the spectral parameter, and a solution of
Eq.(\ref{0.cd1}) is called as $R$ matrix.\\
The solution of the YBE corresponding to integrable quantum field theory
\cite{Yang} is

\begin{equation}\label{0.ys}
\begin{array}{rcl}
{\cal R}(u,c)&=&\displaystyle\frac{u-icP}{u+ic}\\[4mm]
             &=&\displaystyle\frac{1}{u+ic}\left[
             \begin{array}{cccc}
             u-ic& 0 & 0 & 0 \\
             0   & u &-ic& 0 \\
             0   &-ic& u & 0 \\
             0   & 0 & 0 &u-ic \end{array}\right]~,
\end{array}
\end{equation}
where $P$ is the transposition operator in $V\otimes V$,

\begin{equation}
\begin{array}{rcl}
P:~~V_1\otimes V_2&\to& V_2\otimes V_1~,\\
P:~~a\otimes b&\to& b\otimes a~.
\end{array}
\end{equation}
In 6-vertex model, the solution  \cite{Baxter82} has the form

\begin{equation}\label{0.ts}
{\cal R}(u)=\rho\left[\begin{array}{cccc}
                      \sin(\eta+u)&         &        & \\
                                  & \sin u  &\sin\eta& \\
                                  & \sin\eta&\sin u  & \\
                                  &         &        &\sin(\eta+u)
                      \end{array}\right]~,
 \end{equation}
 where
 $$\begin{array}{l}
\rho\sin(\eta+u)=\exp(-i\beta\epsilon_1)~,\\
\rho\sin u=\exp(-i\beta\epsilon_3)~,\\
\rho\sin\eta=\exp(-i\beta\epsilon_5)~,
\end{array}$$
$\epsilon_i$ ($i=1,3,5$) are three distinct energies of 6-vertex model, and
$\eta$ is a free parameter.\\
Another type of solution \cite{Baxter73a} is

\begin{equation}
{\cal R}(u)=\left[\begin{array}{cccc}
                  a(u)&    &    &d(u)\\
                      &b(u)&c(u)&\\
                      &c(u)&b(u)&\\
                  d(u)&    &    &a(u)\end{array}\right]~,
\end{equation}
where

$$
\begin{array}{rcl}
a(u)&=&\theta_0(\eta)\theta_0(u)\theta_1(\eta+u)~,\\
b(u)&=&\theta_0(\eta)\theta_1(u)\theta_0(\eta+u)~,\\
a(u)&=&\theta_1(\eta)\theta_0(u)\theta_0(\eta+u)~,\\
a(u)&=&\theta_1(\eta)\theta_1(u)\theta_1(\eta+u)~,
\end{array}$$
and $\theta_i(u)$ are the elliptic theta functions

$$\begin{array}{rcl}
\theta_0(u)&=&\displaystyle\prod_{i=1}^{\infty}\left(1-2p^{i-1/2}\cos 2\pi u
  +p^{2i-1}\right)(1-p^i)~,\\
\theta_1(u)&=&2p^{1/8}\sin\pi u\displaystyle\prod_{i=1}^{\infty}
\left(1-2p^{i}\cos 2\pi u
  +p^{2i}\right)(1-p^i)~,
  \end{array}$$
$\eta$ and $p$ are free parameters.\\
Solutions of the YBE on high genus Riemann surface ($g>1$) have been found
recently \cite{Perk87,Perk88}.\\
Making use of the basis for $V$

\begin{equation}
\begin{array}{l}
{\cal R}(u)=\sum{\cal R}_{ij}^{kl}E_{ik}\otimes E_{jl}~,\\
E_{ij}=(\delta_{im}\delta_{jn})_{m,n=1,2,\cdots,{\cal D}}~,
\end{array}
\end{equation}
we see that the YBE (\ref{0.cd1}) amounts to ${\cal D}^6$ homogeneous nonlinear
equations with ${\cal D}^4$ unknowns ${\cal R}_{ij}^{kl}$. Here we have
used the notation ${\cal D}\equiv {\rm dim}V$. Generally speaking, it is very
difficult to solve the YBE \cite{Jimbo89,Vega90} save for certain limiting
cases.

\subsection{Classical Yang-Baxter equation}

There is a set of important solutions (quasi-classical solutions) of the
YBE which contains an
extra parameter $\gamma$ (quantization parameter). We expand these
solutions near zero point of the quantization parameter $\gamma$
\begin{equation}
{\cal R}(u,\gamma)=({\rm scalar})\times\left(I+\gamma r(u)+O(\gamma^2)\right)~.
\end{equation}
The $r(u)$ is called as the classical limit of ${\cal R}(u,\gamma)$.
In the zero $\gamma$ limit,  the YBE (\ref{0.cd1}) reduces to the
classical Yang-Baxter equation (CYBE) for $r(u)$
\cite{Kulish82,Drinfeld88,Tyan83},

\begin{equation}\label{0.cd2}
[r_{12}(u),r_{13}(u+v)]+[r_{12}(u),r_{23}(v)]+[r_{13}(u+v),r_{23}(v)]=0~.
\end{equation}

The CYBE that is in Lie bracket form has one less parameter than the YBE.
It is easier to discuss solutions of the CYBE.\\
Let $g$ be a Lie algebra,
and $r(u)$  be a $g\otimes g$-valued function. In terms of a basis
$\{X_\mu\}$ of $g$, we write

\begin{equation}
r(u)=\sum_{\mu,\nu}r^{\mu\nu}(u)X_\mu\otimes X_\nu
\end{equation}
with $c$-valued functions $r^{\mu\nu}(u)$. Let further

\begin{equation}
r_{12}(u)=\sum_{\mu,\nu}r^{\mu\nu}X_{\mu}\otimes X_\nu\otimes I~~~~\in
\left(U(g)\right)^{\otimes 3}~,
\end{equation}
and so on, where $U(g)$ denotes the universal enveloping algebra of $g$.
Then each term in Eq.(\ref{0.cd2}) actually is in  $g^{\otimes
3}$, for example,

\begin{equation}
[r_{12}(u),r_{23}(v)]=\sum r^{\mu\nu}(u)r^{\rho\sigma}(v)X_\mu\otimes[X_\nu,X_
\rho]\otimes X_\sigma~.
\end{equation}
 For arbitrary triplet of representations $(\pi,V_i)~(i=1,~2,~3)$ of
$g$, $(\pi_i\otimes\pi_j)~(r_{ij}(u))$ yields a matrix solution of the CYBE.
The following are two familiar examples of solutions of the  CYBE.\\
\begin{itemize}
\item Rational Solution. \\
For an orthonormal basis $\{X_\mu\}$ of $g$ with a non-degenerate invariant
bilinear form $\Omega$, we have a solution of the CYBE

\begin{equation}\label{0.rs}
\begin{array}{l}
r(u)=\displaystyle\frac{1}{u}\Omega~,\\
\Omega\equiv\displaystyle\sum_{\mu}X_{\mu}\otimes X_{\mu}~,
\end{array}
\end{equation}
namely the rational solution. This solution corresponds
to that given by Eq.(\ref{0.ys}). The $c=0$ limit of Eq.(\ref{0.ys}) is

\begin{equation}
r(u)=\frac{1}{u}P=\frac{1}{2u}\left(1+\sum_{i=1}^3\sigma_i\otimes
\sigma_i\right)~,
\end{equation}
where $\sigma_i$ are Pauli matrices. As a results

\begin{equation} \label{0.su2}
r(u)=\frac{1}{u}\sum_{i=1}^3\left(\frac{\sigma_i}{2}\right)\otimes
\left(\frac{\sigma_i}{2}\right)~,
\end{equation}
is a solution of the CYBE. It is well-known that
$\displaystyle\frac{\sigma_i}{2}$ are
generators of the fundamental representation of the Lie algebra $su(2)$.
Equation (\ref{0.rs})
is a generalization of Eq.(\ref{0.su2}) to general Lie algebra. \\ \\ \\
\item Trigonometric Solution.\\
In Cartan-Weyl basis, the non-degenerate invariant bilinear form $\Omega$ has
the form

\begin{equation}
\begin{array}{rcl}
\Omega&=&\displaystyle\sum_j H_j\otimes H_j+\displaystyle\sum_{\alpha\in
\Delta_+}\left(E_{\alpha}\otimes E_{-\alpha}+E_{-\alpha}\otimes
E_{\alpha}\right)\\
 &\equiv&C_0+C_++C_-~,
 \end{array}
 \end{equation}
 where $\Delta_+$ is the positive root space of  $g$,
 $$\begin{array}{l}
 C_0=\displaystyle\sum_jH_j\otimes H_j~,\\
 C_+=\displaystyle\sum_{\alpha\in\Delta_+}E_\alpha\otimes E_{-\alpha}~,\\
 C_-=\displaystyle\sum_{\alpha\in\Delta_+}E_{-\alpha}\otimes E_{\alpha}~.
 \end{array}$$
One can readily verify that

\begin{equation}
r=C_0+2C_+
\end{equation}
is a trigonometric solution of the CYBE.
\end{itemize}

The solutions of the CYBE can be classified completely by the following
theorem \cite{Belavin82}.\\
{\em The non-degenerate solution $r(u)$ extends
meromorphically to the whole plane
 ${\bf C}$, with all simple poles;  $r(u)$ is classified by
 $\Gamma=\{ {\rm the~set~of ~poles~of~r(u)} \}$:\\
 \begin{itemize}
 \item Elliptic function has rank($\Gamma)=2$ and exists only for $g=sl(n)$.
 \item Trigonometric function has rank($\Gamma)=1$ and exists for each type,
 and can be classified by making use of the Dynkin diagram for affine Lie
algebra.
 \item Rational function has rank($\Gamma)=0$.
 \end{itemize} }

Provided  a solution $r(u)\in
g\otimes g$ of the CYBE, one may ask whether exists a quasi-classical
${\cal R}(u,\gamma)$ having $r(u)$ as its classical limit. It is this
``quantization'' problem that has motivated Drinfel'd and Jimbo to introduce
the theory of quantum group.

\section{Quantum group theory}
\subsection{Hopf algebra}
An algebra over the field ${\bf C}$ is a linear space $A$ which possesses two
linear operations: the multiplication $m$ and the unit $\eta$

\begin{equation}
\begin{array}{rcl}
& & m:~~~A\otimes A\to A~,\\
& & m(a\otimes b)=ab~,~~~~a,b\in A~,\\
& & \eta:~~~{\bf C}\to A~,
\end{array}
\end{equation}
and satisfies the following axioms

\begin{equation}
\begin{array}{rcl}
& & m(m\otimes{\rm id})=m({\rm id}\otimes m)~,\\
& & m({\rm id}\otimes\eta)=m(\eta\otimes {\rm id})={\rm id}~,
\end{array}
\end{equation}
i.e., it is associative (Fig.4) and has unit element (Fig.5). When we deal with
tensor
product representations of the underlying abstract algebra, another linear
operation, namely comultiplication $\Delta$, should be endowed.
The comultiplication satisfies the following relations

\begin{equation}
\begin{array}{rcl}
& &\Delta(ab)=\Delta(a)\Delta(b)~,\\
& &(\Delta\otimes{\rm id})\Delta=({\rm id}\otimes\Delta)\Delta~,
\end{array}
\end{equation}
i.e., $\Delta$ is an algebra homomorphism and co-associative (Fig.6). After
introducing the comultiplication, we can compose two representations ( for
example, add angular momenta). This takes place when two physical systems, each
of them is within a certain representation, interact. \\

\setlength{\unitlength}{5pt}
\thicklines
\begin{picture}(50,26)(0,2)

\put(25,20){\vector(3,1){15}}
\put(20,17){$A\otimes A\otimes A$}
\put(41,24){$A\otimes A$}
\put(41,10){$A\otimes A$}
\put(48,25){\vector(3,-1){15}}
\put(25,16){\vector(3,-1){15}}
\put(48,11){\vector(3,1){15}}
\put(62,17){$A$}
\put(28,22.3){$m$}
\put(30,23){$\otimes$}
\put(32,23.67){id}

\put(28,12){id}
\put(30,11.3){$\otimes$}
\put(32,10.7){$m$}
\put(56,23){$m$}
\put(56,11.5){$m$}
\end{picture}

\hspace{13.5em}Fig.4. Associativity.

\setlength{\unitlength}{5pt}
\thicklines
\begin{picture}(50,30)(0,8)

\put(25,15){\vector(1,0){15}}
\put(63,15){\vector(-1,0){15}}
\put(44,17){\vector(0,1){15}}
\put(18,14){${\bf C}\otimes A$}
\put(41,14){$A\otimes A$}
\put(64,14){$A\otimes {\bf C}$}
\put(43.2,32.5){$A$}
\put(21,17){\line(6,5){19}}
\put(22,17){\line(6,5){19}}
\put(67,17){\line(-6,5){19}}
\put(68,17){\line(-6,5){19}}
\put(28.5,15.7){$\eta\otimes$id}
\put(51.5,15.7){id$\otimes\eta$}
\put(44.7,23){$m$}
\end{picture}

\hspace{14.5em}Fig.5. The unit.

\setlength{\unitlength}{5pt}
\thicklines
\begin{picture}(50,35)(0,5)

\put(40,25){\vector(-3,-1){15}}
\put(20,17){$A\otimes A\otimes A$}
\put(41,24){$A\otimes A$}
\put(41,10){$A\otimes A$}
\put(63,20){\vector(-3,1){15}}
\put(40,11){\vector(-3,1){15}}
\put(63,16){\vector(-3,-1){15}}
\put(62,17){$A$}
\put(28,22.3){$\Delta$}
\put(30,23){$\otimes$}
\put(32,23.67){id}

\put(28,12){id}
\put(30,11.3){$\otimes$}
\put(32,10.7){$\Delta$}
\put(56,23){$\Delta$}
\put(56,11.5){$\Delta$}
\end{picture}

\hspace{13em}Fig.6. Co-associativity.

It is well-known that the operation $\eta$ signals the existence of a
unit in $A$. In the same manner we now define an operation $\epsilon$,
called the counit (Fig.7), $\epsilon: ~A\to {\bf C}$, satisfying

\begin{equation}\label{1.1.4}
({\rm id}\otimes \epsilon)\Delta=(\epsilon\otimes{\rm id})\Delta={\rm id}~.
\end{equation}
If all the above axioms be satisfied, the set $(A,m,\Delta,\eta,\epsilon)$ is
called a bi-algebra. Being related with the fact that every element of a group
has an inverse, certain bi-algebras possess an extra operation, antipode $S$
(Fig.8), $S:~A\to A$, with the properties

\begin{equation}\label{1.1.5}
m(S\otimes{\rm id})\Delta=m({\rm id}\otimes S)\Delta=\eta\circ\epsilon~.
\end{equation}
A bi-algebra with antipode is called a Hopf algebra \cite{Able}. We now
illustrate a useful example of Hopf algebras. \\
Let $g$ denote a Lie algebra and $U(g)$ denote its universal enveloping
algebra; if we define
\begin{itemize}
\item the multiplication $m$ as the ordinary multiplication in $U(g)$,
\item $\Delta(x)=x\otimes 1+1\otimes x~,~~~~\forall x\in g~,$
\item $\eta(\alpha)=\alpha {\bf 1}~,$
\item $\epsilon({\bf 1})=1$ and zero on all other elements,
\item $S(x)=-x~,$
\end{itemize}
$U(g)$ becomes a Hopf algebra.

\setlength{\unitlength}{5pt}
\thicklines
\begin{picture}(50,33)(2,10)

\put(40,15){\vector(-1,0){15}}
\put(48,15){\vector(1,0){15}}
\put(44,32){\vector(0,-1){15}}
\put(18,14){${\bf C}\otimes A$}
\put(41,14){$A\otimes A$}
\put(64,14){$A\otimes {\bf C}$}
\put(43.2,32.5){$A$}
\put(21,17){\line(6,5){19}}
\put(22,17){\line(6,5){19}}
\put(67,17){\line(-6,5){19}}
\put(68,17){\line(-6,5){19}}
\put(28.5,15.7){$\epsilon\otimes$id}
\put(51.5,15.7){id$\otimes\epsilon$}
\put(44.7,23){$\Delta$}
\label{fig4}
\end{picture}

\hspace{13em}Fig.7. The counit.

\setlength{\unitlength}{5pt}
\thicklines
\begin{picture}(50,46)(-5,5)

\put(25,20){\vector(3,1){15}}
\put(22,17){$A\otimes A$}
\put(41,24){$A\otimes A$}
\put(41,10){$A\otimes A$}
\put(48,25){\vector(3,-1){15}}
\put(25,16){\vector(3,-1){15}}
\put(48,11){\vector(3,1){15}}
\put(63.5,17){$A$}
\put(28,22.3){$S$}
\put(30,23){$\otimes$}
\put(32,23.67){id}

\put(28,12){id}
\put(30,11.3){$\otimes$}
\put(32,10.7){$S$}
\put(56,23){$m$}
\put(56,11.5){$m$}

\put(6,18){\vector(1,0){15}}
\put(3,17){$A$}
\put(32.5,33){${\bf C}$}
\put(4,19){\line(0,1){15}}
\put(65,34){\vector(0,-1){14}}
\put(4,34){\vector(1,0){28}}
\put(35,34){\line(1,0){30}}

\put(12.5,18.7){$\Delta$}
\put(18,34.7){$\epsilon$}
\put(49,34.7){$\eta$}

\end{picture}

\hspace{12em}Fig.8. The antipode.

Strictly speaking, the above definitions for
$\Delta,~\eta,~\epsilon$ and $S$ are merely restricted  on the subset
$g$ of $U(g)$. Note that $g$
is not a Hopf algebra because it is not an associative algebra.  However, it
is readily to  see that these operations can be extended in a unique
manner  to all of $U(g)$
so that the Hopf algebra axioms are satisfied everywhere.  The universal
enveloping algebras are commutative Hopf algebras, i.e., we have the equality
$\Delta=P\circ\Delta$.
If a Hopf algebra $(A,m,\Delta,\eta,\epsilon,S)$ is not cocommutative, i.e.,
$\Delta'\equiv P\circ\Delta\not=\Delta$, then it is not difficult to verify
that  $(A,m,\Delta',\eta,\epsilon,S')$ is also a Hopf algebra (so called the
related Hopf algebra of $(A,m,\Delta,\eta,\epsilon,S)$), where we have used the
notation $S'=S^{-1}$.

\subsection{Quantization of Lie bi-algebra}

A Lie algebra $g$ with a co-antisymmetric operation
$\psi:~g\to g\otimes g~~~~ (P\circ\psi=-\psi$) satisfying the so
called 1-cocycle condition
$$\begin{array}{rcl}
& &[X,Y\otimes Z]\equiv [X,Y]\otimes Z+Y\otimes [X,Z]~,~~~~X,Y,Z\in g~,\\
& &\psi\left([X,Y]\right)=[\psi(X),Y]+[X,\psi(Y)]~,
\end{array}$$
is a Lie bi-algebra. A quantization of a Lie bi-algebra
$(A_{(0)},m_{(0)},\Delta_{(0)},\eta_{(0)},\epsilon_{(0)},\psi$)
\cite{Drinfeld86,Jimbo85,Jimbo86b,Manin88,Woronowicz89} is a noncommutative
algebra
$(A,m,\Delta,\eta,\epsilon)$ over the ring ${\bf C}[[\gamma]]$ ($\gamma$ is
the
quantization parameter). The space $A$ is the set of polynomials in  $\gamma$
with coefficients in $A_0$. To factorize $\gamma A$ out is then
equivalent to
set $\gamma=0$, which corresponds to the classical limit. A new
noncommutative multiplication $m$ (denoted as $m(a\otimes b)=a\star b$) is of
the form

\begin{equation}
a\star b=\sum_{i=0}^{\infty}f_i(a,b)\gamma^i~,
\end{equation}
where $f_i:~A\otimes A\to A$. \\
Denote the canonical quotient map $A\to A_0$ by $\pi$, a complete
description of the
quantization $(A,m,\Delta,\eta,\epsilon)$ of $(A_{(0)},m_{(0)},\Delta_{(0)},
\eta_{(0)}, \epsilon_{(0)},\psi)$ is
\begin{itemize}
\item $A/\gamma A\cong A_{(0)}~,$
\item $(\pi\otimes \pi)\circ\Delta=\Delta_{(0)}\circ\pi~,$
\item $m_{(0)}\circ(\pi\otimes\pi)=\pi\circ m~,$
\item $\pi\circ\eta=\eta_{(0)}~,$
\item $\epsilon\circ\pi=\epsilon_{(0)}~,$
\item $\psi(\pi(a))=\pi\displaystyle\left(\frac{1}{\gamma}\left(\Delta(a)-
P\circ\Delta(a)\right)\right)~,~~~~\forall a\in A~.$
\end{itemize}
In the following, we will consider the details of the quantization of the
universal enveloping algebra $U(su(2))$.
There are two subalgebras in $U(su(2))$, i.e., the positive Borel
subalgebra,
$U^+(su(2))$, and  the negative Borel subalgebra, $U^-(su(2))$,
generated by $X^+,~H$ and $X^-,~H$, respectively.
We first discuss the quantization of $U^+(su(2))$. Similar discussions can be
given to $U^-(su(2))$.
It is not difficult to verify that the map  $\psi$ defined by

\begin{equation}
\psi(H)=0~,~~~ ~~~\psi(X^+)=2(H\otimes X^+-X^+\otimes H)~,
\end{equation}
and the ordinary  Lie bracket

\begin{equation}
[X^+,H]=X^+~,
\end{equation}
form a Lie bi-algebra. To quantize this Lie bi-algebra, the first step is
to   find out
the comultiplication on the  set of polynomials in $\gamma$. In the classical
limit the comultiplication $\Delta$ must reduce to the ordinary
comultiplication
$\Delta_{(0)}$ on $U^+(su(2))$ and satisfy

\begin{equation}\label{1.ti1}
\psi(\pi(a))=\pi\left(\frac{1}{\gamma}\left(\Delta(a)-P\circ\Delta
(a)\right)\right)~.
\end{equation}
Also, it must be co-associative. The general form of the comultiplication is

\begin{equation}
\Delta=\sum_{i=0}^{\infty}\frac{\gamma^i}{i!}\Delta_{(i)}~.
\end{equation}
{}From its classical limit, we obtain

\begin{equation}
\begin{array}{rcl}
& & \Delta_0(H)=H\otimes 1+1\otimes H~,\\
& & \Delta_0(X^+)=X^+\otimes 1+1\otimes X^+~.
\end{array}
\end{equation}
The explicit form of Eq.(\ref{1.ti1}) reads

\begin{equation}\label{1.ti2}
\begin{array}{rcl}
& &\psi(H)=0=\Delta_1(H)-P\circ\Delta_{(1)}(H)~,\\
& &\psi(X^+)=2(H\otimes X^+-X^+\otimes H)=\Delta_{(1)}(X^+)-
P\circ\Delta_{(1)}(X^+)~.
\end{array}
\end{equation}
An obvious nontrivial solution of these equations is given by

\begin{equation}\label{1.ti3}
\begin{array}{rcl}
& &\Delta_{(1)}(H)=0~,\\
& &\Delta_{(1)}(X^+)=(H\otimes X^+-X^+\otimes H)~.
\end{array}
\end{equation}
The co-associativity of the comultiplication gives us a recursion relation

\begin{equation}\label{1.ti4}
\sum_{k=0}^{i}\left(\begin{array}{c}
                    i\\
                    k\end{array}\right)\left(\Delta_{(k)}\otimes 1-1\otimes
                    \Delta_{(k)}\right)\Delta_{(i-k)}=0~,
\end{equation}
where we have used the notation $\displaystyle\left(\begin{array}{c}
                                                     n\\
                                                     m\end{array}\right)
=\frac{n!}{m!(n-m)!}~.$\\
{}From Eqs.(\ref{1.ti1}) $\sim$ (\ref{1.ti4}), we get

\begin{equation}
\begin{array}{rcl}
& &\Delta_{(i)}(H)=0~,~~~(i\geq 1)~,\\
& &\Delta_{(i)}(X^+)=H^i\otimes X^++(-1)^iX^+\otimes H^i~.
\end{array}
\end{equation}
Making use of the above results, we obtain the comultiplication $\Delta$

\begin{equation}
\begin{array}{rcl}
& &\Delta(H)=H\otimes 1+1\otimes H~,\\
& &\Delta(X^+)=X^+\otimes q^{-H}+q^{H}\otimes X^+~,
\end{array}
\end{equation}
where the new quantization parameter  $q\equiv e^\gamma$ is introduced. \\
Use is made of axioms of Hopf algebra,
to determine the form of other operations on the quantized universal
enveloping algebra $U_q^+\left(su(2)\right)$. One of these axioms is

\begin{equation}
\Delta(a\star b)=\Delta(a)\star\Delta(b)~,~~~~\forall a,b\in A~.
\end{equation}
In our context this means that the equality

\begin{equation}
\Delta\left([H,X^+]\right)=\left[\Delta(H),\Delta(X^+)\right]~,
\end{equation}
must be hold. It is easy to verify that

\begin{equation}
\left[\Delta(H),\Delta(X^+)\right]=\Delta(X^+)~,
\end{equation}
so that the ordinary $su(2)$ relation $[H,X^+]=X^+$ still holds after
quantization. \\ Equation (\ref{1.1.4}) reads

\begin{equation}
\begin{array}{rcl}
& &(\epsilon\otimes{\rm id})\Delta(H)=\epsilon(H)1+H=H~,\\
& &(\epsilon\otimes{\rm id})\Delta(X^+)=\epsilon(X^+)q^{-H}+
\epsilon(q^{H})X^+=X^+~.
\end{array}
\end{equation}
One of the solutions of these equations is

\begin{equation}
\epsilon(H)=0~,~~~~~~\epsilon(X^+)=0~.
\end{equation}
In the same manner the equations,

\begin{equation}
S(H)=-H~,~~~~~~S(X^+)=-q^{-1}X^+~,
\end{equation}
follow from the explicit form of Eq.(\ref{1.1.5})

\begin{equation}
\begin{array}{rcl}
& &m(S\otimes{\rm id})\Delta(H)=S(H)+H=0~,\\
& &m(S\otimes{\rm id})\Delta(X^+)=S(X^+)q^{-H}+S(q^H)X^+=0~.
\end{array}
\end{equation}
Defining

\begin{equation}\label{1.225}
\begin{array}{l}
t=q^{2H}~,~~~~tt^{-1}=1=t^{-1}t~,\\
e=q^HX^+~,
\end{array}
\end{equation}
we rewrite the Hopf algebra $U^+_q(su(2))$  as

\begin{equation}
\begin{array}{rcl}
& &tt^{-1}=1=t^{-1}t~,\\
& &tet^{-1}=q^2e~,\\
& &\Delta(e)=e\otimes 1+t\otimes e~,~~~~\Delta(t)=t\otimes t~,\\
& &\epsilon(e)=0~,~~~~\epsilon(t)=1~,\\
& &S(e)=-t^{-1}e~,~~~~S(t)=t^{-1}~.
\end{array}
\end{equation}
Similarly, in terms of a new set of definitions

\begin{equation}
\begin{array}{rcl}
& &\bar{t}=q^{2H}~,~~~~\bar{t}\bar{t}^{-1}=1=\bar{t}^{-1}\bar{t}~,\\
& &f=X^-q^{-H}~,
\end{array}
\end{equation}
the Hopf algebra $U_q^-(su(2))$ possesses the form

\begin{equation}
\begin{array}{rcl}
& &{\bar t}{\bar t}^{-1}=1={\bar t}^{-1}{\bar t}~,\\
& &{\bar t}f{\bar t}^{-1}=q^{-2}f~,\\
& &\Delta(f)=f\otimes {\bar t}^{-1}+1\otimes f~,~~~~
              \Delta({\bar t})={\bar t}\otimes {\bar t}~,\\
& &\epsilon(f)=0~,~~~~\epsilon({\bar t})=1~,\\
& &S(f)=-f{\bar t}~,~~~~S({\bar t})={\bar t}^{-1}~.
\end{array}
\end{equation}

\subsection{Quantum double}
For a Hopf algebra $(A,~m,~\Delta,~\eta,~\epsilon,~S)$ the iterated
comultiplication $\Delta^{(n-1)}:~A\to A^{\otimes n}$ is inductively
defined by  $\Delta^{(1)}=\Delta,~\Delta^{(n)}=(\Delta\otimes{\rm
id})\circ\Delta^{(n-1)}$.
Provided two Hopf algebras $A,~B$ and a non-degenerate bilinear form

\begin{equation}
\langle\cdots,\cdots\rangle:~A\times B\longrightarrow {\bf C}~,
\end{equation}
satisfying the following conditions

\begin{equation}\label{1.jc1}
\begin{array}{rcl}
& &\langle a^i,b_jb_k\rangle=\langle\Delta_A(a^i),b_j\otimes
b_k\rangle~,\\[2mm]
& &\langle a^ia^j,b_k\rangle=\langle a^j\otimes a^i,\Delta_B(b_k)
     \rangle ~,\\[2mm]
& &\langle 1^A,b_i\rangle=\epsilon_B(b_i)~,~~~~~
\langle a^i,1_B\rangle=\epsilon_A(a^i)~,\\[2mm]
& &\langle S_A(a^i),S_B(b_j)\rangle=\langle a^i,b_j\rangle~,
\end{array}
\end{equation}
where $\Delta_{A~(B)},~\epsilon_{A~(B)},~S_{A~(B)}$ are the structure
operations of the Hopf algebra $A~(B),~1^{A~(B)}$ is its unit,
we have the following theorem.\\
{\em There exists a unique Hopf algebra $D$ with the following properties:
\begin{itemize}
\item $D$ contains $A,~B$ as Hopf subalgebras, i.e., $A$ and $B$ are
subalgebras of $D$, and
$$\begin{array}{rcl}
& &\Delta_D(a^i)=\Delta_A(a^i)~,~~~\epsilon_D(a^i)=\epsilon_A(a^i)~,~~~
S_D(a^i)=S_A(a^i)~,~~~~~
\forall a^i\in A~,\\
& &\Delta_D(b_i)=\Delta_B(b_i)~,~~~\epsilon_D(b_i)=\epsilon_B(b_i)~,~~~
S_D(b_i)=S_B(b_i)~,~~~~~
\forall b_i\in B~.
\end{array}$$
\item If $\{a^\alpha\}$, $\{b_\beta\}$ are bases of $A$ and $B$, the
product $\{a^\alpha b_\beta\}$ is a basis of $D$.
\item For $a^i\in A$ and $b_j\in B$
$$b_ja^i=\sum\langle a^{i(1)},S(b_{j(1)})\rangle \langle
a^{i(3)},b_{j(3)}\rangle
a^{i(2)}b_{j(2)}~~~\in D~,$$
where
$$\begin{array}{rcl}
  & &\sum a^{i(1)}\otimes a^{i(2)} \otimes a^{i(3)} \equiv
\Delta_A^{(2)}(a^i)~,\\
  & &\sum b_{j(1)}\otimes b_{j(2)} \otimes b_{j(3)} \equiv
\Delta_B^{(2)}(b_j)~.
  \end{array}$$
The Hopf algebra $D$ is called as the quantum double of
$(A,~B,~\langle,\rangle)$ or simply, quantum double of $A$. \end{itemize}}

\subsubsection{$SU_q(2)$ as the quantum double}
For a nonzero complex number $q\in
{\bf C}^\times~({\bf C}\setminus\{0\}),~q^2\not=1$, consider the
two Hopf algebras $U_q^+\left(su(2)\right)$ and $U_q^-\left(su(2)\right)$.\\

Using the properties (\ref{1.jc1}) of the bilinear form
$\langle\cdots,\cdots\rangle$
repeatedly, we have

\begin{equation}
\begin{array}{rcl}
\langle te,\bar{t}\rangle&=&\langle e\otimes t,\bar{t}\otimes\bar{t}\rangle\\
                         &=&\langle e,\bar{t}\rangle \langle t,\bar{t}\rangle
{}~,\\
\langle te,\bar{t}\rangle&=&q^2\langle et,\bar{t}\rangle
                         =q^2\langle t\otimes e,{\bar t}\otimes{\bar t}\rangle
\\
                         &=&q^2\langle e,\bar{t}\rangle\langle t,{\bar
t}\rangle~.
\end{array}
\end{equation}
The equation

\begin{equation}
\begin{array}{rcl}
1&=&\langle t,1\rangle=\langle t,{\bar t}\cdot{\bar t}^{-1}\rangle\\
 &=&\langle t,\bar{t}\rangle\langle t,\bar{t}^{-1}\rangle~,
\end{array}
\end{equation}
leads to

\begin{equation}
\langle t,{\bar t}\rangle=\langle t,{\bar t}^{-1}\rangle^{-1}\not=0~.
\end{equation}
Since $q^2\not= 1$,  we must have

\begin{equation}
<e,{\bar t}>=0~.
\end{equation}
Similar calculations yield

\begin{equation}
<t,f>=0~.
\end{equation}
Making comparison of

\begin{equation}
\begin{array}{rcl}
\langle te,f\rangle&=&\langle e\otimes t,\Delta(f)\rangle
                      =\langle e,f\rangle\langle t,\bar{t}^{-1}\rangle+
                       \langle e,1\rangle\langle t,f\rangle \\
                   &=&\langle e,f\rangle \langle t,\bar{t}^{-1}\rangle ~,\\
\end{array}
\end{equation}
and

\begin{equation}
\begin{array}{rcl}
\langle te,f\rangle&=&q^2\langle et,f\rangle=q^2\langle t\otimes e,
                       \Delta(f)\rangle
                      =q^2\langle t,f\rangle\langle e,\bar{t}^{-1}\rangle+
                       q^2\langle t,1\rangle\langle e,f\rangle \\
                   &=&q^2\langle e,f\rangle  ~,\\
\end{array}
\end{equation}
where $\langle e,f\rangle\not= 0$ for non-degeneracy, we obtain

\begin{equation}
\langle t,{\bar t}\rangle =q^{-2}=\langle t,\bar{t}^{-1}\rangle^{-1}~.
\end{equation}
In general,

\begin{equation}
\begin{array}{rcl}
\langle e^mt^n,f^{m'}{\bar t}^{n'}\rangle&=&\langle e^m,f^{m'}\rangle\langle
t,{\bar t}\rangle^{nn'}\\[2mm]
            &=&\delta_{mm'}q^{m(m-1)/2}[m]!\langle e,f\rangle^m q^{-2nn'}~,
\end{array}
\end{equation}
where $[m]=\displaystyle\frac{q^m-q^{-m}}{q-q^{-1}}$, is a $q$-number,
$[m]!=[m][m-1]\cdots[2][1]$ is used. \\
The inner-product $\langle e,f\rangle$ can always be re-scaled by
the transformation $e \to ce$ ($c\in {\bf C}^\times$).  For latter
convenience  we  choose

\begin{equation}\label{1.j3.3}
\langle e,f\rangle=\frac{-1}{q-q^{-1}}~.
\end{equation}
Now we are ready to examine the quantum double $D$ of
$(U_q^+\left(su(2)\right),~U_q^-\left( su(2)\right),~\langle,
\rangle)$. From the above theorem an element of $D$ is a unique linear
combination of monomials
$e^mt^nf^{m'}{\bar t}^{n'}~(m,~m'\in Z_{\geq 0},~
n,~n'\in Z)$ and the commutators of $e$ with $t$ or $f$ with ${\bar t}$
can be computed easily.

We now compute $ft,~\bar{t}e~,$and $fe$. Since

\begin{equation}
\begin{array}{rcl}
& &\Delta^{(2)}(t)=t\otimes t\otimes t~,\\
& &\Delta^{(2)}(\bar{t})=\bar{t}\otimes \bar{t}\otimes\bar{t}~,
\end{array}
\end{equation}
and

\begin{equation}
\begin{array}{rcl}
& &\Delta^{(2)}(e)=e\otimes 1\otimes 1+t\otimes e\otimes 1+t\otimes t\otimes
e~,\\
& &\Delta^{(2)}(f)=f\otimes{\bar t}^{-1}\otimes {\bar t}^{-1}+1\otimes f\otimes
{\bar t}^{-1}+1\otimes 1\otimes f~,
\end{array}
\end{equation}
we get

\begin{equation}
\begin{array}{rcl}
ft&=&\langle t,S(f)\rangle\langle t,{\bar t}^{-1}\rangle t{\bar t}^{-1}\\
  & &    +\langle t,S(1)\rangle\langle t,\bar{t}^{-1}\rangle tf
      +\langle t,S(1)\rangle\langle t,f\rangle t\cdot 1\\
  &=&q^2 tf~,\\[3mm]
\bar{t}e&=&\langle e,S(\bar{t})\rangle\langle 1,{\bar t}\rangle 1\cdot t\\
        & & +\langle t,S(\bar{t})\rangle\langle 1,\bar{t}\rangle e\bar{t}
      +\langle t,S(\bar{t})\rangle\langle e,\bar{t}\rangle t\bar{t}\\
  &=&q^2 e\bar{t}~,
 \end{array}
 \end{equation}
 and

\begin{equation}
\begin{array}{rcl}
fe&=&\langle e,S(f)\rangle\langle 1,{\bar t}^{-1}\rangle 1\cdot \bar{t}^{-1}\\
        & & +\langle t,S(1)\rangle\langle 1,\bar{t}^{-1}\rangle ef
      +\langle t,S(1)\rangle\langle e,f\rangle t\cdot 1\\
  &=&-\langle e,f\rangle{\bar t}^{-1}+ef+\langle e,f\rangle t~.
\end{array}
\end{equation}
Using Eq.(\ref{1.j3.3}) we have

\begin{equation}
[e,f]=\frac{t-{\bar t}^{-1}}{q-q^{-1}}~.
\end{equation}
Setting $t=\bar{t}$ the quantum group $SU_q(2)$ follows

\begin{equation}\label{1.a1}
\begin{array}{l}
tet^{-1}=q^2e~,~~~~tft^{-1}=q^{-2}f~,\\[2mm]
[e,f]=\displaystyle\frac{t-t^{-1}}{q-q^{-1}}~,\\[2mm]
\Delta(e)=e\otimes 1+t\otimes e~,\\
\Delta(f)=f\otimes t^{-1}+1\otimes f~,\\
\Delta(t)=t\otimes t~,\\
\epsilon(e)=0=\epsilon(f)~,~~~~\epsilon(t)=1~,\\
S(e)=-t^{-1}e~,~~~~S(f)=-ft~,~~~~S(t)=t^{-1}~.
\end{array}
\end{equation}
In the literature there are available various versions of the quantum group.
Setting

\begin{equation}\label{1.a2}
e'=t^{-n}e~,~~~~f'=ft^n~,~~~~t'=t~,
\end{equation}
would result in

\begin{equation}\label{1.a3}
\begin{array}{l}
t'e'{t'}^{-1}=q^2e'~,~~~~t'f'{t'}^{-1}=q^{-2}f'~,\\[2mm]
[e',f']=\displaystyle\frac{t'-{t'}^{-1}}{q-q^{-1}}~,\\[2mm]
\Delta(e')=e'\otimes {t'}^{-n}+{t'}^{1-n}\otimes e'~,\\
\Delta(f')=f'\otimes {t'}^{n-1}+{t'}^n\otimes f~,\\
\Delta(t')=t'\otimes t'~,\\
\epsilon(e')=0=\epsilon(f')~,~~~~\epsilon(t')=1~,\\
S(e')=-t'^{-1}e'~,~~~~S(f')=-f't'~,~~~~S(t')=t'^{-1}~.
\end{array}
\end{equation}
It is easy to see from comparison between Eq.(\ref{1.a3}) and
Eq.(\ref{1.225}) that
$n=\displaystyle\frac{1}{2}$ case of Eq.(\ref{1.a3}) is the version in
Chevalley basis,

\begin{equation}\label{1.a4}
\begin{array}{l}
[H,X^\pm]=X^\pm~,\\[1mm]
[X^+,X^-]=[2H]~,\\[1mm]
\Delta(X^\pm)=X^\pm\otimes q^{-H}+q^H\otimes X^\pm~,\\
\Delta(H)=H\otimes 1+1\otimes H~,\\
\epsilon(X^\pm)=0=\epsilon(H)~,\\
S(X^\pm)=-q^{\mp 1}X^\pm~,~~~~S(H)=-H~.
\end{array}
\end{equation}

\subsubsection{$U_q(g)$ as the quantum double}
Now we come to discuss the general quantum group $U_q(g)$ as the quantum
double. Let $g$ denote an ordinary simple Lie algebra or untwisted affine
Kac-Moody algebra. The
corresponding generalized Cartan matrix $A=(a_{ij})_{1\leq i,j\leq l}$
($a_{ij}=\displaystyle\frac{1}{d_i}\langle\alpha_i,\alpha_j\rangle,~d_i=
\frac{1}{2}\langle\alpha_i,
\alpha_i\rangle)$ is symmetrizable in a sense that there exist nonzero $d_i$
satisfying $d_ia_{ij}=d_ja_{ji}$. For a nonzero complex number
$q$ ($q^{2d_i}\not=1$), in Chevalley basis, define
$U(g)$ as the associative ${\bf C}$-algebra with unity, with $3l$
generators
$$X_i^+~,~~~X_i^-~,~~~H_i~,~~~~~~(1\leq i\leq l)~.$$
Consider the standard Borel subalgebras of $U(g)$,\\
$U^+(g):~{\rm generated~by}~X^+_i~{\rm and}~H_i~~~(1\leq i\leq l)$,

\begin{equation}\label{1.6.15}
\begin{array}{l}
[H_i,H_j]=0~,\\[1mm]
[H_i,X_j^+]=a_{ij}X_i^+~,\\[1mm]
[X_i^+,X_j^+]=0~,~~~~~{\rm if}~a_{ij}=0~,\\[2mm]
\displaystyle\sum_{m=0}^{1-a_{ij}}(-1)^m\left(\begin{array}{c}
                                                   1-a_{ij}\\
                                                   m
                                                   \end{array}
\right)(X_i^+)^{1-a_{ij}-m}X_j^+(X_i^+)^m=0~,~~~~ (i\not=j)~,\\[2mm]
\Delta(H_i)=H_i\otimes 1+1\otimes H_i~,\\
\Delta(X_i^+)=X_i^+\otimes 1+1\otimes X_i^+~,\\
\epsilon(H_i)=0=\epsilon(X_i^+)~,\\
S(H_i)=-H_i~,~~~~S(X_i^+)=-X_i^+~;
\end{array}
\end{equation}
$U^-(g):~{\rm generated~by}~X^-_i~{\rm and}~H_i~~~~(1\leq i\leq l)$

\begin{equation}\label{1.6.151}
\begin{array}{l}
[H_i,H_j]=0~,\\[1mm]
[H_i,X_j^-]=-a_{ij}X_i^-~,\\[1mm]
[X_i^-,X_j^-]=0~,~~~~~{\rm if}~a_{ij}=0~,\\[2mm]
\displaystyle\sum_{m=0}^{1-a_{ij}}(-1)^m\left(\begin{array}{c}
                                                   1-a_{ij}\\
                                                   m
                                                   \end{array}
\right)(X_i^-)^{1-a_{ij}-m}X_j^-(X_i^-)^m=0~,~~~~ (i\not=j)~,\\[2mm]
\Delta(H_i)=H_i\otimes 1+1\otimes H_i~,\\
\Delta(X_i^-)=X_i^-\otimes 1+1\otimes X_i^-~,\\
\epsilon(H_i)=0=\epsilon(X_i^-)~,\\
S(H_i)=-H_i~,~~~~S(X_i^-)=-X_i^-~.
\end{array}
\end{equation}
The quantization of $U^+(g)$ and $U^-(g)$ yields  the
multiplication $m$, comultiplication $\Delta$, counit $\epsilon$ and antipode
$S$ for $U^+_q(g)$
and $U^-_q(g)$ as,

\begin{equation}
\begin{array}{l}
\Delta(H_i)=H_i\otimes 1+1\otimes H_i~,\\
\Delta(X_i^+)=X_i^+\otimes q_i^{-H_i}+q_i^{H_i}\otimes X_i^+~,\\
\epsilon(X_i^+)=0=\epsilon(H_i)~, \\
S(X_i^+)=-q_i^{-1}X_i^+~,~~~~S(H_i)=-H_i~,
\end{array}
\end{equation}
and

\begin{equation}
\begin{array}{l}
\Delta(H_i)=H_i\otimes 1+1\otimes H_i~,\\
\Delta(X_i^-)=X_i^-\otimes q_i^{-H_i}+q_i^{H_i}\otimes X_i^-~,\\
\epsilon(X_i^-)=0=\epsilon(H_i)~, \\
S(X_i^-)=-q_iX_i^-~,~~~~S(H_i)=-H_i~,
\end{array}
\end{equation}
where $q_i=q^{d_i}$.\\
It is convenient to introduce the operators $e_i,~t_i,~$ and $f_i,~
\bar{t}_i~~~~(1\leq i\leq l)$ as

\begin{equation}
\begin{array}{l}
e_i=q_i^{H_i}X_i^+~,~~~~t_i=q_i^{2H_i}~,\\
f_i=X_i^-q_i^{-H_i}~,~~~~\bar{t}_i=q_i^{2H_i}~.
\end{array}
\end{equation}
To avoid confusion the new operators $x_i$ and $y_i$ are introduced by

$$\begin{array}{l}
x_i=e_i~,~~~~(1\leq i\leq l)~,\\
y_i=f_i~,~~~~(1\leq i\leq l)~.\end{array}$$
For $\beta\in I$ ($I=\displaystyle\bigoplus_{i=1}^lZ_{\geq
0}\alpha_i$) define

$$U_\beta^+={\rm
linear~span~of~}\{x_{i_1},~\cdots,~x_{i_r}|\alpha_{i_1}
  +\cdots\alpha_{i_r}=\beta\}~,~~~(\beta\in I)~,$$
$$U_{-\beta}^-={\rm linear~span~of~}\{y_{i_1},~\cdots,~y_{i_r}|\alpha_{i_1}
  +\cdots\alpha_{i_r}=\beta\}~,~~~(\beta\in I)~.$$
Then, $U_{\pm \beta}^\pm$ is finite dimension for each $\beta\in I$.
Similar to the case for $SU_q(2)$, there is a general theorem for the
quantum double $D$ of $U_q^+(q)$.
{\em
\begin{itemize}
\item There exists a unique non-degenerate bilinear pairing $\langle\cdots,
      \cdots\rangle:~  U_q^+(g)\times U_q^-(g)\to {\bf C}$ so that
     $$\begin{array}{l}
      \langle t_i,\bar{t}_j\rangle=q_i^{-a_{ij}}~,~~~\langle
t_i,f_i\rangle=0~,\\[2mm]
      \langle e_i,\bar{t}_i\rangle=0~,~~~\langle e_i,f_i\rangle=-
      \delta_{ij}\displaystyle\frac{1}{q_i-q_i^{-1}}~.
      \end{array}$$

\item The restriction of the pairing $\langle\cdots,\cdots\rangle$ to
$U_{\beta}^+\times
      U_{-\beta}^-$ is non-degenerate for each $\beta\in I$.
\end{itemize} }
It can be shown that $U_q(g)$ follows from the quantum double
$D$ of $U_q^+(g)$ by setting  $t_i={\bar t}_i$

\begin{equation}
D/\langle t_i-{\bar t}_i\rangle\simeq U_q(g)~.
\end{equation}
The resultant  quantum group $U_q(g)$ has the form  \cite{Drinfeld85}--
\cite{Lusztig90c}

\begin{equation}\label{1.6.152}
\begin{array}{rcl}
& &t_it_j=t_jt_i~,~~~~~t_it_i^{-1}=t_i^{-1}t_i=1~,\\
& &t_ie_jt_i^{-1}=q_i^{a_{ij}}e_j~,~~~~t_if_jt_i^{-1}=q_i^{-a_{ij}}f_j~,\\[2mm]
& &[e_i,f_j]=\delta_{ij}\displaystyle\frac{t_i-t_i^{-1}}
               {q_i-q_i^{-1}}~,\\[3mm]
& &\displaystyle\sum_{m=0}^{1-a_{ij}}(-1)^m\left[\begin{array}{c}
                                                   1-a_{ij}\\
                                                   m\end{array}
\right]_{q_i}(e_i)^{1-a_{ij}-m}e_j(e_i)^m=0~,~~~~
                                          (i\not=j)~,\\
& &\displaystyle\sum_{m=0}^{1-a_{ij}}(-1)^m\left[\begin{array}{c}
                                                   1-a_{ij}\\
                                                   m\end{array}
\right]_{q_i}(f_i)^{1-a_{ij}-m}f_j(f_i)^m=0~,~~~~
                                          (i\not=j)~,\\
& &\Delta(t_i)=t_i\otimes t_i~,\\
& &\Delta(e_i)=e_i\otimes 1+t_i\otimes e_i~,~~~~
   \Delta(f_i)=f_i\otimes t_i^{-1}+1\otimes f_i~,\\
& &\epsilon(e_i)=0=\epsilon(f_i)~,~~~~\epsilon(t_i)=1~,\\
& &S(e_i)=-t_i^{-1}e_i~,~~~~S(f_i)=-f_it_i^{-1}~,~~~~S(t_i)=t_i^{-1}~,
\end{array}
\end{equation}
where we have used the notation $\displaystyle\left[\begin{array}{c}
m\\
n\end{array}\right]_q=\frac{[m]_q!}{[n]_q![n-m]_q!}$. \\
In Chevalley basis, the quantum group $U_q(g)$ can be written as

\begin{equation}
\begin{array}{l}
[H_i,H_j]=0~,\\[1mm]
[H_i,X_j^\pm]=\pm a_{ij}X_i^\pm~,\\[1mm]
[X_i^+,X_j^-]=\delta_{ij}[H_i]_{q_i}~,\\[1mm]
[X_i^\pm,X_j^\pm]=0~,~~~~~{\rm if}~a_{ij}=0~,\\[2mm]
\displaystyle\sum_{m=0}^{1-a_{ij}}(-1)^m\left[\begin{array}{c}
                                                   1-a_{ij}\\
                                                   m
                                                   \end{array}
\right]_{q_i}(X_i^\pm)^{1-a_{ij}-m}X_j^\pm(X_i^\pm)^m=0~,~~~~
(i\not=j)~,\\[2mm]
\Delta(H_i)=H_i\otimes 1+1\otimes H_i~,\\
\Delta(X_i^\pm)=X_i^\pm\otimes q_i^{H_i}+q_i^{-H_i}\otimes X_i^+~,\\
\epsilon(H_i)=0=\epsilon(X_i^\pm)~,\\
S(H_i)=-H_i~,~~~~S(X_i^\pm)=-q^{-\rho} X_i^\pm q^{\rho}~,
\end{array}
\end{equation}
where $\rho=\displaystyle\sum_i H_i$.

\subsubsection{Universal $R$-matrix}
For the sake of definiteness, let us assume that $A,~B$ are finite-dimensional
Hopf algebras, and $\{a^i\}$ and $\{b_i\}$ are the bases of the
Hopf algebras $A$ and $B$ respectively. With respect to the non-degenerate
bilinear form $\langle\cdots,\cdots\rangle$,

$$\langle a^i,b_j\rangle=\delta_{ij}~.$$
Suppose that there exists another set of similar bases of the Hopf
algebras $A$ and $B$
 $\{{a'}^{i}\}$ and $\{b'_{i}\}$

$${a'}^{i}=\sum_k a^kM_{ki}~,~~~~b'_{i}=\sum_lN_{il}b_l~,$$

\begin{equation}
\delta_{ij}=\langle {a'}^{i},b'_{j}\rangle=\sum_{kl}M_{ki}N_{jl}\langle a^k,
b_l\rangle=\sum_kN_{jk}M_{ki}~,
\end{equation}
i.e., $N$ is the inverse of $M$, $N=M^{-1}$.\\
Then we get

\begin{equation}
\sum_i{a'}^{i}\otimes b'_{i}=\sum_{ikl}M_{ki}N_{il}a^k\otimes b_l
=\sum_ia^i\otimes b_i~,
\end{equation}
and the element,

\begin{equation}
{\cal R}=\sum_i a^i\otimes b_i~~~\in A\otimes B\subset D\otimes D~,
\end{equation}
is {\it independent} of choice of the dual bases. The element ${\cal
R}$ is the so-called universal $R$ matrix. There is a theorem for the
universal $R$ matrix.
{\em
\begin{itemize}
\item ${\cal R}\Delta(x)=\Delta'(x){\cal R}~,~~~~\forall x\in D~,~~~~\Delta'
      =P\circ\Delta~$,
\item $(\Delta\otimes{\rm id}){\cal R}={\cal R}_{13}{\cal R}_{23}~,~~~~
      ({\rm id}\otimes\Delta){\cal R}={\cal R}_{13}{\cal R}_{12}~.$
\item $(\epsilon\otimes{\rm id}){\cal R}=1=({\rm id}\otimes \epsilon){\cal R}~,
{}~~~~(S\otimes{\rm id}){\cal R}={\cal R}^{-1}=({\rm id}\otimes S^{-1}){\cal
R}~,$
\end{itemize}}
where
$$\begin{array}{rcl}
& &{\cal R}_{12}=\displaystyle\sum_i a^i\otimes b_i\otimes 1~,\\
& &{\cal R}_{13}=\displaystyle\sum_i a^i\otimes 1\otimes b_i~,\\
& &{\cal R}_{23}=\displaystyle\sum_i 1\otimes a^i\otimes b_i~.
\end{array}$$
In particular, ${\cal R}$ is invertible in $D\otimes D$. The pair $(D,{\cal
R})$
is the so-called  quasi-triangular Hopf algebra. There is a
relationship of the quasi-triangular Hopf algebra to the YBE. \\
Making Comparison of

\begin{equation}
\begin{array}{rcl}
({\rm id}\otimes\Delta'){\cal R}&=&{\cal R}_i^{(1)}\otimes\Delta'({\cal
R}_i^{(2)})
                           ={\cal R}_i^{(1)}\otimes{\cal R} \Delta({\cal R}
                           _i^{(2)}){\cal R}^{-1}\\
                         &=&({\rm id}\otimes{\cal R})\left({\cal R}_i^{(1)}
                         \otimes\Delta
                           ({\cal R}_i^{(2)})\right)({\rm id}\otimes{\cal
R}^{-1})
                           ={\cal R}_{23}({\rm id}\otimes\Delta){\cal R}{\cal
R}
                           _{23}^{-1}\\
                           &=&{\cal R}_{23}{\cal R}_{13}{\cal R}_{12}
                           {\cal R}_{23}^{-1}
  \end{array}
  \end{equation}
with

\begin{equation}
\begin{array}{rcl}
({\rm id}\otimes\Delta'){\cal R}&=&({\rm id}\otimes P)({\rm
id}\otimes\Delta){\cal R}
                            =({\rm id}\otimes P){\cal R}_{13}{\cal R}_{12}\\
                         &=&{\cal R}_{12}{\cal R}_{13}~,
\end{array}
\end{equation}
yields

\begin{equation}
{\cal R}_{12}{\cal R}_{13}{\cal R}_{23} ={\cal R}_{23} {\cal R}_{13} {\cal
R}_{12}~.
\end{equation}
This is precisely the YBE. \\
Now we are discussing the explicit form of the universal
$R$ matrix for $SU_q(2)$. Notice that the Hopf algebras
$A=U_q^+\left(su(2)\right)$ and
$B=U_q^-\left(su(2)\right)$ are not finite dimension. However if we suppose
that
$q$ is a primitive $p$-th root of unity for an odd integer $p>1$, it is not
difficult to verify that

$$e^p=0=f^p~,~~~~t^p=1=\bar{t}^p~,$$
so that the quotient algebras

\begin{equation}
\bar{A}=A/\langle e^p,t^p-1\rangle~,~~~~\bar{B}=B/\langle
f^p,\bar{t}^p-1\rangle~,
\end{equation}
are finite dimension, and

$$a^{mn}=e^mt^n~,~~~~b_{mn}=f^m\bar{t}^n~,~~~~(0\leq m,n\leq p)~,$$
are linear bases.\\
Since

\begin{equation}
\langle a^{mn},b_{m'n'}\rangle=\delta_{mm'}q^{m(m-1)/2}[m]!\left(\frac{-1}
{q-q^{-1}}\right)^mq^{-2nn'}~,
\end{equation}
$\{b_{mn}\}$ is not a dual basis of $\{a^{mn}\}$. The basis dual to
$\{a^{mn}\}$ should be given by

\begin{equation}
a_{mn}^*=\frac{q^{-m(m-1)/2}}{[m]!}\left(-(q-q^{-1})\right)^m\frac{1}{p}
\sum_{n'=0}^{p-1}q^{2nn'}b_{mn'}~.
\end{equation}
As a result, the explicit form of the universal $R$ matrix is

\begin{equation}
\begin{array}{rcl}
& &{\cal R}=\bar{\cal C}{\cal K}~,\\
& &\bar{\cal C}=\displaystyle\sum_{m=0}^{p-1}
   \frac{q^{-m(m-1)/2}}{[m]!}\left(-(q-q^{-1})\right)^m
   e^m\otimes f^m~,\\
& &{\cal K}=\displaystyle\frac{1}{p}\sum_{n,n'=0}^{p-1}q^{2nn'}t^n
   \otimes\bar{t}^{n'}~.
\end{array}
\end{equation}
For the pair $(\pi,V),~(\pi',V')$ of representations of $D$, if

$$t\cdot v=q^\mu v~,~~~~~\bar{t}\cdot v'=q^\nu v'~,~~~~~(v\in V,~v'
\in V',~\mu,\nu\in Z) ~,$$
we have

\begin{equation}
\begin{array}{rcl}
{\cal K}\cdot v\otimes v'&=& q^{-\mu\nu/2}v\otimes v'\\
                         &=&q^{-2H\otimes H}v\otimes v'~.
\end{array}
\end{equation}
Therefore we can write the universal $R$ matrix as

\begin{equation}
{\cal R}=\sum_{m=0}^{p-1}\frac{q^{-m(m-1)/2}}{[m]!}\left(-(q-q^{-1})\right)^m
(e^m\otimes f^m)\cdot q^{-2H\otimes H}~.
\end{equation}
In terms of the $q$-exponential
$$\exp_qz=\sum_{m=0}^\infty\frac{q^{-m(m-1)/2}}{[m]!}z^m~,$$
with the condition that $e^p=0=f^p$,
the universal $R$ matrix can be formally written as

\begin{equation}
{\cal R}=\exp_q\left(-(q-q^{-1})e\otimes f\right)q^{-2H\otimes H}~.
\end{equation}
This expression for universal $R$ matrix is independent of $p$, and in
fact it is also valid for generic $q$.

The explicit form of the universal $R$ matrix except for the case
$SU_q(2)$ is more complicated \cite{Reshetikhin91}--\cite{Drinfeld90}.

\section{Representation theory}

As the same as the representation theory of group, a linear representation
of a quantum group $A$ is a pair $(\pi,V)$ consisting of a vector space $V$
and a homomorphism

$$\pi:~~A\to {\rm End}(V)~,~~~~~\pi(ab)=\pi(a)\pi(b)~,$$
where, as usual, End$(V)$ denotes the space of all linear maps from $V$ to
itself.
We often drop $\pi$ and write $\pi(a)v$ as $av$ ($a\in A,~v\in V)$. A
representation $(\pi,V)$ of a quantum group $A$ is called irreducible if $V$
has not  non-trivial submodules, i.e., if $W(\subset V)$ is a subspace such
that
$A\cdot W\subset W$, either $W=\{0\}$, or $W=V$. If the vector space $V$
is finite dimension, the representation $(\pi,V)$ is a finite-dimensional
representation.  For two representations $(\pi,V),~(\pi',V')$, an intertwiner
is the map $\phi:~V\to V'$, which commutes with the action of $A$, namely
$\phi\circ\pi(a)=\pi'(a)\circ\phi$ for all $a\in A$ (Fig.9). We say
that $(\pi,V)$ and $(\pi',V')$ are equivalent if there exists an intertwiner
that is an isomorphism. In other words, if there exist such bases of
$V,~V'$ that the matrices representing $\pi(a)$ and $\pi'(a)$ are
identical  for all $a\in A$.
In this section we discuss the inequivalent irreducible finite-dimensional
representations of quantum group. When dealing with representations , we
only use the algebra structure, ignoring  the Hopf algebra structure
$(\Delta,\epsilon,S)$.
The Hopf algebra structure enters into the picture when the relationship
between various representations is to be involved.

\setlength{\unitlength}{5pt}
\begin{picture}(50,25)
\put(30,18){\vector(1,0){15}}
\put(27,17){$V$}
\put(46,17){$V'$}
\put(36.5,19){$\phi$}
\put(28,16){\vector(0,-1){10}}
\put(47,16){\vector(0,-1){10}}
\put(30,4){\vector(1,0){15}}
\put(27,3){$V$}
\put(46,3){$V'$}
\put(36.5,5){$\phi$}
\put(22,10){$\pi(a)$}
\put(48,10){$\pi'(a)$}
\end{picture}

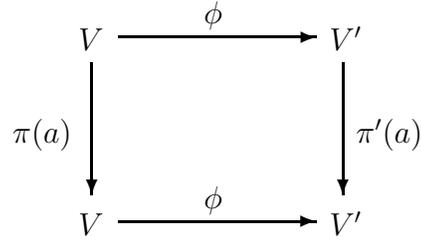
\hspace{7em}Fig.9. Intertwiner of representations.

\subsection{Representations for generic $q$}
\subsubsection{Representations of $SU_q(2)$}

It is well-know that the quantum group $SU_q(2)$ is generated by the
generators: $e,~f$ and $t^\pm$, and is subject to

\begin{equation}\label{2.cha1}
\begin{array}{rcl}
& &tt^{-1}=1=t^{-1}t~,\\[2mm]
& &tet^{-1}=q^{2}e~,~~~~tft^{-1}=q^{-2}f~,\\[2mm]
& &[e,f]=\displaystyle\frac{t-t^{-1}}{q-q^{-1}}~.
\end{array}
\end{equation}

Defining $v$ be the non-zero eigenvector of $e$ in the representation
$(\pi,V)$

$$ev=\lambda v,~~~~~v\in V,~\lambda\in {\bf C}~,$$
and making use of Eq.(\ref{2.cha1}), we obtain

$$e(t^{\pm 1}v)=q^{\mp 2}\lambda(t^{\pm 1}v)~.$$
Since $t$ is invertible, $t^{\pm 1} v\not=0$, and  $q^{\mp 2}\lambda$
is also an eigenvalue of $e$ with the eigenstate $t^{\pm 1} v$.
If $\lambda\not=0$, we would have infinitely many eigenvalues and
eigenstates
$(\lambda,v),~(q^{\mp 2}\lambda,t^{\pm 1} v),~(q^{\mp 4}\lambda,t^{\pm 2}v),
{}~\cdots$. Since $q$ is not a root of unity, both the eigenvalues and the
eigenstates are distinct to each other.
This fact would otherwise be contradictory to the finite dimensionality of
$V$ for the finite-dimensional representation $(\pi,V)$, unless
$\lambda=0$.  The discussion for $f$ is similar. \\
Thus,  {\em if $(\pi,V)$ is a finite-dimensional representation of $SU_q(2)$,
then $e,~f$ act nilpotently on $V$}.

Let  $(\pi,V)$ be finite dimension and irreducible, and set $V_{\rm
high}=\{v\in V|ev=0\}$. By the above statement we know that $V_{\rm
high}\not=\{0\}$. Since $tV_{\rm high}=V_{\rm high}$,  we can find
such a non-zero vector $v_0\in V_{\rm high}$ that $tv_0=\lambda v_0$
with some nonzero $\lambda$ ($\in {\bf C}^\times$). $f$ also acts
nilpotently, so
that we can find the smallest non-negative integer $l$ satisfying
$$f^jv_0\not=0~~~(0\leq j\leq l),~~~~f^{l+1}v_0=0~.$$
Setting

\begin{equation}
f^{(k)}=\frac{f^k}{[k]!}~,~~~~~~v_k=f^{(k)}v_0~,
\end{equation}
we obtain

\begin{equation}\label{2.chang1}
\begin{array}{rcl}
[e,f^{(k)}]&=&\displaystyle\frac{1}{[k]}[e,f]f^{(k-1)}+
                    \frac{1}{[k]}f[e,f^{(k-1)}]\\[4mm]
                 &=&\displaystyle\frac{[k-1]!}{[k]!}[e,f]f^{(k-1)}
                    +\frac{[k-2]!}{[k]!}f[e,f]f^{(k-2)}\\[4mm]
                 & & \displaystyle  +\frac{[k-2]!}{[k]!}f^2[e,f^{(k-2)}]\\[4mm]
                 &\vdots&\\[3mm]
                 &=& \displaystyle\sum_{i=1}^k
                    \frac{[k-i]!}{[k]!}\left(f\right)^i[e,f]f^{(k-i)}
                    \\[4mm]
                 &=&\displaystyle\frac{f^{(k-1)}}{[k](q-q^{-1})}
                    \sum_{i=1}^k(q^{-2(k-i)}t-q^{2(k-i)}t^{-1})\\[4mm]
                 &=&f^{(k-1)}\displaystyle\frac{q^{-k+1}t-q^{k-1}t^{-1}}
                 {q-q^{-1}}~~~~(k\geq 1)~,
\end{array}
\end{equation}
where we have used Eq.(\ref{2.cha1}) repeatedly.\\
Making use of Eq.(\ref{2.chang1}), we have

\begin{equation}
0=[e,f^{(l+1)}]v_0=f^{(l)}\frac{q^{-l}t-q^{l}t^{-1}}{q-q^{-1}}v_0
=\frac{q^{-l}\lambda-q^{l}\lambda^{-1}}{q-q^{-1}}v_l~.
\end{equation}
Since $v_l\not=0$, we get $\lambda=\pm q^l$. Use is made of the
algebra  relations, Eq.(\ref{2.cha1}), and the definition for $v_k$, to
get

\begin{equation}
\begin{array}{rcl}
& &tv_k=\pm q^{l-2k}v_k~,\\[2mm]
& &ev_k=[e,f^{(k)}]v_0=\pm[l-k+1]v_{k-1}~,\\[2mm]
& &fv_k=\displaystyle\frac{f^{k+1}}{[k]!}v_0=[k+1]v_{k+1}~,
\end{array}
\end{equation}
where $v_{-1}=0,~v_{l+1}=0$.\\
The above equations can be written in the standard notation as

\begin{equation}\label{2.j1.7}
\begin{array}{rcl}
& &\pi^\sigma_l(e)v_k^l=\sigma[l-k+1]v_{k-1}^l~,\\[2mm]
& &\pi^\sigma_l(f)v_k^l=[k+1]v_{k+1}^l~,\\[2mm]
& &\pi^\sigma_l(t)v_k^l=\sigma q^{l-2k}v_{k}^l~,
\end{array}
\end{equation}
where $l\in Z_{\geq 0}$ and $\sigma=\pm 1$, $v_j^l=0$ if $j>l$ or $j<0$.
Therefore, the representation ($\pi^\sigma_l,V^\sigma(l)$) is a
representation of dimension $(l+1)$.
Obviously $(\pi^+,V^+(l)$) is a $q$-analog to the spin $l/2$ representation of
$su(2)$. However, $(\pi^-,V^-(l)$) has no classical counterpart. This is
attribute to
the presence of an automorphism $\kappa$ of the quantum group $SU_q(2)$  given
by
$$\kappa(e)=-e,~~~ \kappa(f)=f,~~~\kappa(t)=-t~.$$
If $(\pi,V)$ is a representation, one can always ``twist it by sign'' to get
another representation  $(\pi\circ\kappa,V)$, e.g.,
$\pi^-_l=\pi^+_l\circ\kappa$. We
shall simply write $(\pi_l^+,V^+(l))$ as $(\pi_l,V(l))$. \\
For the representation $(\pi,V)$, we have
already found a non-zero submodule $V^\sigma(l)=\displaystyle\bigoplus^l_{k=0}
{\bf C}v_k^l\subset V$. Because $V$ is irreducible, we must have
$V^\sigma(l)=V$. \\
Thus {\em an irreducible finite-dimensional representation of
$SU_q(2)$ is equivalent to that of $(\pi_l^\sigma,V^\sigma(l))$.}

 If $(\pi,V)$ is a representation, the eigenspace of $t$, $V_\mu=\{v\in
V|tv=q^\mu v\}(\mu\in Z)$ is called as the weight space of weight $\mu$,
and $v\in V_\mu$ is called as a weight vector of weight $\mu$. If $V$ is a
direct sum of weight spaces $\displaystyle\bigoplus_\mu V_\mu$, so is a
submodule $W$: $W=\displaystyle\bigoplus_\mu(W\bigcap
V_\mu)$. In other words, if $w=\displaystyle\sum_\mu
v^{(\mu)},~v^{(\mu)}\in V_\mu$, it is true that $v^{(\mu)}\in W$ for all
$\mu$. In fact, applying $t^k$ we have

\begin{equation}
t^kw=\sum_\mu q^{k\mu}v^{(\mu)}~,~~~~k=0,1,2,\cdots~.
\end{equation}
Treating $v^{(\mu)}$ as unknowns and $t^kw\in W$ as knowns, we can solve
the linear equation set, whose coefficient matrix is a Verdermonde type
$(q^{k\mu})$, and non-singular. This means $v^{(\mu)}\in W$.

Obviously $V=V^\sigma(l)$  is a direct sum of weight spaces
$V_{l-2k}={\bf C}v_k^l$, and $V^\sigma(l)={\bf C}v_0^l\oplus
{\bf C}v_1^l\oplus\cdots\oplus
{\bf C}v_l^l$. Acting $e$ on it, we find that all $v_j^l$ belong to $W$. It
follows that $W=V$. Therefore {\em all representations
$(\pi^\sigma_l,V^\sigma(l))~(l=0,1,2,\cdots,~\sigma=\pm)$ are
irreducible}.
Finally {\em $(\pi_{l}^\pm,V^\pm(l))$ are
inequivalent,}  because the sets of eigenvalues of $t$ $\{\pm q^l,\pm
q^{l-2},\cdots,\pm q^{-l}\}$ are distinct to each other.

{}From the above discussion the  classification theorem follows.
{\em
\begin{itemize}
\item All representations
$(\pi^\sigma_l,V^\sigma(l))~(l=0,1,2,\cdots,~\sigma=\pm)$ are
irreducible and  inequivalent to each other.
\item An irreducible finite-dimensional representation of $SU_q(2)$ is
equivalent to that of $(\pi^\sigma_l,V^\sigma(l))$.
\end{itemize}}

Generally, a representation ($\pi,V)$ with dim$V<\infty$ is reducible,
i.e., there is a submodule $W$ in $V$. Without loss of generality, one may
suppose that the submodule $W$ be
irreducible, and  it is the highest weight module with highest weight $l$.
For a reducible $W$, there exists a submodule $W_1\subset W
\subset V$. Repeating this procedure, always one can get an irreducible
submodule $W\subset V$. An irreducible finite-dimensional
representation is characterized by a Casimir operator $C$

\begin{equation}
C=\frac{(qt-1)(1-q^{-1}t^{-1})}{(q-q^{-1})^2}+fe~,
\end{equation}
with the non-zero eigenvalue $\displaystyle\frac{(\sigma
q^{l+1}-1)(1-\sigma^{-1}q^{-(l+1)})}
{(q-q^{-1})^2}$ (here again, $q$ is not a root of unity).
It is easy to verify  that the Casimir operator $C$ is an element of the
center of $SU_q(2)$.  In other words, $C$ commutes with $e,~f$ and $t^{\pm
1}$. Assuming

\begin{equation}
C'=C-\frac{(\sigma'q-1)(1-(\sigma')^{-1}q^{-1})}{(q-q^{-1})^2}~,
\end{equation}
and acting $C'$ on an irreducible representation, we have
$$\frac{\sigma q^{l+1}+\sigma^{-1}q^{-(l+1)}-\sigma' q-(\sigma')^{-1}q^{-1}}
{(q-q^{-1})^2}~.$$
This eigenvalue is zero, if and only if
$$\frac{q^{l+1}+q^{-(l+1)}}{q+q^{-1}}=\frac{\sigma'}{\sigma}~\in\{1,-1\}~.$$
However, the following equation is not true
$$\frac{q^{l+1}+q^{-(l+1)}}{q+q^{-1}}=1~\longleftrightarrow~
q(q^l-1)=q^{-(l+1)}(q^l-1)~,$$
if the dimension of the representation is greater than $2$
($l\geq 1$ as the dimension of the representation is $(l+1)$).
Similarly, the following equation is not true
$$\frac{q^{l+1}+q^{-(l+1)}}{q+q^{-1}}=-1~\longleftrightarrow~
q(q^l+1)=q^{-(l+1)}(q^l+1)~,$$
if $q$ is not a root of unity. Therefore {\em $C'$ acts in every irreducible
finite-dimensional representation by a scalar, if the dimension of the
representation is greater than $2$}.

If $W$ is codimension $1$ and dim$W\geq 2$, we consider the
representation of $SU_q(2)$ in $V/W$, which is one dimensional: $e,~f$ act
by $0$
and $t$ by a scalar $\sigma$. The acting of $C'$ in $V$ puts $W$ into $W$ by a
nonzero scalar, and in fact, it puts $V$ into $W$ as it acts by $0$ in $V/W$
(by choice of $\sigma'$). Thus there exists such a $1$-dimensional submodule
$W' =$ker$C'$, that $V=W\oplus W'$.
Furthermore, $W'$ is invariant under $SU_q(2)$, because $C'$ belongs to the
center.

If dim$W$=1 and dim$V$=2, the only no-trivial case is for
$\sigma=\sigma'$ with the weight of the representation in $W$ equal to the
weight of the
representation in $W/V$. Thus,  there exists a basis $(v_1,v_2)$ in $V$, in
which $t$ has matrix: $\displaystyle\left(\begin{array}{cc}
                                         \sigma&\alpha\\
                                         0     &\sigma\end{array}\right),~~
                                         \alpha\in {\bf C}$,
and
$$t\left(ev_1\right)=\sigma q^2ev_1~,$$
leading to  $t(e v_1)=0$;
$$t\left(ev_2\right)= q^2e(\sigma v_2+\alpha v_1)=\sigma q^2
ev_2~,$$
leading to $t(ev_2)=0$ and $e=0$.\\
Similarly, $f=0$. Then the relation $[e,f]=\displaystyle
\frac{t-t^{-1}}{q-q^{-1}}$ implies  $t=t^{-1}$, and $\alpha=0$, i.e.,
$t$ is diagonolized. Therefore we obtain $V=W\oplus W'$.

For $W$ with arbitrary codimension, define

\begin{equation}
\begin{array}{rcl}
& &{\cal V}=\{f\in{\cal L}(V,W)\vert f_{|W}~{\rm is~a~scalar~operator}\}~,\\
& &{\cal W}=\{f\in{\cal L}(V,W)\vert f_{|W}=0\}~.
\end{array}
\end{equation}
${\cal W}$ is a submodule of codimension $1$ in ${\cal V}$.
Let $V^*={\rm Hom}(V,C)$ be the dual vector space of $V$. For $v\in V$ and
$v^*\in V^*$, we write $v^*(v)$ as $\langle v^*,v\rangle$. If
$f\in$End$(V)$, $^tf\in$End$(V^*)$ is defined  by
$\langle^tf(v^*),v\rangle=\langle v^*,f(v)\rangle$ as usual. Therefore
$^t(f\circ g)=^tg \circ ^tf$. The dual representation
$(\pi^*, V^*)$ is defined by $\pi^*=^t\pi\circ S$:
$$SU_q(2)\stackrel{S}{\to}SU_q(2)\stackrel{^t\pi}{\to}{\rm End}(V^*)~.$$
We make
$SU_q(2)$ act in ${\cal L}(V,W)$ after identifying ${\cal L}(V,W)$
with $W\otimes V^*$ and putting: $\bar{\pi}=(\pi\otimes\pi^*)\circ\Delta$.
For a fixed  basis $(y_1,~y_2,~\cdots,~y_p)$ of $W$, we can write
$\phi\in {\cal L}(V,W)$  as $\phi=\sum y_i\otimes x_i^*$ for some
$x_i^*\in V^*$ in a unique manner. ${\cal V}$ and ${\cal W}$ are invariant
under $\bar{\pi}$.
Repeating the arguments for $W$ with codimension $1$, we come to know that
there exists such a submodule ${\cal W'}$  that ${\cal V=W+W'}$. Let
$\phi=\sum y_i\otimes x_i^*$ be a nonzero element in ${\cal W'}$, it acts in
$W$ by a
nonzero scalar and Ker$\phi=\cap_i$Ker$x_i^*$ verifies $V=$Ker$\phi+W$.
Also ,Ker$\phi$ is invariant under $SU_q(2)$ (because ${\cal W'}$ is such
an invariant).\\ The above lead to another important theorem for
the representations of quantum group, the complete reducibility theorem.\\
{\em  A finite-dimensional
representation $(\pi,V)$ of $SU_q(2)$ is completely reducible.}\\
This is equivalent to the statement that $V$ is a direct sum of
irreducible
representations. Therefore all  finite-dimensional representations of
$SU_q(2)$ are direct sums of some copies of the
$(\pi^\sigma_{l_i},V^\sigma(l_i))$.

A  quantum group $A$ is a Hopf algebra. The comultiplication $\Delta$ maps
$A$ into $A\otimes A$ and still satisfies the same algebra relations. As a
result, one may consider tensor products of irreducible finite-dimensional
representations
of $A$. For two representations $(\pi_i,V_i)~(i=1,~2)$, we define the tensor
product representation by the composition of the maps
$$A\stackrel{\Delta}{\to}A\otimes
A\stackrel{\pi_1\otimes\pi_2}{\longrightarrow}
{\rm End}(V_1)\otimes {\rm End}(V_2) \subset {\rm End}(V_1\otimes V_2)~. $$
The tensor product of the two highest weight representations constitutes a new
representation by the action of  comultiplication. Generally, the
new representation is reducible. Now, we are discussing the tensor products
for
$SU_q(2)$ in more detail. Provided the two highest weight representations
$$V(m)={\bf C}v_0^m\oplus {\bf C}v_1^m\oplus \cdots\oplus {\bf C}v_m^m~,$$
and
$$V(n)={\bf C}v_0^n\oplus {\bf C}v_1^n\oplus \cdots\oplus {\bf C}v_n^n~,$$
we investigate the irreducible decomposition of  $V=V(m)\otimes V(n)$. Let

\begin{equation}
w^l=\sum_{j=0}^s a_jv_j^m\otimes v_{s-j}^n~~~\in V(m)\otimes V(n)~,
\end{equation}
where $s=0,~1,~
\cdots,~{\rm min}(m,n),~l=m+n-2s$. The action of $\Delta(e)$ on $w^l$ yields

\begin{equation}
\Delta(e)(w^l)=\sum_{j=1}^s\left(a_j[m-j+1]+a_{j-1}q^{m-2(j-1)}[n-s+j]\right)
v_{j-1}^m\otimes v_{s-j}^n~,
\end{equation}
and

\begin{equation}
w_0^l=\sum_{j=0}^s a_0(-1)^jq^{j(m+1-j)}\frac{[n-s+j]![m-j]!}{[m]![n-s]!}
v_j^m\otimes v_{s-j}^n
\end{equation}
yields a set of highest weight vectors.\\
The $q$-version of the binomial formula has the form

\begin{equation}\label{zzz}
(A+B)^k=\sum_{i=0}^k q^{i(k-i)}\left[\begin{array}{c}
                                     k\\
                                     i\end{array}\right]A^{k-i}B^i~,
\end{equation}
where $A$ and $B$ are elements in a non-commutative algebra satisfying
$BA=q^2AB$.
Substituting  $A=1\otimes f$ and $B=f\otimes t^{-1}$ into Eq.(\ref{zzz}),
yields

\begin{equation}
\begin{array}{rcl}
\Delta\left(f^{(k)}\right)&=&\displaystyle\frac{1}{[k]!}\left(\Delta
(f)\right)^k=\frac{1}{[k]!}\left(
1\otimes f+f\otimes t^{-1}\right)^k\\[3mm]
&=&\displaystyle\sum_{i=0}^k q^{i(k-i)}f^{(i)}\otimes f^{(k-i)}t^{-i}~.
\end{array}
\end{equation}
Acting $\Delta\left(f^{(k)}\right)$ on the set of highest weight vectors
$w_0^l$, we obtain an expression for the general weight vectors $w_k^l$ in
$V(l)\subset V(m)\otimes V(n)$,

\begin{equation}
\begin{array}{rcl}
w_k^l&=&\Delta\left(f^{(k)}\right)w_0^l\\[3mm]
     &=&\displaystyle\left(\sum_{i=0}^kq^{i(k-i)}f^{(i)}\otimes
        f^{(k-i)}t^{-i}\right)\sum_{j=0}^sa_0
        (-1)^jq^{j(m+1-j)}\frac{[n-s+j]![m-j]!}{[m]![n-s]!}
v_j^m\otimes v_{s-j}^n\\[5mm]

&=&\displaystyle\sum_{i=0}^k\sum_{j=0}^sa_0(-1)^jq^{i(k-i)+j(m+1-j)-i(n-2(s-j))}
        \frac{[n-s+j]![m-j]!}{[m]![n-s]!}\times\\[4mm]
      & &~~~~\times  \left[\begin{array}{c}
                                               i+j\\
                                               j\end{array}\right]
                                          \left[\begin{array}{c}
                                                k+s-i-j\\
                                                s-j\end{array}\right]
              v_{i+j}^m\otimes v_{s+k-i-j}^n\\[7mm]
     &=&\displaystyle\sum_{j=\max(0,k-n+s)}^{\min(m,k+s)}\sum_{\nu=\max(j-k,0)}
         ^{\min(j,s)}a_0(-1)^\nu q^{\nu(l+1-k)+j(k-j+m-l)}\times\\[4mm]
     & &~~~~\times    \left[\begin{array}{c}
               j\\
               \nu\end{array}\right]
\left[\begin{array}{c}
               k+s-j\\
               s-\nu\end{array}\right]
\left[\begin{array}{c}
               n-s+\nu\\
               n-s\end{array}\right]
\left[\begin{array}{c}
               m\\
               \nu\end{array}\right]^{-1} v_j^m\otimes v_{s+k-j}^n~.
\end{array}
\end{equation}
Therefore, we obtain a set of irreducible submodules
$V(m+n),~V(m+n-2),~\cdots,~ V(|m-n|)$ and the Clebsch-Gordan rule
\cite{K88}--\cite{Hou90b}:\\
{\em For any $m,~n\in Z_{\geq 0}$ we have
$V^+(m)\otimes V^+(n)\cong V^+(m+n)\oplus\cdots\oplus V^+(|m-n|)$.}

\subsubsection{Representations of $U_q(g)$,   the general case}

It is well-known that the general quantum group $U_q(g)$ satisfies
the following algebra relations

\begin{equation}
\begin{array}{rcl}
& &t_it_i^{-1}=1=t_i^{-1}t_i~,~~~~t_it_j=t_jt_i~,\\
& &t_ie_jt^{-1}_i=q_i^{a_{ij}}e_j~,\\
& &t_if_jt^{-1}_i=q_i^{-a_{ij}}f_j~,\\[2mm]
& &\left[e_i,f_j\right]=\delta_{ij}\displaystyle\frac{t_i-t_i^{-1}}
{q_i-q_i^{-1}}~,\\[3mm]
& &\displaystyle\sum_{\nu=0}^{1-a_{ij}}(-1)^\nu\left[\begin{array}{c}
                                                     1-a_{ij}\\
                                                      \nu\end{array}\right]
                                                      _{q_i}
    \left(e_i\right)^{1-a_{ij}-\nu}e_j\left(e_i\right)^\nu=0~,~~~
    {\rm for}~i\not= j~,\\[3mm]
& &\displaystyle\sum_{\nu=0}^{1-a_{ij}}(-1)^\nu\left[\begin{array}{c}
                                                     1-a_{ij}\\
                                                      \nu\end{array}\right]
                                                      _{q_i}
    \left(f_i\right)^{1-a_{ij}-\nu}f_j\left(f_i\right)^\nu=0~,~~~
    {\rm for}~i\not= j~.
\end{array}
\end{equation}
Let $T$ denote the subgroup of invertible elements of $U_q(g)$
generated by $t_i$'s, and $C[T]$ denote its group algebra with basis
$$t_\alpha=t_1^{n_1}t_2^{n_2}\cdots t_l^{n_l},~~~~\alpha=
\sum_{i=1}^ln_i\alpha_i\in Q ~~~(Q=\bigoplus_{i=1}^lZ\alpha_i)~,$$
$U_qn_\pm$ denote the subalgebras generated by $e_i$'s, $f_i$'s with basis
$$\begin{array}{l}
(e)_r=e_1^{m_1} e_2^{m_2} \cdots e_l^{m_l}~,\\
(f)_r=f_1^{m_1} f_2^{m_2} \cdots f_l^{m_l}~,\\
r=\displaystyle\sum_{i=1}^lm_i\alpha_i\in I~~~(I=\bigoplus_{i=1}^lZ_{\geq
0}\alpha_i)~,
\end{array}$$
$U_qb_\pm$  denote the subalgebras generated by $e_i$'s, $t^{\pm 1}_i$'s and
$f_i$'s, $t^{\pm 1}_i$'s with basis
$\left((e)_r\cdot t_\alpha\right)_{r\in I,~\alpha\in Q}$. So,
$U_qb_\pm\simeq U_qn_\pm\otimes C[T]$ as vector spaces. The quantum
group $U_q(g)$ with
basis $\left((e)_r\cdot ( f)_{r'}\cdot t_\alpha\right)
_{r,r'\in I,~\alpha\in Q}$ has a triangular decomposition
$$U_q(g)\simeq U_qn_-\otimes C[T]\otimes U_qn_+$$
as vector spaces and is a free $U_qb_+$-module \cite{Rosso88}.

Let $(\pi,V)$ be representation of $U_q(g)$ in the finite-dimensional
vector space
$V$. Similar to the case of $SU_q(2)$, one may  proof that:
{\em
\begin{itemize}
\item The generator $e_i,~f_i$,  ($1\leq i\leq l)$ is nilpotent.
\item If the representation is irreducible, the $t_i$'s are  diagonalizable
and $V=\bigoplus V_\mu$, where
$$V_\mu=\{v\in V\vert\forall i~t_iv=q_i^{\mu_i}v\}~,~~~~\mu=(\mu_1,~\mu_2,~
\cdots,\mu_l)~,$$
$\mu$'s are the weights of the representation.
\end{itemize}}
Since a 1-dimensional representation is irreducible, we may denote it
by $(\pi_\sigma,{\bf C}_\sigma)$. If $(\pi,V)$ is an irreducible
finite-dimensional representation, $(\pi\otimes\pi_\sigma)\circ
\Delta$ gives an irreducible representation in $V\otimes {\bf C}_\sigma$.
Therefore {\em if $(\pi,V)$ is a irreducible finite-dimensional representation
with highest weight $\mu$, we can associate a
$1$-dimensional representation $(\pi_\sigma,{\bf C}_\sigma)$ and an
irreducible representation with dominant weight
$\tilde{\mu}$ that satisfies $\mu_i\in Z_{\geq 0}$}.\\
A nonzero vector
$v\in V$ is referred as a highest weight vector, if $e_iv=0~~(1\geq i\geq l)$.
For the highest weight vector $v_+$, with weight $\mu=(\mu_1,~\mu_2,~\cdots,~
\mu_l)$, we can construct a cyclic $U_q(g)$-module $V$ spanned by
$$v_0~{\rm and}~f_{i_1}f_{i_2} \cdots f_{i_p} v_0~,~~~~
i_1,~i_2,~\cdots,~i_p\in\{1,~2,~\cdots,~l\}~.$$
$V$ is an indecomposable $U_q(g)$-module, with a unique maximal
proper submodule. Taking the quotient by the maximal proper submodule, we
have an irreducible module with highest weight $\mu:~V_\mu$. For the case that
$\mu$ is of the dominant weight $\tilde{\mu}$, let $w=(f_i)^{\mu_i+1}v_0$.
Using
the algebra relations of $U_q(g)$, we have
$$t_jw=q_j^{-a_{ij}(\mu_i+1)}(f_i)^{\mu_i+1}t_jv_+=q_j^{-a_{ij}(\mu_i+1)}
q_j^{\mu_i}w~.$$
For $w\not= 0$, it is a weight vector with weight ${\mu_i-
(\mu_i+1)a_{ij}}\not=\mu_i$. For $i\not= j,~e_j$ and
$f_i$ commute to each other, and $e_jw=0$. For $i=j$, using the relation
(\ref{2.chang1}) and
$$t_iv_+=q_i^{\mu_i}v_+~,$$
we get $e_iw=0$. As a result, if $w\not= 0$, it would be a highest weight
vector.
Because $V_{\tilde{\mu}}$ is irreducible, such a vector can not exist.
We must have
$$f_i^{\mu_i+1}v_+=0~.$$
For $1\geq i\geq l$, it is easy to verify that the subvectorspace spanned by
$v_0,~f_iv_0,~\cdots,~f_i^{\mu_i}v_0$ is invariant under the action of the
subalgebra $L_i$ generated by $e_i,~f_i$ and $t_i^\pm$. \\
Therefore {\em for each $1\geq i\geq l$, $V_{\tilde{\mu}}$ contains a nonzero
finite-dimensional $L_i$-module.}

For $U_q(g)$, $1-a_{ij}\in\{1,~\cdots,~4\}$, if $1-a_{ij}=1$,
$e_i e_j=e_j e_i$. For $1-a_{ij}\geq 2$, define $e_{i,j}=
e_i e_j-q_i^{a_{ij}}e_j e_i$. If $1-a_{ij}=2$,
one defining relation yields
$$e_i e_{i,j}-q_i^{2+a_{ij}}e_{i,j}e_i=0~.$$
If $1-a_{ij}=3$, put
$$e_{i,i,j}=e_i e_{i,j}-q_i^{2+a_{ij}}e_{i,j} e_i~,$$
and we have
$$e_ie_{i,i,j}=q_i^{4+ a_{ij}}e_{i,i,j}e_i~.$$
For $1-a_{ij}=4$, put
$$e_{i,i,i,j}=e_i e_{i,i,j}-q_i^{4+a_{ij}}e_{i,i,j} e_i~,$$
and then
$$e_ie_{i,i,i,j}=q_i^{6+a_{ij}}e_{i,i,i,j}e_i~.$$
The same relations can be obtained for the $f_i$'s.
Let $V'$ be the sum of the finite-dimensional $L_i$-submodules, obviously
$V'\not=\{0\}$.
If $W$ is an invariant finite-dimensional $L_i$-submodule, the
vector space spanned by $e_j W,~f_j W,~t_jW;~e_{i,j} W,~f_{i,j} W;~
e_{i,i,j}W,~f_{i,i,j}W;$
and $e_{i,i,i,j}W,~f_{i,i,i,j}W,$ (where $j\in\{1,~2,~\cdots,~l\}
\setminus \{i\}$) is finite-dimensional and invariant under $L_i$. So
$U_q(g)(W)\subset V'$. Then we have $V'=V_{\tilde{\mu}}$. \\
We now obtain the basic results about representations of $U_q(g)$ for generic
$q$:
{\em \begin{itemize}
\item If $\mu$ is dominant weight, $V_{\mu}$ is irreducible
and finite-dimensional.
\item An irreducible finite-dimensional representation of $U_q(g)$ is
equivalent to $V_{\tilde{\mu}}$, up to ``twist by sign''.
\end{itemize}}
For the representations of general $U_q(g)$, there is another
important theorem, the complete reducibility theorem.\\
{\em Finite-dimensional representations of $U_q(g)$ are completely
reducible.}

\subsection{Representations for $q$ being a root of unity}

When the quantization  parameter $q$ becomes a root of unity ($q^p=\pm 1$),
representation theory
of quantum group changes its phases drastically. This is one of the most
intriguing situations which are absent in the classical case ($q=1$), and
is important also for applications.
We are now discussing the representations for quantum group with the
quantization parameter $q$ is a root of unity. Now the quantum group is not
semi-simple. Let us begin with $SU_q(2)$.

\subsubsection{Representations of $SU_q(2)$}
{\bf The continuity of $V(l)$}

The basic problem for $q^p=\pm 1$  is that
$e^p=0=f^p$, which generates null vectors in some representations.
The representations $(\pi^\sigma_l,V^\sigma(l))$ are well defined also
for $q$ being a root of unity, however they are no longer irreducible in
general. Many representations appearing in the decomposition of tensor
products of irreducible representations will be reducible,  but not fully
reducible. It is convenient to construct a different basis for the
representation with highest weight $2j$  \cite{Lusztig90a} as

\begin{equation}
\vert j,m\rangle\equiv v^l_{j-m}~,~~~~(j=\frac{l}{2})~.
\end{equation}
Then, we have

\begin{equation}\label{2.a2}
\begin{array}{l}
e\vert jm\rangle=[j+m+1]\vert j,m+1\rangle~,\\
f\vert jm\rangle=[j-m+1]\vert j,m-1\rangle~,\\[2mm]
\displaystyle\frac{e^a}{[a]!}\vert jm\rangle=\displaystyle
\frac{[j+m+a]!}{[a]![j+m]!}\vert j,m+a\rangle~,\\[2mm]
\displaystyle\frac{f^a}{[a]!}\vert jm\rangle=\displaystyle
\frac{[j-m+a]!}{[a]![j-m]!}\vert j,m-a\rangle~.
\end{array}
\end{equation}
For $q^p=\pm 1$, $[p]=[2p]=\cdots=[kp]=0$. However, the operators
$\displaystyle \frac{e^p}{[p]!}$  and $\displaystyle\frac{f^p}{[p]!}$
are still well defined, as can be
seen by setting $a=p$ in Eq.(\ref{2.a2}) for generic $q$, and then taking the
limit $q^p\to \pm 1$.

It is well-known that the acting of the Casimir operator $C$ on the highest
weight vector $\vert jj\rangle$ yields

\begin{equation}
C\vert jj\rangle=\left[j+\frac{1}{2}\right]^2\vert jj\rangle~.
\end{equation}
For generic $q$ the eigenvalues of $C$ are different for
different values of $j$. For $q^p=\pm 1$, it is easy to see that
the Casimir operator takes identical values for highest weights ${2j}$ and
${2j'}$ related by one of the transformations

\begin{equation}\label{2.tta}
j'=j+np~,~~~~j'=p-1-j+np~.
\end{equation}
The Casimir operator is no longer sufficient to label representations $V(2j)$.
Some $V(2j)$ and $V(2j')$, where $j,~j'$ satisfy Eq.(\ref{2.tta}),  can be
mixed up and get connected under the action of $e$.

It is useful to introduce the $q$-dimension

\begin{equation}
D_q(j)=\sum_{{\rm states~ in~}V(2j)}t=[2j+1]~,
\end{equation}
so that

\begin{equation}\label{2.kk1}
D_q(j)=D_q(j-kp)=-D_q(p-1-j+kp)~.
\end{equation}
Both the  symmetry properties of the Casimir operator $C$ and the fact that
$e^p=0=f^p$ suggest that if we try to decompose the tensor product
$\left(V(1)\right)^{\otimes n}$ for sufficient high $n$ into irreducible
representations, odd things begin to happen. $\left(V(1)\right)^{\otimes n}$
contains in its decomposition reducible but not fully reducible
representations. For example if $q^3=\pm 1$ we can try to decompose
$V(1)\otimes
 V(1)\otimes V(1)$. For generic values of $q$ this tensor product
decomposes into $V(3)\oplus V(1)$. For $q^3=\pm 1$ the weight states
of the $j=\displaystyle\frac{3}{2}$ and one of the
$j=\displaystyle\frac{1}{2}$ representations are mixed up to a reducible,
but not fully reducible representation, because of the presence of null
vectors. It is not difficult to verify that the state
$$
\begin{array}{rcl}
\vert\alpha\rangle&=&\displaystyle f\vert
\frac{1}{2}\frac{1}{2}\rangle\otimes\vert\frac{1}{2}
               \frac{1}{2}\rangle\otimes\vert\frac{1}{2}
               \frac{1}{2}\rangle \\[2mm]
             &=&f\vert v^3_0\rangle
\end{array}$$
is annihilated by
$e$. $\vert \alpha\rangle$ is a highest-weight state of $V(3)$ with
zero norm.
Out of the two other states in $V(1)\otimes V(1)\otimes V(1)$
only one of them is orthogonal to $\vert\alpha\rangle$.
The other state $\vert\beta\rangle$ is not orthogonal to
$\vert\alpha\rangle$, and thus
$$\langle f\alpha\vert\beta\rangle=\langle\alpha\vert e
\vert\beta\rangle\not=0~,$$
i.e., $V(3)$ and $V(1)$ are mixed up (Fig.10). Notice $f^3=0$, implying
that the arrow leaving from $f^2\vert v^3_0\rangle$  to $f^3\vert
v^3_0\rangle$
disappears, while $f^3\vert v^3_0\rangle$  can still be reached from
$\vert v^3_0\rangle$  by applying $\displaystyle\frac{f^3}{[3]!}$. This is
replaced by a new arrow connecting $f\vert\beta\rangle$ and
$\displaystyle\frac{f^3}{[3]!}\vert v^3_0\rangle$. Finally, we obtain

\begin{equation}
ef\vert\beta\rangle=\vert\beta\rangle+f\vert v^3_0\rangle~.
\end{equation}
Thus $(V(1))^3$ decomposes into a big (type-I) representation which is a
mixture of $V(3)$ with $V(1)$,  and a small type-II representation $V(1)$.
Because $\displaystyle D_q\left(\frac{3}{2}\right)+D_q\left(\frac{1}{2}\right)
=0$, type-I representation has $q$-dimension zero. The type-I
representation is
indecomposable, but it is not irreducible, because it contains a
sub-representation $V(3)$.

\setlength{\unitlength}{5pt}
\begin{picture}(50,35)(-5,0)
\put(15,3){$f^3\vert v^3_0\rangle$}
\put(20,6){\vector(0,1){5}}
\put(15,12.5){$f^2\vert v^3_0\rangle$}
\put(20,15.5){\vector(0,1){5}}
\put(17,20.5){\vector(0,-1){5}}
\put(10,22){$\vert\alpha\rangle=f\vert v^3_0\rangle$}
\put(17,30){\vector(0,-1){5}}
\put(16,31.5){$\vert v^3_0\rangle$}

\put(32,22){$\vert\beta\rangle$}
\put(35,15.5){\vector(0,1){5}}
\put(32,20.5){\vector(0,-1){5}}
\put(31,12.5){$f\vert \beta\rangle$}
\put(31,25){\vector(-2,1){10}}
\put(31,15.5){\vector(-2,1){10}}
\put(31,11){\vector(-2,-1){10}}

\put(47,22){$\vert\gamma\rangle$}
\put(50,15.5){\vector(0,1){5}}
\put(47,20.5){\vector(0,-1){5}}
\put(46,12.5){$f\vert \gamma\rangle$}

\end{picture}

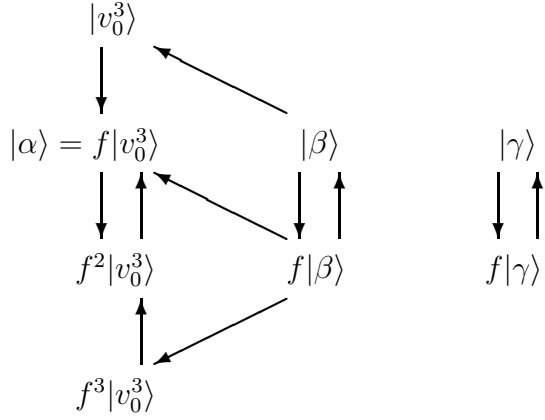
\hspace{6em}Fig.10. Structure of $(V(1))^{\otimes 3}$ for $q=e^{i\pi/3}$.

The analysis of more general situations can be accomplished in the
same way. In the irrational case, $\left(V(1)\right)^{\otimes n}$ decomposes
into
a sum of representations $V(2j)$. For $q^p=\pm 1$,
we notice that $V(1)$ has positive $q$-dimension. Solving the equation
$D_q(j)=[2j+1]=0$, we find that the $q$-dimension vanishes, whenever
$j=\displaystyle\frac{p-1}{2}+kp$. Making tensor products
$V(1)\otimes V(1)\otimes\cdots\otimes V(1)$ consecutively, we eventually
attain the representation with
$j=\displaystyle\frac{p-1}{2}$ with vanishing $q$-dimension, and others
remained positive $q$-dimension. Further tensoring another copy of $V(1)$
results in pairing of
representations. We can limit the representations of $SU_q(2)$ to those
with the smallest possible $j$ and positive $q$-dimension. This is achieved
by requiring that in the tensor product of the fundamental representations we
only keep
those highest weight vectors that are annihilated by $e$ and not
in the imagine of $e^{p-1}$. This restricts the representations to those
with $j<\displaystyle\frac{p-1}{2}$. In this way we find an alcove in the space
of weights where the $q$-dimension is strictly positive with the lowest
possible
value of $j$. The representations with $j<\displaystyle\frac{p-1}{2}$ are
characterized by the fact that they are highest weight representations and
the highest weight
vector $\vert jj\rangle$ cannot be written as $e^{p-1}\vert{\rm
anything}\rangle$.

For a given values of
the Casimir operator $C$ there are sets of indices like

\begin{equation}
\{j_k>j_{k-1}>\cdots>j_1~,~~0\leq j_1<\frac{1}{2}(p-1)\}~,
\end{equation}
which are related by symmetries (\ref{2.tta}).
If $i$ is odd, $j_i=j_1$ mod $p$, and if $i$ is even, $j_i=p-1-j_1$
mod $p$. If $k=1$, $V(2j_1)$ cannot be mixed up, and it is still an
irreducible highest weight representation (type II). If $k>1$,
mixing of the representations occurs, which is to be analyze.

We construct $V(2j_k)$ by acting  $f$ and $\displaystyle\frac{f^p}{[p]!}$
upon the highest weight vector $\vert j_k,j_k\rangle$.
It is easy to see that

\begin{equation}
\begin{array}{rcl}
e\vert j_k,j_{k-1}\rangle&=&[j_k+j_{k-1}+1]\vert j_k,j_{k-1}+1\rangle\\
                          &=&[np]\vert j_k,j_{k-1}+1\rangle=0~.
\end{array}
\end{equation}
Generally, we have

\begin{equation}
\begin{array}{rcl}
e\vert j_k,j_{k-(2i+1)}\rangle&=&[j_k+j_{k-(2i+1)}+1]\vert j_k,j_{k-(2i+1)}+1
\rangle\\
       &=&[n'p]\vert j_k,j_{k-(2i+1)}+1\rangle=0~.
\end{array}
\end{equation}
So that $V(2j_k)$ is not irreducible. It contains states that are annihilated
by $e$ at $j_{k-1},~j_{k-3}=j_{k-1}-p,~j_{k-5}=j_{k-1}-2p,~\cdots.$
It is also easy to see that

\begin{equation}
\begin{array}{rcl}
f\vert j_k,j_{k-2i}+1\rangle&=&[j_k-j_{k-2i}]\vert j_k,j_{k-2i}\rangle\\
&=&[mp]\vert j_k,j_{k-2i}\rangle=0~.
\end{array}
\end{equation}
Then it also contains states that are annihilated by $f$ at
$j_{k-2}+1=j_k-p+1,~j_{k-4}=j_{k}-2p+1,~\cdots.$

The state $\vert j_k,j_{k-1}\rangle $ is annihilated by $e$ and $e^p/[p]!$.
Under the action of $f$ and $f^p/[p]!$ it generates an irreducible
sub-representation, of which all states have zero norm. At spin
$j_{k-1}$, there must exist such a state $\vert \beta\rangle$ that
$e\vert \beta\rangle=\vert j_k,j_{k-1}+1\rangle$. Acting $f$ and
$f^p/[p]!$ upon $\vert\beta\rangle $ generates the states $\vert
j_{k-1},m\rangle$ satisfying,

\begin{equation}\label{2.a16}
f\vert j_{k-1},m\rangle=[j_{k-1}-m+1]\vert j_{k-1},m-1\rangle~.
\end{equation}
{}From the relation

\begin{equation}\label{2.110}
\langle j_k,j_{k-1}+1\vert e\vert j_{k-1},j_{k-1}\rangle=[j_k-j_{k-1}]\langle
j_k,
j_{k-1}\vert j_{k-1},j_{k-1}\rangle~,
\end{equation}
and the normalization

\begin{equation}\label{2.a34}
\langle j_k,j_{k-1}\vert j_{k-1},j_{k-1}\rangle=\frac{1}{[j_{k}-j_{k-1}]}
\left[\begin{array}{c}
      2j_{k}\\
      j_{k}+j_{k-1}+1
      \end{array}\right]~,
\end{equation}
it is follows that

\begin{equation}\label{2.111}
e\vert j_{k-1},j_{k-1}\rangle=\vert j_k,j_{k-1}+1\rangle~.
\end{equation}
In general, the acting of $e$ on $\vert j_{k-1},m\rangle$ has the form

\begin{equation}\label{2.112}
e\vert j_{k-1},m\rangle=[j_{k-1}+m+1]|j_{k-1},m+1>+\displaystyle\left[
                       \begin{array}{c}
				   j_k-m-1\\
				   j_{k-1}-m
                                   \end{array}\right]\vert j_k,m+1\rangle~.
\end{equation}
The above equation can be proved inductively. First, for $m=j_{k-1}$,
Eq.(\ref{2.112}) reduces to Eq.(\ref{2.111}),  so that it is satisfied.
Provided that Eq.(\ref{2.112}) is also satisfied for $m=M$, then
\begin{equation}
\begin{array}{rcl}
[2M]\vert j_{k-1},M\rangle
&=&\displaystyle\frac{t-t^{-1}}{q-q^{-1}}\vert j_{k-1},
M\rangle\\[2mm]
&=&\left(ef-fe\right)\vert j_{k-1},M\rangle\\[2mm]
&=&[j_{k-1}-M+1]e\vert j_{k-1},M-1\rangle-[j_{k-1}+M+1][j_{k-1}-M]
\vert j_{k-1},M\rangle\\[4mm]
& &-\left[\begin{array}{c}
                            j_k-M-1\\
                            j_{k-1}-M
                            \end{array}\right][j_k-M]\vert j_k,M\rangle~.
\end{array}
\end{equation}
Thus we obtain

\begin{equation}
e\vert j_{k-1},M-1\rangle=[j_{k-1}+M]\vert j_{k-1},M\rangle +
\left[\begin{array}{c}
                            j_k-M\\
                            j_{k-1}-M+1
                            \end{array}\right]\vert j_k,M\rangle~.
\end{equation}
Therefore Eq.(\ref{2.112}) is also satisfied when $m=M-1$.
Using the normalization (\ref{2.a34}), we have \\ \\
$\langle j_k,-j_{k-1}-1\vert f\vert j_{k-1},-j_{k-1}\rangle$
\begin{equation}
\begin{array}{rcl}
  &=&[j_k-j_{k-1}]\langle j_k,-j_{k-1}\vert j_{k-1},-j_{k-1}\rangle\\[2mm]
  &=&[j_k-j_{k-1}]\langle j_k,-j_{k-1}\vert \displaystyle
     \frac{f^{2j_{k-1}}}{[2j_{k-1}]!}\vert j_{k-1},
     j_{k-1}\rangle\\[4mm]
  &=&[j_k-j_{k-1}]\left[\begin{array}{c}
                        j_k+j_{k-1}\\
                        2j_{k-1}
                        \end{array}\right]
     \langle j_k,j_{k-1}\vert j_{k-1},j_{k-1}\rangle\\[4mm]
  &=&\left[\begin{array}{c}
                        j_k+j_{k-1}\\
                        2j_{k-1}
                        \end{array}\right]
     \left[\begin{array}{c}
                        2j_k\\
                        j_k+j_{k-1}+1
                        \end{array}\right]~,
\end{array}
\end{equation}
and

\begin{equation}
f\vert j_{k-1},-j_{k-1}\rangle=\left[
                       \begin{array}{c}
				   j_k+j_{k-1}-1\\
				   2j_{k-1}
				   \end{array}\right]\frac{[j_k+j_{k-1}]}{[j_k-j_{k-1}]}
                                   \vert j_k,-j_{k-1}-1\rangle~.
\end{equation}
{}From the above discussion, we see that $e$ connects $V({2j_k})$ and
$V({2j_{k-1}})$ (Fig.11); $V({2j_{k-1}})$ is the highest weight representation
generated from $\vert \beta\rangle$ if we factorize out $V({2j_{k}})$. From
Eq.(\ref{2.112}), we know that the states in $V(2j_{k-1})$
annihilated by $e$ can be constructed  at spin $j_{k-2}=j_k-p$,
$j_{k-4}=j_k-2p$, $\cdots$, by making appropriate linear combinations of
$\vert j_k,m\rangle$ and $\vert j_{k-1},m\rangle $. All states of $(V(2{j_k}),
V(2{j_{k-1}}))$ belonging to ${\rm Ker}~e$ are in ${\rm Im}~e^{p-1}$.
The only states in $(V(2j_k),~V(2j_{k-1}))$ that are annihilated
simultaneously by $e$ and $\displaystyle\frac{e^p}{[p]!}$ are
$\vert \alpha\rangle$ and $\vert \alpha'\rangle=\displaystyle
\frac{f^{j_k-j_{k-1}}}{[j_k-j_{k-1}]!}\vert
\alpha\rangle$, as can be verified by using
formulas (\ref{2.a2}) and (\ref{2.a16}). This fact ensures that there is no
common vector among $(V(2{j_k}),~V(2{j_{k-1}}))$ and its orthogonal.
Therefore, we can repeat the same arguments in the orthogonal of
$(V({2j_k}),V({2j_{k-1}}))$.

\setlength{\unitlength}{4pt}
\begin{picture}(50,120)(-22,0)
\multiput(15,4)(0,1.5){7}{\line(0,1){1}}
\multiput(20,4)(0,1.5){7}{\line(0,1){1}}

\multiput(12.5,14)(0,20){6}{\line(1,0){10}}
\multiput(12.5,19)(0,20){6}{\line(1,0){10}}
\multiput(12.5,24)(0,20){5}{\line(1,0){10}}

\multiput(15,24)(0,1.5){7}{\line(0,1){1}}
\multiput(20,24)(0,1.5){7}{\line(0,1){1}}

\multiput(15,44)(0,1.5){7}{\line(0,1){1}}
\multiput(20,44)(0,1.5){7}{\line(0,1){1}}

\multiput(15,64)(0,1.5){7}{\line(0,1){1}}
\multiput(20,64)(0,1.5){7}{\line(0,1){1}}

\multiput(15,84)(0,1.5){7}{\line(0,1){1}}
\multiput(20,84)(0,1.5){7}{\line(0,1){1}}

\multiput(15,104)(0,1.5){7}{\line(0,1){1}}
\multiput(20,104)(0,1.5){7}{\line(0,1){1}}

\multiput(15,19)(0,20){6}{\vector(0,-1){5}}
\multiput(20,14)(0,20){6}{\vector(0,1){5}}
\multiput(15,24)(0,40){3}{\vector(0,-1){5}}
\multiput(20,39)(0,40){2}{\vector(0,1){5}}

\multiput(45,4)(0,1.5){7}{\line(0,1){1}}
\multiput(50,4)(0,1.5){7}{\line(0,1){1}}

\multiput(42.5,14)(0,20){5}{\line(1,0){10}}
\multiput(42.5,19)(0,20){5}{\line(1,0){10}}
\multiput(42.5,24)(0,20){4}{\line(1,0){10}}

\multiput(45,24)(0,1.5){7}{\line(0,1){1}}
\multiput(50,24)(0,1.5){7}{\line(0,1){1}}

\multiput(45,44)(0,1.5){7}{\line(0,1){1}}
\multiput(50,44)(0,1.5){7}{\line(0,1){1}}

\multiput(45,64)(0,1.5){7}{\line(0,1){1}}
\multiput(50,64)(0,1.5){7}{\line(0,1){1}}

\multiput(45,84)(0,1.5){7}{\line(0,1){1}}
\multiput(50,84)(0,1.5){7}{\line(0,1){1}}

\multiput(45,19)(0,20){5}{\vector(0,-1){5}}
\multiput(50,14)(0,20){5}{\vector(0,1){5}}
\multiput(45,44)(0,40){2}{\vector(0,-1){5}}
\multiput(50,19)(0,40){2}{\vector(0,1){5}}

\multiput(42.5,14)(0,20){5}{\vector(-4,1){20}}
\multiput(42.5,19)(0,20){5}{\vector(-4,1){20}}

\put(-12,18){$j_{k-5}=j_{k-1}-2p$}
\put(-11.5,38){$j_{k-4}=j_{k}-2p$}
\put(-11,58){$j_{k-3}=j_{k-1}-p$}
\put(-10.5,78){$j_{k-2}=j_{k}-p$}

\put(7,118){$j_k$}
\put(6,98){$j_{k-1}$}

\put(16,120){$\vert\alpha\rangle$}
\put(20,100){$\vert\alpha'\rangle$}
\put(46,100){$\vert\beta\rangle$}
\end{picture}

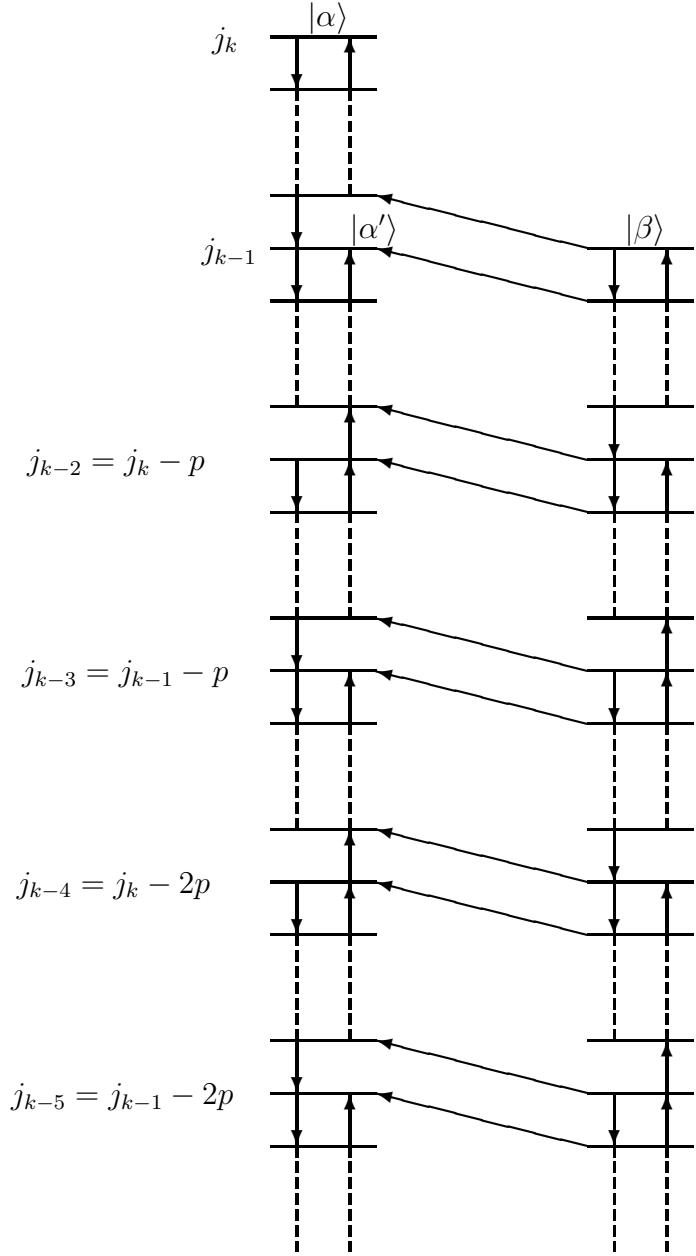
\hspace{7em}Fig.11. Pairing of representations $(V(2j_k),V(2j_{k-1}))$.

In this way we get pairs of
representations mixed with each other in larger structures (type I)
$(V(2{j_k}),V({2j_{k-1}}))$; $(V(2{j_{k-2}}),~V({2j_{k-3}}))$;
$\cdots$. From Eq.(\ref{2.kk1}), it is follows that

\begin{equation}
\begin{array}{ccc}
D_q(j_k)&=&-D_q(j_{k-1})~,\\
D_q(j_{k-2})&=&-D_q(j_{k-3})~,\\
\vdots&\vdots& \vdots~,
\end{array}
\end{equation}
And the type-I representations have $q$-dimension zero.
Depending on the number of $V(1)$'s we end up with a certain
number of $V({2j_1})$ that cannot be mixed up and are still irreducible
highest weight representations (type II).

Type-II representations are described by their highest weight vector
$\vert \alpha_j\rangle$, for which $k\vert \alpha_j\rangle=q^{2j}\vert
\alpha_j \rangle$, $0\leq
j<\displaystyle\frac{1}{2}p-1$. $V(2j)$ is indeed isolated if,  moreover
$\vert \alpha_j\rangle$ does not belong to a larger $V(2{j'})$
representation. On basis of
the above analysis, this implies that
$\vert \alpha_j\rangle\in {\rm Im}~e^{p-1}$. Notice that all highest
weights of $V(2{j'})~(j'\geq\displaystyle\frac{1}{2}p-1)$ also belong to
Im $e^{p-1}$. The highest weights of type-II representations are
thus completely characterized by the condition

\begin{equation}\label{2.a19}
\vert \alpha_j\rangle\in\frac{{\rm Ker}~e}{{\rm Im}~e^{p-1}}~.
\end{equation}
Another feature of these states is

\begin{equation}
\vert \alpha_j\rangle\in\frac{{\rm Ker}~f^{2j+1}}{{\rm Im}~f^{p-1-2j}}~.
\end{equation}
For $(V(1))^{\otimes n}$ case, their number reads

\begin{equation}
\Omega_j^{(n)}=\Gamma_j^{(n)}-\Gamma_{p-1-j}^{(n)}+\Gamma_{j+p}^{(n)}
-\Gamma_{p-1-j+p}^{(n)}+\cdots~,
\end{equation}
where

\begin{equation}
\Gamma_j^{(n)}=\left[\begin{array}{c}
                     n\\
                     \displaystyle\frac{n}{2}-j
                     \end{array}\right]-\left[
                     \begin{array}{c}
                     n\\
                     \displaystyle\frac{n}{2}+j+1
                     \end{array}\right]~.
\end{equation}
A special situation is for
$j_1=\displaystyle\frac{1}{2}(p-1)$,  since $p-1-j_1=j_1$,  whereupon
the representations $V(2{j_k})$ are still irreducible, and do not pair.
Conventionally, we shall also call them as type I. Then, the representations
space splits into:\\
{\em
\begin{itemize}
\item Type-I representations which have $q$-dimension zero, and are either
mixed up or the kind $V({(np-1)})$.
\item Type-II representations which have a nonzero $q$-dimension, and are
still isomorphic to representations of $U\left(su(2)\right)$.
\end{itemize}}

{\bf Cyclic representations}

It is well-known that the type-II representations $V(2j)~(0\leq j< p/2-1)$
remain irreducible, but have not exhausted the irreducible finite-dimensional
representations, which contain continuous parameters in general. The
continuous parameters arisen from the fact that $SU_q(2)$ has a large
center at roots of unity compared to the generic case.
For definiteness, in dealing with cyclic representations, we
merely consider $q=e^{2\pi i/p}$ with odd $p$
\cite{Concini89}--\cite{Pasquier90}.

Let ${\cal Z}$ denote the center of
$SU_q(2)$

\begin{equation}
{\cal Z}=\{u\in SU_q(2)|au=ua~~~ \forall a\in SU_q(2)\}~.
\end{equation}
Except for the well-known central element, the Casimir operator

\begin{equation} \label{2.mmm}
C=\frac{(qt-1)(1-q^{-1}t^{-1})}{(q-q^{-1})^2}+fe~,
\end{equation}
it is straightforward to show that $e^p,~f^p,~t^p\in{\cal Z}$.\\
Let

\begin{equation}
x=\left((q-q^{-1})e\right)^p~,~~~~ y=\left((q-q^{-1})f\right)^p~,~~~~
z=t^p~,
\end{equation}
and ${\cal Z}_0$ denote the subalgebra of ${\cal Z}$ generated by
$x,~y,~z^\pm$. Then, ${\cal Z}$ {\sl is generated by $C$ and ${\cal Z}_0$.}\\
Rewriting Eq.(\ref{2.mmm}) as

\begin{equation}
fe=C-\frac{(qt-1)(1-q^{-1}t^{-1})}{(q-q^{-1})^2}~,
\end{equation}
we generally have

\begin{equation}
\begin{array}{rcl}
f^pe^p
&=&f^{p-1}\left(fe\right) e^{p-1}\\[3mm]
&=&f^{p-1}e^{p-1}

\left(C-\displaystyle\frac{(q^{2(p-1)+1}t-1)(1-q^{-2(p-1)-1}t^{-1})}{(q-q^{-1})^2}\right)
   \\[3mm]
&=&f^{p-2}e^{p-2}

\left(C-\displaystyle\frac{(q^{2(p-2)+1}t-1)(1-q^{-2(p-2)-1}t^{-1})}{(q-q^{-1})^2}\right)\times\\[3mm]
&
&\times\left(C-\displaystyle\frac{(q^{2(p-1)+1}t-1)(1-q^{-2(p-1)-1}t^{-1})}{(q-q^{-1})^2}\right)\\[3mm]
&\vdots&\\[3mm]
&=&\displaystyle\prod_{i=0}^{p-1}\left(C-\frac{(q^{2i+1}t-1)(1-q^{-2i-1}t^{-1})}
{(q-q^{-1})^2}\right)\\[3mm]
&=&\displaystyle\prod_{i=0}^{p-1}\left(C-\frac{(q^it-1)(1-q^{-i}t^{-1})}
{(q-q^{-1})^2}\right)~.
\end{array}
\end{equation}
Introducing the formal operators $u^{\pm 1}$ through the following relation

\begin{equation}
C=\frac{(u-1)(1-u^{-1})}{(q-q^{-1})^2}~,
\end{equation}
we can rewrite $f^p e^p$ as

\begin{equation}
\begin{array}{rcl}
\left((q-q^{-1})f\right)^p\left((q-q^{-1})e\right)^p
&=&\displaystyle\prod_{i=0}^{p-1}\left(u+u^{-1}-q^it-q^{-i}t^{-1}\right)\\
&=&\displaystyle\prod_{i=0}^{p-1}\left(u-q^it\right)\left(1-q^{-i}t^{-1}u^{-1}
   \right)~.
\end{array}
\end{equation}
which, upon utilizing the following identities

\begin{equation}
\begin{array}{l}
\displaystyle\prod_{i=0}^{p-1}(A-q^{\pm i}B)=A^p-B^p~~~~(AB=BA)~,\\
\displaystyle\prod_{i=0}^{p-1}(A+q^{\pm i}B)=A^p+B^p~~~~(AB=BA)~,
\end{array}
\end{equation}
reduces to

\begin{equation}
\begin{array}{rcl}
\left((q-q^{-1})f\right)^p\left((q-q^{-1})e\right)^p
&=&\left((\sigma u^p-1)(\sigma -u^{-p})-(\sigma t^p-1)(\sigma -t^{-p})\right)\\
&=&\displaystyle\prod_{i=0}^{p-1}\left(\sigma u-q^i\right)\left(\sigma -
q^{-i}u^{-1}\right)-(\sigma t^p-1)(\sigma -t^{-p})\\
&=&\displaystyle\prod_{i=0}^{p-1}\left(C(q-q^{-1})^2-\sigma (q^i+q^{-i})
+2\right)-(\sigma t^p-1)(\sigma -t^{-p})~.
\end{array}
\end{equation}
This leads to the relation

\begin{equation}\label{2.j4.1}
\psi_p(C)=xy+(\sigma z-1)(\sigma -z^{-1})~,
\end{equation}
where $\psi_p(\mu)=
\displaystyle\prod_{i=0}^{p-1}\left(\mu (q-q^{-1})^2-\sigma (q^i+q^{-i})
+2\right)$. \\
Therefore, {\em the elements $x,~y,~z$ are algebraically independent, while
$C$ is algebraic over ${\cal Z}_0$.}

${\cal Z}_0$ is a Hopf subalgebra, in fact,

\begin{equation}
\Delta(x)=x\otimes 1+z\otimes x~,~~~\Delta(y)=y\otimes z^{-1}+1\otimes y~,~~~
\Delta(z)=z\otimes z~.
\end{equation}
However ${\cal Z}$ is not a Hopf algebra.

Now, let Rep$SU_q(2)$ denote the set of the equivalence classes
of the irreducible finite-dimensional representations
of $SU_q(2)$, Spec${\cal Z}$ denote the hypersurface defined
by  Eq.(\ref{2.j4.1}),

 $${\rm Spec}{\cal Z}=\{(x,~y,~z,~C\in {\bf C}^4)|z\not=0,~
\psi_p(C)-xy-(\sigma z-1)(\sigma -z^{-1})=0\}~,$$
and further

$${\rm Spec}{\cal Z}_0=\{(x,~y,~z)|z\not=0\}={\bf C}^2\times {\bf
C}^\times~.$$ For an irreducible finite-dimensional representation
 $\pi\in{\rm Rep}SU_q(2)$ the central elements $x,~y,~z,~C$
act as scalar in
$\pi$ (Schur's lemma), resulting in the maps

\begin{equation}
{\rm Rep}SU_q(2)\stackrel{X}{\longrightarrow}{\rm Spec}{\cal Z}\stackrel{\tau}
{\longrightarrow}{\rm Spec}{\cal Z}_0~.
\end{equation}
$X$ maps a representation (class) $\pi$ to the values of
$x,~y,~z,~C$ in $\pi$, and the map $\tau$ is a projection, so that
$\tau^{-1}(s)$ consists of $p$ points for general $s\in {\rm Spec}{\cal
Z}_0$. The hypersurface ${\rm Spec}{\cal Z}$ has singularities, given
by the set of $p-1$ points

\begin{equation}
D=\left\{(0,~0,~\sigma,~\frac{\sigma(q^j+q^{-j})-2}{(q-q^{-1})^2}\vert
\sigma=\pm 1,~j=1,~2,~\cdots,~
\frac{p-1}{2}\right\}~.
\end{equation}
De Concini and Kac shown that:
{\em
\begin{itemize}
\item $X$ is surjective.
\item If $\chi \not\in D$, there is only one irreducible finite-dimensional
representation $\pi\in{\rm Rep}
SU_q(2)$ with $X(\pi)=\chi$ and {\rm dim}$\pi=p$.\\
 Let $X,~Z$
denote the following $p\times p$ matrices

\begin{equation}
X=\left(\begin{array}{ccccc}
     0      &  0    &\cdots&    0&1\\
     1      &  0    &\cdots&    0&0\\
     0      &  1    &\cdots&    0&0\\
     \vdots &\vdots &\ddots&\vdots&\vdots\\
     0      &  0    &\cdots&1    &0
	   \end{array}\right)~,~~~~
Z=\left(\begin{array}{ccccc}
      1  &  0    &     0 & \cdots &0\\
      0  &  q    &     0 & \cdots &0\\
      0  &  0    &    q^2& \cdots &0\\
   \vdots&\vdots & \vdots&\ddots  &\vdots\\
      0  &  0    &     0 &\cdots &q^{p-1}
                \end{array}\right)~,
\end{equation}

\begin{equation}
ZX=qXZ~,~~~~Z^p=X^p=1~.
\end{equation}
The $p$-dimensional representation is then given by

\begin{equation}\label{2.jnn}
\begin{array}{rcl}
& &e=\displaystyle\frac{1}{q-q^{-1}}x_1(a_1Z-a_1^{-1}Z^{-1})X~,\\[3mm]
&
&f=\displaystyle\frac{1}{q-q^{-1}}x_1^{-1}(a_2Z^{-1}-a_2^{-1}Z)X^{-1}~,\\[3mm]
& &t=\displaystyle\frac{a_1}{a_2}Z^2~,
\end{array}
\end{equation}
where either $a_1^{2p}\not=1$ or $a_2^{2p}\not=1$, $a_1,~a_2,~x_1\in
{\bf C}^\times$ and

$$x=(a_1^p-a_1^{-p})x_1^p~,~~~y=(a_2^p-a_2^{-p})x_1^{-p}~,~~~
z=\left(\frac{a_1}{a_2}\right)^p~,~~~
C=\frac{\left(qa_1a_2+(qa_1a_2)^{-1}-2\right)}{(q-q^{-1})^2}~,$$
satisfy Eq.(\ref{2.j4.1}).
\item Let $\chi \in D$, then there are exactly two irreducible finite
dimensional representations $\pi^\pm_{j-1},~
\pi_{p-j-1}^\pm\in{\rm Rep}SU_q(2)$ in $X^{-1}(\chi)$, where
$\pi_l^\pm$ signifies the usual highest weight representation
(\ref{2.j1.7}) specialized to $q^p=1$. We have

$${\rm dim}\pi_{j-1}^\pm+{\rm dim}\pi^\pm_{p-j-1}=p~.$$
\end{itemize}}
In conclusion, the independent continuous parameters of the irreducible
finite-dimensional representations of $SU_q(2)$ is three; dim$\pi\leq
p~~(\pi\in {\rm Rep}SU_q(2))$ and the equality holds for general $\pi$.

\subsubsection{Representations of $U_q(g)$, the general case}
{\bf The continuity of regular representations}

For a general quantum group $U_q(g)$, we want to find the
regular representations and the conditions
on the restricted tensor product. Use is made of the Weyl's character
formula to find the regular alcove in
the space of weight. For a representation
with highest weight $\mu$ the expression for the $q$-dimension is

\begin{equation}\label{2.3.39}
D_q(\mu)=\prod_{\alpha>0}\frac{[\langle\mu+\rho,\alpha\rangle]}
{[\langle\rho,\alpha\rangle]}~,
{}~~~~~\rho=\frac{1}{2}\sum_{\alpha>0}\alpha~.
\end{equation}
If $w_i$ is the Weyl reflection with respect
to the simple root $\alpha_i$,

\begin{equation}
\prod_{\alpha>0}\frac{[\langle w_i(\mu)+w_i(\rho), \alpha\rangle]}
{[\langle\rho,\alpha\rangle]}
=-\prod_{\alpha>0}\frac{[\langle\mu+\rho, \alpha \rangle]}{[\langle
\rho,\alpha\rangle]}~,
\end{equation}
and for an element in the Weyl group,

\begin{equation}
\prod_{\alpha>0}\frac{[\langle w(\mu+\rho), \alpha\rangle]}{[\langle
\rho,\alpha\rangle]}
=\epsilon(w)\prod_{\alpha>0}\frac{[\langle\mu+\rho, \alpha\rangle]}
{[\langle\rho,\alpha\rangle]}~,
\end{equation}
where $\epsilon$ is the parity of $w$. Then, we obtain

\begin{equation}
D_q\left(w(\mu+\rho)-\rho+p\sum_{i=1}^ln_i\alpha_i\right)=\epsilon(w)
D_q(\mu)~.
\end{equation}
The $\alpha_i~(i=1,\cdots,l)$ are the simple roots of the classical
algebra $g$. Every positive root can be written as $\alpha=\sum
n_i\alpha_i,~n_i\geq 0$, and $\displaystyle\sum_in_i\equiv{\rm
level}(\alpha)$.

For the highest root $\theta$, the level plus one is
the dual Coxeter number of the algebra
$\hat{g}=(\theta,\theta+2\rho)/\theta^2$. The highest root is normalized
to be length $2$. The largest value of
$(\mu+\rho,\alpha),~\alpha>0$, of a weight $\mu$ is obtained for the highest
root $\theta$, $\langle\rho,\theta\rangle=\hat{g}-1$. The denominator of
Eq.(\ref{2.3.39}) can be written as

\begin{equation}
\prod_{\alpha>0}\left[\langle\rho,\alpha\rangle\right]=
\prod_{l(\alpha)=1}^{\hat{g}-1}
\left[l(\alpha)\right]^{Nl(\alpha)}~,
\end{equation}
where $l(\alpha)$ is the level of $\alpha$ and $Nl(\alpha)$ is the number of
positive roots with the same level. For $q^p=\pm 1,~p>\hat{g}$, the
$q$-dimension
of the generating representations of $U_q(g)$ are positive.
For the representations in increasing values of $\langle\mu,\theta\rangle$
the $q$-dimension remains  positive until $\langle\mu,
\theta\rangle=p-\hat{g}$, the $q$-dimension vanishes for
$\langle\mu,\theta\rangle=p-\hat{g}+1$. Beyond this values it can be
positive, negative or zero. And null vectors, reducible but not fully
reducible
representations etc. begin to appear. In analogy with the case of
$SU_q(2)$, let $p=k+\hat{g}$, the regular irreducible representations
acquire the highest weights that are not in

\begin{equation}
{\rm Im}~(e_\theta)^{p-\hat{g}+1}~.
\end{equation}
The first with vanishing $q$-dimension appears as
$\langle\mu,\theta\rangle=k+1$ with  $\vert\mu\rangle
=(f_\theta)^{k+1}\vert\alpha\rangle$. Thus, the condition
$\vert\mu\rangle\not\in
{\rm Im}~(e_\theta)^{k+1}$ makes the representations restricted to those
with $\langle\mu,\theta\rangle\leq k$. Because
$\langle\mu,,\theta\rangle\geq\langle\mu,\alpha\rangle,~(\alpha>0)$,
we then obtain $(f_\alpha)^{k+1}\vert\mu\rangle=0,~\forall\alpha.$\\ \\
{\bf Cyclic representations}

Similar to the case of $SU_q(2)$, the new type of the irreducible
finite-dimensional representations  appears for general $U_q(g)$, when
$q$ is a root of unity. It is worth finding out the center elements of
$U_q(g)$ first.
To describe the center we need to prepare the braid group actions. For
each $i=1,~\cdots,~l$ the automorphisms $T_i$ of
$U_q(g)$ can be introduced by the following formulas,

\begin{equation}
\begin{array}{rcl}
&&T_i(e_i)=-f_it_i~,\\[3mm]
&&T_i(e_j)=\displaystyle\sum_{m=0}^{-a_{ij}}(-1)^{m-a_{ij}}q_i^{-m}
\frac{e_i^{-a_{ij}-m}}{[-a_{ij}-m]_{q_i}!}e_j
\frac{e_i^{m}}{[m]_{q_i}!}~,~~~{\rm if~}i\not=j~;
\end{array}
\end{equation}

\begin{equation}
\begin{array}{rcl}
&&T_i(f_i)=-t_i^{-1}e_i~,\\[3mm]
&&T_i(f_j)=\displaystyle\sum_{m=0}^{-a_{ij}}(-1)^{m-a_{ij}}q_i^{m}
\frac{f_i^{m}}{[m]_{q_i}!}f_j
\frac{f_i^{-a_{ij}-m}}{[-a_{ij}-m]_{q_i}!}~,~~~{\rm if~}i\not=j~;
\end{array}
\end{equation}

\begin{equation}
T_i(t_j)=t_jt_i^{-a_{ij}}~.~~~
{}~~~~~~~~~~~~~~~~~~~~~~~~~~~~~~~~~~~~~~~~~~~~~~~~~
\end{equation}
Now, let $w_0$ be the longest element of the Weyl group, and fix a
reduced decomposition
$$w_0=s_{i_1}s_{i_2}\cdots s_{i_\nu}~.$$
It is well-known that the sequence
$$\alpha_{i_1},~s_{i_1}(\alpha_{i_2}),~s_{i_1}s_{i_2}(\alpha_{i_3}),\cdots,
s_{i_1}s_{i_2}\cdots s_{i_\nu}(\alpha_{i_\nu})$$
coincides with the set
of positive roots. For example, if ${ g}= sl(3)$ and
$w_0=s_1s_2s_1$, we have $\alpha_1,~\alpha_1+\alpha_2,~\alpha_2$. This
allows us to define the root vectors $X_{\alpha}$ by

\begin{equation}
\begin{array}{llll}
X_{\alpha_{i_1}}=e_{i_1}~,~~~&X_{\alpha_{i_2}}=T_{i_1}(e_{i_2})~,~~~
&X_{\alpha_{i_3}}=T_{i_1}T_{i_2}(e_{i_3})~,~~~&\cdots~,\\[2mm]
X_{-\alpha_{i_1}}=f_{i_1}~,~~~&X_{-\alpha_{i_2}}=T_{i_1}(f_{i_2})~,~~~
&X_{-\alpha_{i_3}}=T_{i_1}T_{i_2}(f_{i_3})~,~~~&\cdots~.
\end{array}
\end{equation}
As having done for $SU_q(2)$ one can verify
that $e_i^p,~f_i^p,~t_i^p$ belong to the center ${\cal Z}$. Since $T_i$
are automorphisms, $(X_\alpha)^p~(\alpha\in\Delta$, the set of roots) also
belong to ${\cal Z}$. Let Rep$U_q(g)$ denote the equivalence classes of
the irreducible finite-dimensional representations  of $U_q(g)$ and
${\cal Z}_0$ denote the subalgebra generated by $(X_\alpha)^p~(\alpha\in
\Delta),~ t_i^{\pm p}~(1\leq i\leq l)$. There exists a natural map

$${\rm Rep}U_q(g)\stackrel{X}{\longrightarrow}{\rm Spec}{\cal Z}~,$$
where Spec${\cal Z}$ means the set of algebra homomorphisms
$\chi:~{\cal Z}\to {\bf C}$.\\
The following theorem can then be proved.
{\em
\begin{itemize}
\item  $(X_\alpha)^p~(\alpha\in\Delta),~t_i^p~(1\leq i\leq l)$ are algebraic
independent, and ${\cal Z}$ is algebraic over ${\cal Z}_0$.
\item  $X$ is surjective with finite fiber. For general $\chi\in {\rm
Spec}{\cal Z},~X^{-1}(\chi)$ consists of a single representation of
dimension $p^l$.
\end{itemize}}
In particular the theorem indicates that: the number of  continuous
parameters
of the  irreducible finite-dimensional representations is equal to the
dimension of the classical
algebra $g$. For $\pi\in{\rm Rep}U_q(g)$, the dimension of $\pi$ is less or
equal to $p^l$.

\section{Hamiltonian system}

In this section, through a concrete example -- the symmetric top system,
we discuss quantum symmetry in Hamiltonian systems.
To investigate the motion of rigid body, it is convenient to introduce the
components
of the angular momentum $J_i$ resolved
with respect to the axes in the body frame \cite{Sudarshan,Goldstein}.
The total kinetic energy of the rigid body can be expressed
in  terms of $J_i$'s by

\begin{equation}\label{5.1.1}
T=\frac{1}{2}I_{mn}^{-1}J_mJ_n~,
\end{equation}
where $I=||I_{ik}||$ is both symmetric and positive definite, $I_{ik}$ are
components of inertia tensor of the rigid body. In the usual Lagrangian
approach, equations of motion of the rigid body are

\begin{equation}\label{5.1.2}
\dot J_i=\epsilon_{ijm}J_jI_{mn}^{-1}J_n~,
\end{equation}
that possess a simple and elegant form. The interesting point is that the
equations are solely described by $J_i$'s. It is straightforward to
verify   that $J_i$'s give the Poisson bracket (PB) realization of the
Lie algebra $su(2)$

\begin{equation}\label{5.1.3}
[J_i,J_j]_{PB}=-\epsilon_{ijk}J_k~.
\end{equation}
If we rewrite the equations of motion in the  form

\begin{equation}\label{5.1.4}
\dot J_i=-\epsilon_{mnj}J_j\frac{\partial J_i}{\partial J_m}
                          \frac{\partial T}{\partial J_n}~,
\end{equation}
they can be put into the standard Hamiltonian formalism

\begin{equation}\label{5.1.5}
\dot J_i=[J_i,T]_{PB}~.
\end{equation}
Because only the variables $J_i$'s appear, the above equations may be directly
computed by just using the Lie algebra $su(2)$ and the derivation property.
Thus, as far as the equations of motion are concerned, all that we need are
the expression of the Hamiltonian as a function of the $J_i$'s, and the
Poisson brackets among the $J_i$'s. $J_i$'s do not form a complete set of
dynamical variables of rigid body rotation. However, there is a family of
classical systems, each of which is completely described by the  variables
$J_i$'s, with proper physical interpretation. The
equations of motion can thus be written in the Hamiltonian form but using a
generalized Poisson bracket (GPB). The basic GPB's among $J_i$'s are
postulated to have the form

\begin{equation}\label{5.1.6}
[J_i,J_j]_{GPB}=-\epsilon_{ijk}J_k~,
\end{equation}
and the GPB of any two functions $f({\bf J})$ and $g({\bf J})$ is computed
by the derivation  property

\begin{equation}\label{5.1.7}
[f({\bf J}),g({\bf J})]_{GPB}=-\epsilon_{mni}J_i
                                \frac{\partial f}{\partial J_m}
                                \frac{\partial g}{\partial J_n}~.
\end{equation}
Such classical dynamical systems are called as the classical pure-spin systems,
which differ from the systems of rotating rigid body:
\begin{itemize}
\item for the former, all dynamical variables are defined as suitable functions
      of the $J_i$'s;
\item the three-dimensional space with the $J_i$'s as coordinates, each with
      a specified range, forms the generalized phase space of the systems;
\item the solution of the equations of motion that equate each
      time
      derivative $\dot J_i$ with the GPB of $J_i$'s with a Hamiltonian
      $H({\bf J})$, amounts to a complete solution of the motion and suffices
      to determine the ''phase'' of the system at
      any time in terms of its ''phase'' at an earlier time.
\end{itemize}
For a pure-spin system, because the $J_i$'s form a complete set of
dynamical variables, there are no other invariants than the Casimir's,
and there is only one of them, to wit, $J^2=\displaystyle\sum_{i=1}^3
J_iJ_i$. This Casimir is invariant
under all generalized canonical transformations. It thus follows that each
transformation of such kind
acts as a canonical mapping of surface of sphere onto themselves
in the three-dimensional space of $J_i$'s, which preserving the radii and the
basic
GPB's. Therefore, the phase space for a pure-spin system is composed of some
set of
surfaces of spheres centered on the origin in the three-dimensional space.
For example, it could be the entire three-dimensional space, with each $J_i$'s
varying independently from $-\infty$ to $+\infty$, or it could be only the
surface of some sphere, with $J^2$ having some given numerical value
characteristic of the system and with only two independent $J_i$.
Various intermediate possibilities are easily
conceivable. To preserve the symmetry of the formulae, even if $J^2$ is
constrained to have some given value, we treat all three $J_i$ as independent
variables for partial differentiation. As long as the
partial differentiation are merely associated with the computations of
GPB's, the value of $J^2$ follows at the end of all
calculations.

\subsection{Classical symmetric top system}
The standard Hamiltonian of a symmetric top is

\begin{equation}\label{5.e233}
H=\frac{J_1^2+J_2^2}{2I}+\frac{J_3^2}{2I_3}~.
\end{equation}
By the above discussion, the phase space of the symmetric top
system can be of the form

\begin{equation}\label{5.e197}
M_0:~~~~~~J_{1}^{2}+J_{2}^{2}+J_{3}^{2}=J_{0}^{2}~.
\end{equation}
In the phase space $M_0$, we define a symplectic form by

\begin{equation}\label{5.e235}
\Omega_{0}=\displaystyle\frac{1}{2J_0^{2}}\sum_{i,j,k=1}^3 \epsilon_{ijk}
          J_{i} dJ_{j}\wedge dJ_{k}~.
\end{equation}
For the one-parameter group of the rotations around the $i$-th axis there
exist  Hamiltonian vector fields $X_{J_i}$ with respect to $\Omega_0$

\begin{equation}\label{5.1.11}
X_{J_i}=\sum_{j,k=1}^3\epsilon_{ijk}J_j\frac{\partial}{\partial J_k}~.
\end{equation}
By using the relation (\ref{5.e197}), it is not difficult to show that the
following relations are satisfied

\begin{equation}\label{5.1.12}
X_{J_i}\rfloor\Omega_0=-dJ_i,~~~~~~~\Omega_0(X_{J_i},X_{J_j})
=[J_i,J_j]_{GPB}~.
\end{equation}
The generalized Poisson brackets among the variables $J_{i}$ read

\begin{equation}\label{5.2.8}
[J_i,J_j]_{GPB}=-X_{J_i} J_{j}=-\epsilon_{ijk} J_{k}~,
\end{equation}
or

\begin{equation}\label{5.e202}
[J_{3},J_{\pm}]_{GPB}=\pm i J_{\pm}~,~~~
[J_{+},J_{-}]_{GPB}=i2J_{3}~,
\end{equation}
where $J_\pm=J_1\pm iJ_2$.
{}From the above Lie algebra $su(2)$, it is not difficult to show that

\begin{equation}\label{5.2.10}
\begin{array}{l}
\dot J_1=[J_1,H]_{GPB}=\displaystyle\frac{I-I_3}{II_3}J_2J_3~,\\[3mm]
\dot J_2=[J_2,H]_{GPB}=\displaystyle\frac{I_3-I}{II_3}J_1J_3~,\\[3mm]
\dot J_3=[J_3,H]_{GPB}=0~.
\end{array}
\end{equation}
Thus, we obtained the equations of motion for the symmetric top in terms of
the symplectic geometry.

Following the above method, we are going to discuss the symmetric
top system with deformed Hamiltonian and symplectic structure
\cite{Chang92b}--\cite{Chang92a}. To begin with, we  write  the deformed
Hamiltonian of the
symmetric top as,

\begin{equation}\label{5.2.31}
H_q=\frac{I-I_3}{2II_3}J_3^{\prime 2}+\frac{1}{2I}\left(J_1^{\prime 2}+
J_2^{\prime 2}+
\frac{(\sinh\gamma J'_3)^{2}}{\gamma \sinh\gamma}\right)~.
\end{equation}
The observables $J_i^\prime$'s are related with $J_i$'s by

\begin{equation}\label{5.F1}
\begin{array}{l}
J_1^\prime=\displaystyle\frac{1}{\sqrt{\gamma\sinh\gamma}}
\sqrt{\displaystyle\frac{\sinh\gamma(J_0+J_3)\sinh\gamma(J_0-J_3)}
                        {(J_0+J_3)(J_0-J_3)}}J_1~,\\[4mm]
J_2^\prime=\displaystyle\frac{1}{\sqrt{\gamma\sinh\gamma}}
\sqrt{\displaystyle\frac{\sinh\gamma(J_0+J_3)\sinh\gamma(J_0-J_3)}
                        {(J_0+J_3)(J_0-J_3)}}J_2~,\\[4mm]
J_3^\prime=J_3~.
\end{array}
\end{equation}
In terms of the $J_i^\prime$'s, Eq.(\ref{5.e197}) is

\begin{equation}\label{5.e205}
M_0^q:~~~~~J_{1}^{\prime 2}+J_{2}^{\prime 2}+\displaystyle
\frac{(\sinh\gamma J_{3}^\prime)^{2}}
    {\gamma \sinh\gamma}=J_q^{2}~,
\end{equation}
where  $J_q=\displaystyle\frac{\sinh\gamma J_{0}}{\sqrt{\gamma\sinh\gamma}}$.

On the deformed sphere $M_0^q$, the symplectic form \cite{Fei91,Fei92} is
defined by

\begin{equation}\label{5.e206}
\Omega_q=\displaystyle\frac{1}{J^{2}_q}
(J_1^\prime dJ_2^\prime \wedge dJ_3^\prime +J_2^\prime dJ_3^\prime
\wedge dJ_1^\prime
+\displaystyle\frac{\tanh\gamma J_3^\prime}{\gamma} dJ_1^\prime
\wedge dJ_2^\prime)~.
\end{equation}
The Hamiltonian vector fields $X_{J_{i}^\prime}$ on $M_0^q$
now possess the form

\begin{equation}
\begin{array}{l}
X_{J_1^\prime}=-J_2^\prime \displaystyle\frac{\partial}{\partial J_3^\prime}+
\displaystyle\displaystyle\frac{\sinh 2\gamma J_3^\prime}
{2\sinh\gamma}\displaystyle\frac{\partial}{\partial J_2^\prime}~,\\[6mm]
X_{J_2^\prime}=-\displaystyle\frac{\sinh 2 \gamma J_3^\prime}
{2 \sinh \gamma}\displaystyle\frac{\partial}{\partial
J_1^\prime}+J_1^\prime \displaystyle\frac{\partial}{\partial
J_3^\prime}~,\\[6mm]
X_{J_3^\prime}=-J_1^\prime\displaystyle\frac{\partial}{\partial J_2^\prime}
+J_2^\prime\displaystyle\frac{\partial}{\partial J_1^\prime}~.
\end{array}
\end{equation}
It is not difficult to verify that the Hamiltonian vector fields satisfy the
relations

\begin{equation}
\begin{array}{l}
 X_{J_i^\prime}\rfloor \Omega_q=-dJ_i^\prime~,\\[4mm]
 [X_{J_i^\prime},X_{J_j^\prime}]=-X_{[J_i^\prime,J_j^\prime]_{GPB}}~,\\[4mm]
 \Omega_q(X_{J_i^\prime},X_{J_j^\prime})=[J_i^\prime,J_j^\prime]_{GPB}~.
\end{array}
\end{equation}
Then, we get the basic generalized Poisson brackets,

\begin{equation}\label{5.e209}
\begin{array}{l}
[J_1^\prime,J_2^\prime]_{GPB}=-\displaystyle\displaystyle\frac
{\sinh 2 \gamma J_3^\prime}{2\sinh\gamma}~,\\[4mm]
[J_2^\prime,J_3^\prime]_{GPB}=-J_1^\prime~,
{}~~~~~~[J_3^\prime,J_1^\prime]_{GPB}=-J_2^\prime~;
\end{array}
\end{equation}
or

\begin{equation}
[J_+^\prime,J_-^\prime]_{GPB}=i \displaystyle
\frac{\sinh 2\gamma J_3^\prime}{\sinh\gamma}~,~~~
[J_3^\prime,J_{\pm}^\prime]_{GPB}=\pm i J_{\pm}^\prime~.
\end{equation}
This algebra is the quantum algebra $SU_{q,\hbar\rightarrow 0}(2)$
\cite{Chang90a}--\cite{Flato}.

We now introduce two open sets $U_\pm$ on the phase space $M_0^q$,

\begin{equation}
U_{\pm}=\left\{ x \in M_0^q \mid J_q \pm
\displaystyle\frac{\sinh \gamma J_3^\prime}
{\sqrt{\gamma \sinh \gamma}}\not= 0\right\}~,
\end{equation}
and two complex functions $ z_+$ and $ z_-$ on $ U_+ $ and $ U_-$,
respectively,

\begin{equation}\label{5.e212}
z_{\pm}=(J_1^\prime\mp i J_2^\prime)\left(J_q\pm\displaystyle
\frac{\sinh\gamma J_3^\prime}
{\sqrt{\gamma \sinh\gamma}}\right)^{-1}~.
\end{equation}
In $U_+\bigcap U_-$ we have

\begin{equation}
z_{+} z_{-} =1~.
\end{equation}

In order to construct a Hopf algebra structure for
the  quantum algebra $SU_{q,\hbar\rightarrow 0}(2)$, we first search for a set
of classical
operators $\tilde{J}_{i}^\prime$ which give the Lie bracket realization of the
quantum algebra by using the pre-quantization method \cite{Sniatychi}.
{}From the definition of complex coordinates $z_{+}$ and
$z_{-}$ introduced on $M_0^q$ of Eq.(\ref{5.e212}), we get the expressions for
$J_{i}^\prime$'s, in terms of  $z_{+}$ and $z_{-}$,

\begin{equation}\label{5.e214}
\begin{array}{l}
J_1^\prime=-J_q\displaystyle\displaystyle\frac{z_{\pm}+\overline{z}_{\pm}}
{1+z_{\pm}\overline{z}_{\pm}}~,~~~
J_2^\prime=\mp i J_q\displaystyle\displaystyle\frac{z_{\pm}-\overline{z}_{\pm}}
{1+z_{\pm}\overline{z}_{\pm}}~,\\[5mm]
\displaystyle\frac{\sinh\gamma J_3^\prime}{\sqrt{\gamma \sinh\gamma}}
=\mp J_q\displaystyle\frac{1-
z_{\pm}\overline{z}_{\pm}}{1+z_{\pm}\overline{z}_{\pm}}~.
\end{array}
\end{equation}
Then, we can write the q-deformed symplectic
form (\ref{5.e206}) as

\begin{equation}
\begin{array}{rcl}
\Omega_q\vert_{ U_{\pm}}&=&2 i J_q\left(J^{2}_q\gamma^{2}
\displaystyle\frac{(1-
z_{\pm}\overline{z}_{\pm})^2}{(1+z_{\pm}\overline{z}_{\pm})^{2}}+
\frac{\gamma}{\sinh\gamma}\right)^{-\frac{1}{2}}
\displaystyle\frac{d\overline{z}_{\pm}\wedge dz_{\pm}}
{(1+z_{\pm}\overline{z}_{\pm})^2}\\[6mm]
&=&-i Q_{\pm}d\overline{z}_{\pm}\wedge dz_{\pm}~,
\end{array}
\end{equation}
where
\begin{equation}
Q_{\pm}=-2 J_q
\left(J^{2}_q\gamma^{2}\displaystyle\frac{(1-z_{\pm}\overline{z}_{\pm})^2}
{(1+z_{\pm}\overline{z}_{\pm})^{2}}+
\frac{\gamma}{\sinh\gamma}\right)^{-\frac{1}{2}}(1+z_{\pm}\overline{z}_{\pm})^{-2}~.
\end{equation}
Since $\Omega_q$ is closed, it should be locally exact on the open set
$U_+$ and $U_-$, i.e.,

\begin{equation}\label{5.2.45}
\Omega_q\vert_{U_{\pm}}=d \theta_{\pm}~.
\end{equation}
Here the symplectic one forms $ \theta_{\pm}$ read

\begin{equation}\label{5.2.46}
\begin{array}{rcl}
\theta_{\pm}&=&-\displaystyle\frac{i}{\gamma z_{\pm}}\left(\sinh^{-1} \left(J_q
\sqrt{\gamma \sinh\gamma}\frac{1-z_{\pm}\overline{z}_{\pm}}
{1+z_{\pm}\overline{z}_{\pm}}\right)-\sinh^{-1}\left(J_q
\sqrt{\gamma \sinh\gamma}\right)\right)dz_{\pm}\\[4mm]
&=&-i p_{\pm}dz_{\pm}~,
\end{array}
\end{equation}
where
\begin{equation}\label{5.2.47}
p_{\pm}=\displaystyle\frac{1}{\gamma z_{\pm}}\left(\sinh^{-1}\left(J_q
\sqrt{\gamma \sinh\gamma}\displaystyle\frac{1-z_{\pm}\overline{z}_{\pm}}
{1+z_{\pm}\overline{z}_{\pm}}\right)-\sinh^{-1}\left(J_q
\sqrt{\gamma \sinh\gamma}\right)\right)~.
\end{equation}
The Hamiltonian vector fields of $J_{i}^\prime$'s now possess the form

\begin{equation}\label{5.e220}
\begin{array}{l}
X_{J_1^\prime}=i \sqrt{\displaystyle\displaystyle\frac{\gamma}{\sinh\gamma}}
\displaystyle\displaystyle\frac{\cosh\gamma J_3^\prime}{2}
\left((z_{\pm}^{2}-1)\displaystyle\displaystyle\frac{\partial}
{\partial z_{\pm}}+(1-\overline{z}_{\pm}^{2})
\displaystyle\displaystyle\frac{\partial}{\partial \overline{z}_{\pm}}\right)~,
\\[6mm]
X_{J_2^\prime}=\mp\sqrt{\displaystyle\displaystyle\frac{\gamma}{\sinh\gamma}}
\displaystyle\displaystyle\frac{\cosh\gamma J_3^\prime}{2}
\left((z_{\pm}^{2}+1)\displaystyle\displaystyle\frac{\partial}{\partial
z_{\pm}}+
(1+\overline{z}_{\pm}^{2})
\displaystyle\displaystyle\frac{\partial}{\partial \overline{z}_{\pm}}\right)~,
\\[6mm]
X_{J_3^\prime}=\mp i
\left(\overline{z}_{\pm}\displaystyle\displaystyle\frac{\partial}
{\partial \overline{z}_{\pm}}-z_{\pm}\displaystyle\displaystyle\frac{\partial}
{\partial \overline{z}_{\pm}}\right)~.
\end{array}
\end{equation}
Let us rewrite expressions (\ref{5.e214}) in terms of the variables $z_{\pm}$
and $p_{\pm}$,

\begin{equation}\label{5.e221}
\begin{array}{rcl}
J_1^\prime&=&-\displaystyle\frac{1}{\sqrt{\gamma\sinh\gamma}}
             \left(\cosh\left(\frac{\gamma}{2}z_{\pm}p_{\pm}\right)
             z_{\pm}\sinh\left(\frac{\gamma}{2}\left(z_{\pm}p_{\pm}+2b\right)
             \right)\right.\\[4mm]
&&~~~~~~\displaystyle\left.-\cosh\left(\frac{\gamma}{2}\left(z_{\pm}p_{\pm}+2b\right)

\right)\frac{1}{z_{\pm}}\sinh\left(\frac{\gamma}{2}z_{\pm}p_{\pm}\right)
        \right)~,\\[4mm]
J_2^\prime&=&\displaystyle\frac{\mp i}{\sqrt{\gamma\sinh\gamma}}
             \left(\cosh\left(\frac{\gamma}{2}z_{\pm}p_{\pm}\right)
             z_{\pm}\sinh\left(\frac{\gamma}{2}\left(z_{\pm}p_{\pm}+2b\right)
             \right)\right.\\[4mm]
&&~~~~~~\displaystyle+\left.\cosh\left(\frac{\gamma}{2}\left(z_{\pm}p_{\pm}+2b\right)
        \right)\frac{1}{z_{\pm}}\sinh\left(\frac{\gamma}{2}z_{\pm}p_{\pm}
        \right)\right)~,\\[4mm]
J_3^\prime&=&\mp\left(z_{\pm}p_{\pm}+b\right)~.
\end{array}
\end{equation}
Here, for convenience, we have used the relation

\begin{equation}\label{5.e267}
\sinh\gamma b=J_q\sqrt{\gamma\sinh\gamma}~.
\end{equation}
The Hamiltonian vector fields of $z$ and $p$ are

\begin{equation}\label{5.e268}
X_{z_{\pm}}=-i Q_{\pm}^{-1}\displaystyle\displaystyle\frac{\partial}
{\partial \overline{z}_{\pm}}~,~~~
X_{p_{\pm}}=i\displaystyle \displaystyle\frac{\partial}{\partial z_{\pm}}-
 i Q_{\pm}^{-1}\displaystyle\frac{\partial p_{\pm}}{\partial z_{\pm}}
\displaystyle\displaystyle\frac{\partial}{\partial \overline{z}_{\pm}}~,
\end{equation}
where the relation $\displaystyle\frac{\partial p_{\pm}}{\partial\overline{z}_
{\pm}}=Q_{\pm}$ has been used.\\
We get their pre-quantization operator representations as,

\begin{equation}\label{5.e224}
\tilde{z}_{\pm}=-Q_{\pm}^{-1}\displaystyle\displaystyle\frac{\partial}
{\partial \overline{z}_{\pm}}-z_{\pm}~,~~~
\tilde{p}_{\pm}=\displaystyle\displaystyle\frac{\partial}{\partial z_{\pm}}-
Q_{\pm}^{-1}\displaystyle\frac{\partial p_{\pm}}{\partial z_{\pm}}
\displaystyle\displaystyle\frac{\partial}{\partial \overline{z}_{\pm}}~.
\end{equation}
It is not difficult to verify that the commutator of $ \tilde{z}$ and $
\tilde{p}$ is

\begin{equation}\label{5.e225}
[\tilde{z}_{\pm},\tilde{p}_{\pm}]=1~.
\end{equation}
By means of the formulas of Eq.(\ref{5.e224}), the pre-quantization operators
with
respect to Eq.(\ref{5.e221}) can be expressed as

\begin{equation}\label{5.e226}
\begin{array}{rcl}
\tilde{J}_{1}^\prime&=&-\displaystyle\frac{1}{\sqrt{\gamma\sinh\gamma}}
        \left(\cosh\left(\frac{\gamma}{2}\tilde{z}_{\pm}\tilde{p}_{\pm}\right)
        \tilde{z}_{\pm}\sinh\left(\frac{\gamma}{2}\left(\tilde{z}_{\pm}
        \tilde{p}_{\pm}+2b\right)\right)\right.\\[4mm]
&&~~~~~~\left.\displaystyle-\cosh\left(\frac{\gamma}{2}\left(\tilde{z}_{\pm}
        \tilde{p}_{\pm}+2b\right)\right)\frac{\gamma\tilde{p}_{\pm}}{2}
        \sum_{n=0}^{\infty}\left(\left(2n+1\right)!\right)^{-1}\left(

\frac{\gamma}{2}\tilde{z}_{\pm}\tilde{p}_{\pm}\right)^{2n}\right)~,\\[4mm]
\tilde{J}_{2}^\prime&=&\displaystyle\frac{\mp i}{\sqrt{\gamma\sinh\gamma}}
        \left(\cosh\left(\frac{\gamma}{2}\tilde{z}_{\pm}\tilde{p}_{\pm}\right)
        \tilde{z}_{\pm}\sinh\left(\frac{\gamma}{2}\left(\tilde{z}_{\pm}
        \tilde{p}_{\pm}+2b\right)\right)\right.\\[4mm]
&&~~~~~~\displaystyle\left.+\cosh\left(\frac{\gamma}{2}\left(\tilde{z}_{\pm}
        \tilde{p}_{\pm}+2b\right)\right)\frac{\gamma\tilde{p}_{\pm}}{2}
        \sum_{n=0}^{\infty}\left(\left(2n+1\right)!\right)^{-1}\left(

\frac{\gamma}{2}\tilde{z}_{\pm}\tilde{p}_{\pm}\right)^{2n}\right)~,\\[4mm]
\tilde{J}_{3}^\prime&=&\mp\left(\tilde{z}_{\pm}\tilde{p}_{\pm}+b\right)~.
\end{array}
\end{equation}
It is straightforward to verify that they yield the Lie bracket realization of
the  quantum algebra $SU_{q,\hbar\rightarrow 0}(2)$ by owing to
Eq.(\ref{5.e225}),

\begin{equation}\label{5.e227}
[\tilde{J}_{1}^\prime,\tilde{J}_{2}^\prime]= -i\displaystyle
 \displaystyle\frac{\sinh 2 \gamma\tilde{ J}_{3}^\prime}
{2 \gamma}~,~~~~
[\tilde{J}_{2}^\prime,\tilde{J}_{3}^\prime ]= -i \tilde{J}_{1}^\prime~,~~~~
[\tilde{J}_{3}^\prime,\tilde{J}_{1}^\prime ]= -i \tilde{J}_{2}^\prime~,
\end{equation}
or

\begin{equation}\label{5.e228}
[\tilde{J}_{+}^\prime,\tilde{J}_{-}^\prime]=-\displaystyle
\frac{\sinh 2 \gamma \tilde{J}_{3}^\prime}
{\gamma}~,~~~
[\tilde{J}_{3}^\prime,\tilde{J}_{\pm}^\prime]=\mp \tilde{J}_{\pm}^\prime~.
\end{equation}
Keeping the operators of Eq.(\ref{5.e226}) and commutators of Eq.(\ref{5.e227})
in mind,
we can define the Hopf  algebra structure  of the  classically
realized quantum algebra $SU_{q,\hbar\rightarrow 0}(2)$ as follows

\begin{equation}
\begin{array}{l}
\Delta(\tilde{J}_3')=\tilde{J}_3'\otimes 1+1\otimes \tilde{J}_3' ~,\\
\Delta(\tilde{J}_\pm')=\tilde{J}_\pm'\otimes e^{-\gamma\tilde{J}_3'}
+e^{\gamma\tilde{J}_3'}
\otimes \tilde{J}_3' ~,\\
\epsilon(\tilde{J}_\pm') =\epsilon(\tilde{J}_3') =0~,\\
S(\tilde{J}_\pm')=-q^{\pm 1}\tilde{J}_\pm'~,~~~~S(\tilde{J}_3')=-
\tilde{J}_3' ~.
\end{array}
\end{equation}

It is straightforward to calculate the equations of motion for the classical
symmetric top system with deformed Hamiltonian and deformed symplectic
structure
by using the quantum group $SU_{q,\hbar\to 0}(2)$

\begin{equation}\label{5.2.60}
\begin{array}{rcl}
\dot J_1^\prime&=&[J_1^\prime,H_q]_{GPB}=\displaystyle\frac{I-I_3}{II_3}
J^\prime_2J^\prime_3~,\\[3mm]
\dot J_2^\prime&=&[J^\prime_2,H_q]_{GPB}=\displaystyle\frac{I_3-I}{II_3}
J^\prime_1J^\prime_3~,\\[3mm]
\dot J^\prime_3&=&[J^\prime_3,H_q]_{GPB}=0~.
\end{array}
\end{equation}
Because the deformed Hamiltonian system $(M^q_0,\Omega_q,H_q)$
obeys the same equations of motion with the standard Hamiltonian system
$(M_0,\Omega_0,H)$, the Hamiltonian systems $(M_0,\Omega_0,H)$ and
$(M_0^q,\Omega_q,H_q)$
describe the same physical motions.
It is worth noting that the symmetry of the standard Hamiltonian system
$(M_0,\Omega_0,H)$ is the Lie group $SU(2)$, however,
the symmetry possessed by the deformed Hamiltonian system
$(M_0^q,\Omega_q,H_q)$ is the quantum group $SU_{q,\hbar\to 0}(2)$.
It is well-known that the equations of motion for the standard symmetric top
can be
solved exactly. Therefore the deformed Hamiltonian system
$(M_0^q,\Omega_q,H_q)$
is the model that can be solved exactly.

\subsection{Quantum symmetric top system}
By means of geometric quantization, we now discuss quantum symmetry in quantum
symmetric top system. Let us begin with writing down the deformed
Hamiltonian of the symmetric top \cite{Chang92c,Chang92a}

\begin{equation}\label{5.ssss}
\hat{H_q}=\frac{\hbar^2(I-I_3)}{2II_3}J_3^{\prime 2}+\frac{\hbar^2}{2I}
    \left(\frac{\gamma}{\sinh\gamma}\hat{J}_{+}^\prime \hat{J}_{-}^\prime
    +[\hat{J}_{3}^\prime ][\hat{J}_{3}^\prime +1]\right)~.
\end{equation}
The geometric quantization
of the deformed Hamiltonian system $(M_0^q,\Omega_q,H_q)$ is described by the
pre-quantization line bundle $L_q$ and the polarization $F$ \cite{Sniatychi}.
For the case under consideration  such a
quantum line bundle $L_{q}$ exists if and only if  $(2\pi)^{-1}\Omega_q$
defines an integral de Rham cohomology class, i.e., the de Rham cohomology
class $\{-(2\pi)^{-1}\Omega_q\}$ of $-(2\pi)^{-1}\Omega_q$ should be
integrable. Integrating the right hand side of Eq.(\ref{5.e206}) over the
symplectic manifold  $M_0^q$, we have

\begin{equation}
\int_{M_0^q}\Omega_q=\displaystyle\frac{1}{J_q^{2}}\left.\left(2V+
\pi \left(J_q^{2}\displaystyle\frac{\tanh\gamma J_{3}^\prime }{\gamma}
-\displaystyle\frac{\gamma J_{3}^\prime
-\tanh\gamma J_{3}^\prime}{\gamma^{2}\sinh\gamma}\right)\right)
\right\vert^{\frac{\sinh\gamma J_{3}^\prime}
{\sqrt{\gamma\sinh\gamma}}=+J_q}
 _{\frac{\sinh\gamma J_{3}^\prime}{\sqrt{\gamma\sinh\gamma}}=-J_q}~,
\end{equation}
where V is the volume of the manifold $M_0^q$,

\begin{equation}
\begin{array}{rcl}
V&=&\displaystyle\int_{M_0^q}dJ_{1}^\prime dJ^\prime_{2}dJ^\prime_{3}\\[4mm]
 &=&\left.\left(\pi J_q^{2}J^\prime_{3}-\displaystyle\frac{\pi}{2}
    \displaystyle\frac{\sinh 2\gamma J^\prime_{3}-
    2\gamma J^\prime_{3}}{2\gamma^{2}\sinh\gamma}\right)
    \right\vert^{\frac{\sinh\gamma
J^\prime_{3}}{\sqrt{\gamma\sinh\gamma}}=+J_q}
    _{\frac{\sinh\gamma J^\prime_{3}}{\sqrt{\gamma\sinh\gamma}}=-J_q}~.
\end{array}
\end{equation}
Then, we have

\begin{equation}\label{5.2.37}
\int_{M_0^q}\Omega_q=-4\pi\displaystyle\frac{\sinh^{-1}
(\sqrt{\gamma\sinh\gamma}J_q)}{\gamma}~.
\end{equation}
Setting

\begin{equation}
\frac{\sinh^{-1}(\sqrt{\gamma\sinh\gamma}J_q)}{\gamma}=J~,
\end{equation}
we get

\begin{equation}
-(2\pi)^{-1}\int_{M_0^q}\Omega_q=4\pi (2\pi)^{-1}J=2J~,
\end{equation}
which must be an integer if $\{-(2\pi)^{-1}\Omega_q\}$ is integrable.
Therefore, $2J\in Z$ and J is an integer or half integer.
It is clear now that $J_q$ takes some special values according to J,

\begin{equation}\label{5.e274}
J_q=\displaystyle\frac{\sinh\gamma J}{\sqrt{\gamma\sinh\gamma}}~.
\end{equation}
Comparing Eq.(\ref{5.e267}) with Eq.(\ref{5.e274}), we know that here b
should be integer or half integer.

For a suitable polarization let us consider the linear
frame fields $ X_{z_{\pm}}$,
\begin{equation}
X_{z_{\pm}}=- i Q_{\pm}^{-1}\displaystyle\frac{\partial}{\partial
\overline{z}_{\pm}}~.
\end{equation}
For each $x \in U_{+}\cap U_{-}$, we have
$$  X_{z_{-}}=-z_{+}^{-2}X_{z_{+}}~.$$
Thus, $X_{z_+}$ and  $X_{z_-}$ span a complex dis\-tri\-bu\-tion $F$
on $M_0^q$ and $F$ is a polari\-za\-tion of symplectic manifold
\hspace{0.5mm} ($M_0^q$, $\Omega_q$).\hspace{0.5mm} Further,

\begin{equation}\label{5.2.42}
i\Omega_q\left(X_{z_\pm},\overline{X}_{z_\pm}\right)=-\frac{1}{2J_q}
   \sqrt{J_q^{2}\gamma^{2}\frac{(1-z_\pm\overline{z}_\pm)^2}
   {(1+z_\pm\overline{z}_\pm)^2}+\displaystyle\frac{\gamma}{\sinh \gamma}}
   ~(1+z_\pm\overline{z}_\pm)^2 >0~,
\end{equation}
implying that $F$ is a complete strongly admissible positive polarization of
($M_0^q$, $\Omega_q$).

To get the quantum operator expressions for $J_{i}^\prime$ ($i=1,~2,~3$),
we start with the quantum operators of $p$ and $z$. For the polarization
preserving functions $p$ and $z$ Eq.(\ref{5.e268}) yields

\begin{equation}
[X_{p_\pm},X_{z_\pm}]=0~ ,
\end{equation}
whose quantum counterparts are

\begin{equation}
\hat{p}_\pm=-\frac{\partial}{\partial z_\pm}~,~~~
\hat{z}_\pm=z_\pm~,
\end{equation}
where the terms with derivative $\displaystyle\frac{\partial}{\partial
\overline{z}}$ have been omitted as the section space is covariantly constant
along the
polarization F, i.e., the quantum representation space
is the holomorphic section space. Thus, the quantum commutator of
$\hat{p}$ and $\hat{z}$ is

\begin{equation}\label{5.e278}
[\hat{z}_\pm,\hat{p}_\pm]=1~.
\end{equation}
Making use of Eqs.(\ref{5.e221}) and (\ref{5.e278}), we obtain the quantum
operators with suitable
ordering

\begin{equation}\label{5.e280}
\begin{array}{rcl}
\hat{J}_{1}^\prime&=&-\displaystyle\frac{1}{\sqrt{\gamma\sinh\gamma}}
       \left(\cosh\left(\frac{\gamma}{2}z_\pm\frac{\partial}{\partial z_\pm}
       \right)z_\pm\sinh\left(\frac{\gamma}{2}\left(-z_\pm\frac{\partial}
       {\partial z_\pm}+2J\right)\right)\right.\\[4mm]
&&~~~~~~\displaystyle+\cosh\left.\left(\frac{\gamma}{2}\left(-z_\pm
       \frac{\partial}{\partial z_\pm}+2J\right)\right)\frac{1}{z_\pm}
       \sinh\left(\frac{\gamma}{2}z_\pm\frac{\partial}{\partial z_\pm}\right)
       \right)~,\\[4mm]
\hat{J}_{2}^\prime&=&\displaystyle\frac{\mp i}{\sqrt{\gamma\sinh\gamma}}
       \left(\cosh\left(\frac{\gamma}{2}z_\pm\frac{\partial}{\partial z_\pm}
       \right)z_\pm\sinh\left(\frac{\gamma}{2}\left(-z_\pm\frac{\partial}
       {\partial z_\pm}+2J\right)\right)\right.\\[4mm]
&&~~~~~~\displaystyle\left.-\cosh\left(\frac{\gamma}{2}\left(-z_\pm\frac{\partial}
       {\partial z_\pm}+2J\right)\right)\frac{1}{z_\pm}\sinh\left(
       \frac{\gamma}{2}z_\pm\frac{\partial}{\partial
z_\pm}\right)\right)~,\\[4mm]
\hat{J}_{3}^\prime&=&\displaystyle\mp\left(-z_\pm\frac{\partial}{\partial
z_\pm}+J\right)~. \end{array}
\end{equation}
Equation (\ref{5.e280}) yields a realization of the quantum group $SU_q(2)$

\begin{equation}
\begin{array}{rcl}
[\hat{J}_1^\prime,\hat{J}_2^\prime]&=&-\displaystyle\frac{i\sinh\gamma}{2\gamma}
[2\hat{J}_3^\prime]_q~,\\[4mm]
[\hat{J}_2^\prime,\hat{J}_3^\prime]&=&-i \hat{J}^\prime_1~,~~~~
[\hat{J}^\prime_3,\hat{J}^\prime_1]=-i\hat{J}^\prime_2~;
\end{array}
\end{equation}
or

\begin{equation}\label{5.2.48}
[\hat{J}_+^\prime,\hat{J}^\prime_{-}]=-\displaystyle\frac{\sinh(\gamma)}{\gamma}
[2\hat{ J}^\prime_{3}]_q~,~~~
[\hat{J}^\prime_{3},\hat{J}^\prime_\pm]=\mp\hat{ J}^\prime_\pm~.
\end{equation}
The Hopf algebra structure is of the form

\begin{equation}
\begin{array}{l}
\Delta(\hat{J}_3')=\hat{J}_3'\otimes 1+1\otimes \hat{J}_3' ~,\\
\Delta(\hat{J}_\pm')=\hat{J}_\pm'\otimes e^{-\gamma\hat{J}_3'}
+e^{\gamma\hat{J}_3'}
\otimes \hat{J}_3' ~,\\
\epsilon(\hat{J}_\pm') =\epsilon(\hat{J}_3') =0~,\\
S(\hat{J}_\pm')=-q^{\pm 1}\hat{J}_\pm'~,~~~~S(\hat{J}_3')=-
\hat{J}_3' ~.
\end{array}
\end{equation}
For the quantum symmetric top system, making use of the quantum group
$SU_{q}(2)$ symmetry, we write the Heisenberg equation as

\begin{equation}\label{5.2.50}
\begin{array}{rcl}
i\hbar\dot{\hat{J}^\prime}_1&=&[\hat{J}^\prime_1,\hat{H_q}]=
                 \displaystyle\frac{i\hbar^2(I-I_3)}{2II_3}
                 \left(\hat{J}^\prime_2\hat{J}^\prime_3+\hat{J}^\prime_3
                 \hat{J}_2^\prime\right)~,\\[3mm]
i\hbar\dot{\hat{J}^\prime}_2&=&[\hat{J}^\prime_2,\hat{H_q}]=
                 \displaystyle\frac{i\hbar^2(I_3-I)}{2II_3}
                 \left(\hat{J}^\prime_1\hat{J}^\prime_3+\hat{J}^\prime_3
                 \hat{J}^\prime_1\right)~,\\[3mm]
i\hbar\dot{\hat{J}^\prime}_3&=&[\hat{J}^\prime_3,\hat{H_q}]=0~,
\end{array}
\end{equation}
or

\begin{equation}\label{5.2.50.2}
\begin{array}{rcl}
\dot{\hat{J}^\prime}_1&=&\displaystyle\frac{\hbar(I-I_3)}{2II_3}
                 \left(\hat{J}^\prime_2\hat{J}^\prime_3+\hat{J}^\prime_3
                 \hat{J}^\prime_2\right)~,\\[3mm]
\dot{\hat{J}^\prime}_2&=&\displaystyle\frac{\hbar(I_3-I)}{2II_3}
                 \left(\hat{J}^\prime_1\hat{J}_3^\prime+\hat{J}^\prime_3
                 \hat{J}^\prime_1\right)~,\\[3mm]
\dot{\hat{J}^\prime}_3&=&0~,
\end{array}
\end{equation}
which are the same as those for the quantum symmetric top system with
standard  Hamiltonian. Thus, the quantum symmetric top system with
deformed  Hamiltonian  and the quantum symmetric top system with
standard  Hamiltonian obey the same Heisenberg
equation in quantum mechanics.

The quantum counterparts \cite{Zachos} of Eq.(\ref{5.F1}) are

\begin{equation}\label{5.F3}
\begin{array}{rcl}
\hat{J}_+^\prime&=&\displaystyle\frac{1}{\sqrt{\gamma\sinh\gamma}}
\sqrt{\displaystyle\frac{\sinh\gamma(\hat{J}_0+\hat{J}_3)
\sinh\gamma(\hat{J}_0-\hat{J}_3+1)}
                        {(\hat{J}_0+\hat{J}_3)(\hat{J}_0-\hat{J}_3+1)}}
\hat{J}_+~,\\[4mm]
\hat{J}_-^\prime&=&\displaystyle\frac{1}{\sqrt{\gamma\sinh\gamma}}
\hat{J_-}\sqrt{\displaystyle\frac{\sinh\gamma(\hat{J}_0+\hat{J}_3)
\sinh\gamma(\hat{J}_0-\hat{J}_3+1)}
{(\hat{J}_0+\hat{J}_3)(\hat{J}_0-\hat{J}_3+1)}}~,\\[4mm]
\hat{J}_3^\prime&=&\hat{J}_3~.
\end{array}
\end{equation}
It is well-known that the stationary states of the symmetric top can be given
by the Wigner $D$-functions, $D^{J}_{MK}$ \cite{Biedenharn81},

\begin{equation}\label{5.2.17}
D^{J}_{MK}(\alpha,\beta,\gamma)=e^{iM\alpha+iK\gamma}d^J_{MK}(\beta)~,
\end{equation}
where
\begin{equation}\label{5.253}\begin{array}{rcl}
d^J_{MK}(\beta)&=&\left((J+M)!(J-M)!(J+K)!(J-K)!\right)^{1/2}\times\\
               &
&\displaystyle\times\sum_\nu\left((-1)^\nu(J-M-\nu)!(J+K-\nu)!(\nu+M-K)!~\nu!~
                  \right)^{-1}\times\\
               & &~~~~~\times\left(\cos\beta/2\right)^{2J+K-M-2\nu}
                    \left(-\sin\beta/2\right)^{M-K-2\nu}
\end{array}
\end{equation}
($\alpha$, $\beta$ and $\gamma$ are the Euler angles). In terms of the Euler
angles,  $\hat{J}_i$'s can be expressed as

\begin{equation}\label{5.F2}
\begin{array}{rcl}
\hat{J_+}&=&e^{i\alpha}\left(i\cot\beta\displaystyle\frac{\partial}
{\partial\alpha}+\frac{\partial}{\partial\beta}-\frac{i}{\sin\beta}
\frac{\partial}{\partial\gamma}\right)~,\\[3mm]
\hat{J_-}&=&e^{-i\alpha}\left(i\cot\beta\displaystyle\frac{\partial}
{\partial\alpha}-\frac{\partial}{\partial\beta}-\frac{i}{\sin\beta}
\frac{\partial}{\partial\gamma}\right)~,\\[3mm]
\hat{J_3}&=&-i\displaystyle\frac{\partial}{\partial\alpha}~,
\end{array}
\end{equation}
resulting in the differential equations

\begin{equation}\label{5.e253}
\begin{array}{rcl}
\hat{J}_\pm D^{J}_{MK}&=&\sqrt{(J\mp M)(J\pm M-1)}D^{J}_{M\pm 1,K}~,\\[3mm]
\hat{J}_3D^{J}_{MK}&=&K D^{J}_{MK}~,
         ~~~~\hat{J}_zD^{J}_{MK}=M D^{J}_{MK}~,
\end{array}
\end{equation}
where $\hat{J}_z$ is the projection of ${\bf \hat{J}}$ onto the $z$-axis
of lab-fixed coordinate system.
Using Eqs.(\ref{5.F2}) and (\ref{5.F3}), one can verify that
$\hat{J}_i^\prime$'s satisfy
the following differential equations,

\begin{equation}
\begin{array}{rcl}
\hat{J}_\pm^\prime D^{J}_{MK}&=&\sqrt{[J\mp M][J\pm M-1]}
D^{J}_{M\pm 1,K}~,\\[3mm]
\hat{J}_3^\prime D^{J}_{MK}&=&K D^{J}_{MK}~,
         ~~~~\hat{J}_z^\prime D^{J}_{MK}=M D^{J}_{MK}~.
\end{array}
\end{equation}
{}From the above equations, it is not difficult to verify that the
eigenvalues of the deformed Hamiltonian (\ref{5.ssss}) are

\begin{equation}\label{5.2.52}
E_{JK}^q=\frac{\hbar^2}{2I}[J][J+1]+\frac{I-I_3}{2II_3}\hbar^2K^2~.
\end{equation}

\section{Integrable lattice model}
The integrable lattice model \cite{Baxter82}, defined on a two-dimensional
square
lattice (Fig.1), can be divided into two types: vertex model
\cite{Baxter73a}--\cite{Baxter73c} and
Solid-On-Solid (S.O.S) model \cite{Andrews84}--\cite{Date87}. State variables
of vertex model are located
on the edges. We associate the Boltzmann weight with each vertex
configuration defined by the state variables on the four
edges joining together at the vertex (Fig.2).  The degrees
of freedom of S.O.S model are located on the sites and interact
through ``interaction-round-face'' around each plaquette.
The Boltzmann weight is
assigned to each unit face depending on the state variable configuration
round the face (Fig.12).

\setlength{\unitlength}{5pt}
\thicklines
\begin{picture}(50,15)(0,7)
\multiput(35,10)(0,8){2}{\line(1,0){8}}
\multiput(35,10)(8,0){2}{\line(0,1){8}}
\put(32,9){$a_i$}
\put(32,17.7){$a_l$}
\put(44,9){$a_j$}
\put(44,17.7){$a_k$}

\end{picture}

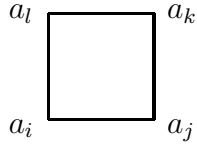
\hspace{8em}Fig.12. Boltzmann weight $w(a_i,a_j,a_k,a_l)$

\hspace{11.4em}of the S.O.S model.

A precise
study of the thermodynamics of those models indicates  that they undergo a
second phase transition at a certain critical temperature $T=T_c$. This fact,
combined with the short range of interactions, implies that in a suitable
critical continuum limit, the system be locally scale, rotation and
translation invariant \cite{Cardy87}.
Considerable progress has been made recently in understanding the
structure of those models, from their connection with
quantum groups and conformal field theory.  Integrability
of those models is ensured by local Boltzmann weights satisfying the YBE.
Previously known models have been generalized in several directions.
In particular, the celebrated
six-vertex model and the related XXZ spin 1/2 quantum chain have been
recognized to be the first hirarches, involving either higher spin
representations of $su(2)$ or higher rank algebras or both.  This
progress has been made possible by the algebraic formulation of the Yang-
Baxter integratibility condition in the quantum group form.  Concept
of quantum group becomes a major theme of the study \cite{Saleur90,Jimbo90k}.

\subsection{Vertex model}
Let us consider a two-dimensional square lattice, whose state variables
are located
on the edges. We associate the Boltzmann weight with each vertex configuration.
The configuration is defined by the state variables say, $i,~j,~k,~l$ on the
four
edges joining together at the vertex. An example is the
$6$-vertex model, for which we set a definite direction by an arrow on each
edge of the
lattice. Four edges meet at each lattice point, and so there are $12$
distinct types of combinations of arrows (Fig.13). We only consider the
configurations
with equal number of incoming and outcoming arrows.
For type $j$ configuration $(j=1,~2,~\cdots,~6)$, we assign energy
$\epsilon(j)$ which is assumed
to be invariant under simultaneous inversion of the direction
of all arrows, so that

\begin{equation}
\epsilon(1)=\epsilon(2)~,~~~~ \epsilon(3)=\epsilon(4)~,~~~~\epsilon(5)=
\epsilon(6)~.
\end{equation}

\setlength{\unitlength}{5pt}
\thicklines
\begin{picture}(50,15)(-3,1)
\multiput(0,10)(13,0){6}{\line(1,0){8}}
\multiput(4,6)(13,0){6}{\line(0,1){8}}

\multiput(0,10)(26,0){3}{\vector(1,0){2}}
\multiput(6,10)(26,0){2}{\vector(1,0){1}}
\put(67,10){\vector(-1,0){1}}
\multiput(15,10)(26,0){3}{\vector(-1,0){1}}
\multiput(21,10)(26,0){2}{\vector(-1,0){2}}
\put(71,10){\vector(1,0){1}}
\put(60,10){\vector(-1,0){2}}

\multiput(4,6)(0,5){2}{\vector(0,1){2}}
\multiput(17,8)(0,5){2}{\vector(0,-1){1}}
\multiput(30,8)(0,5){2}{\vector(0,-1){1}}
\multiput(43,6)(0,5){2}{\vector(0,1){2}}
\put(56,8){\vector(0,-1){1}}
\put(56,12){\vector(0,1){1}}
\put(69,6){\vector(0,1){2}}
\put(69,14){\vector(0,-1){2}}

\put(4,3){$1$}
\put(17,3){$2$}
\put(30,3){$3$}
\put(43,3){$4$}
\put(56,3){$5$}
\put(69,3){$6$}
\end{picture}

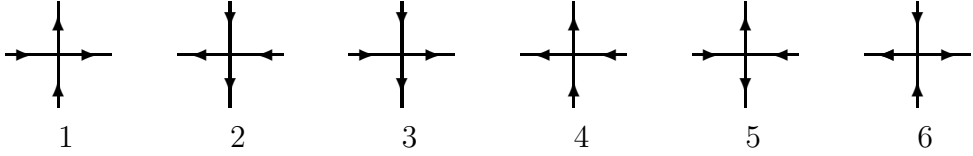
\hspace{10em}Fig.13. The six-vertex model.

This completely defines the $6$-vertex model. With each configuration on the
entire lattice,
we associate a total energy $E$ with

\begin{equation}
E=\sum_{i=1}^6 n_i\epsilon(i)~,
\end{equation}
where $n_i$ is the number of type $i$ vertex in the given configuration.
The partition function $Z_N$ and the free energy per site $f$ are then given by

\begin{equation}
\begin{array}{rcl}
& &Z_N=\sum\exp(-\beta E)~,~~~~\beta=1/k_BT~,\\[2mm]
& &f=k_BT\displaystyle\lim_{N\to\infty}N^{-1}\log Z_N~,
\end{array}
\end{equation}
where $k_B$ is the Boltzmann constant, $T$ is the (Kelvin) temperature and $N$
is the
number of lattice site and the summation is taken over all configurations of
arrows.

For a given configuration of lattice, we
consider a horizontal row of the lattice and the adjacent vertex edges.
Let $\alpha=\{\alpha_i,\cdots,\alpha_n\}$ be the state variables on the
lower row of the vertical edges, $\alpha'=\{\alpha'_i,\cdots,\alpha'_n\}$ be
the
state variables on the upper row and $\beta=\{\beta_i,\cdots,\beta_n\}$ be the
state variables on the horizontal edges.
We adopt a Hamiltonian picture, in which ``time'' flows upward
on the lattice, and the various configurations of vertical links are considered
being independent possible states of the system at a given time. Time evolution
is carried out by the row-to-row transfer matrix ${\cal V}^{(n)}$ (Fig.3),
whose
matrix elements are ${\cal V}^{(n)}_{\alpha,\alpha'}$ is defined by

\begin{equation}\label{3.2.11}
{\cal V}^{(n)}_{\alpha,\alpha'}=\sum_{\beta_1\cdots\beta_n}
                       w(\beta_1,\alpha_1,\beta_2,\alpha_1')
                                   w(\beta_2,\alpha_2,\beta_3,\alpha_2')
                \cdots w(\beta_n,\alpha_n,\beta_1,\alpha_n')~,
\end{equation}
where $w(\beta_i,\alpha_i,\beta_i',\alpha_i')$ is Boltzmann weight
of the vertex
$$w(\beta_i, \alpha_i,\beta_i', \alpha_i')=
\exp\left(-\epsilon(\beta_i,\alpha_i,\beta_i',\alpha_i')/k_BT\right)~.$$
For the general case, the degrees of freedom denoted by $\alpha$ and $\beta$
may be of a rather general nature and take their values in a discrete set.
If the vertical ones take $d$ independent values, the space spanned
by those states is $V^{\otimes n}$, where $V$ is the one-body vector space:
$V={\bf C}^d$ for the $d$-state model.
In terms of the transfer matrix ${\cal V}^{(n)}$, the partition function $Z_N$
and the free
energy per site $f$ are given by

\begin{equation}
\begin{array}{rcl}
& &Z_N=\displaystyle\sum_{\alpha^{(1)}\alpha^{(2)}\cdots \alpha^{(m)}}
   {\cal V}^{(n)}_{\alpha^{(1)}\alpha^{(2)}} {\cal V}^{(n)}_{\alpha^{(2)}
   \alpha^{(3)}}\cdots {\cal V}^{(n)}_{\alpha^{(m-1)}\alpha^{(m)}}
   {\cal V}^{(n)}_{\alpha^{(m)}\alpha^{(1)}}
   ={\rm Tr}\left({\cal V}^{(n)}\right)^m~,\\
& &f=-k_BT\displaystyle\lim_{N\to\infty}N^{-1}\log Z_N~.
\end{array}
\end{equation}
The consistency condition of the transfer matrix ${\cal V}^{(n)}$ yields
the YBE (Fig.14),\\ \\
$\displaystyle\sum_{\gamma\beta''\bar{\beta}''}
w(\beta,\alpha,\beta'',\gamma)
w(\bar{\beta},\gamma,\bar{\beta}'',\alpha')
w(\bar{\beta}'',\beta'',\beta',\bar{\beta}')$
\begin{equation}\label{3.2.18}
=\sum_{\gamma\beta''\bar{\beta}''}
w(\bar{\beta},\beta,\beta'',\bar{\beta}'')
w(\beta'',\alpha,\beta',\gamma)w(\bar{\beta}'',\gamma,\bar{\beta}',\alpha')~.
\end{equation}

\setlength{\unitlength}{4pt}
\thicklines
\begin{picture}(50,32)(-18,0)
\multiput(3,5)(30,0){2}{\line(1,1){21}}
\multiput(23,5)(30,0){2}{\line(-1,1){21}}
\multiput(10,5)(37,0){2}{\line(0,1){21}}

\put(27,15){$=$}

\multiput(3,2)(30,0){2}{$\beta$}
\multiput(10,2)(37,0){2}{$\alpha$}
\multiput(23,2)(30,0){2}{$\beta'$}

\multiput(3,26.7)(30,0){2}{$\bar{\beta}$}
\multiput(10,26.7)(37,0){2}{$\alpha'$}
\multiput(23,26.7)(30,0){2}{$\bar{\beta}'$}

\multiput(7,15)(41,0){2}{$\gamma$}
\multiput(12,19)(30,0){2}{$\bar{\beta}''$}
\multiput(12,10)(30,0){2}{$\beta''$}
\end{picture}

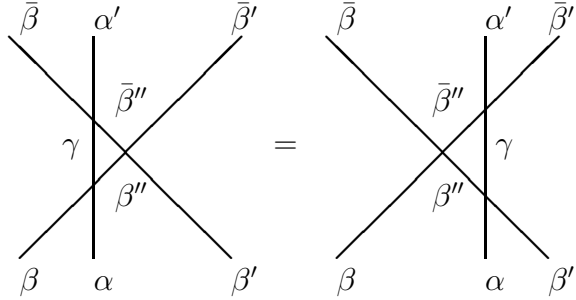
\hspace{8em}Fig.14. YBE for the vertex model.

The quantities
$w(\beta,\alpha,\beta',\alpha')$ can be straightforwardly
interpreted as an operator ${\cal R}$ in the tensor product space
$V^{\otimes 2}$.   Thus  Eq.(\ref{3.2.18}) can be rewritten in the form

\begin{equation}\label{3.cd1}
{\cal R}_{12}(u){\cal R}_{13}(u+v){\cal R}_{23}(v)={\cal R}_{23}(v)
   {\cal R}_{13}(u+v){\cal R}_{12}(u)~.
\end{equation}
Corresponds to the  6-vertex model, the solution of the YBE has the form

\begin{equation}\label{3.I8}
\begin{array}{rcl}
{\cal R}(u) &=&\displaystyle\frac{1}{4}\sum_{i=0}^3w_i\sigma^i
               \otimes \sigma^i\\[7mm]
            &=&\rho\left[\begin{array}{cccc}
                         a&   &   & \\
                          & b & c & \\
                          & b & c & \\
                          &   &   &a
                         \end{array}\right]~,

 \end{array}
 \end{equation}
 where
$$\begin{array}{rcl}
& &a=\sin(\eta+u)=\displaystyle\frac{w_0+w_3}{4\rho}=\frac{1}{\rho}
   \exp(-i\beta\epsilon(1))~,\\[3mm]
& &b=\sin
u=\displaystyle\frac{w_0-w_3}{4\rho}=\frac{1}{\rho}\exp(-i\beta\epsilon(3))~,\\[3mm]
&
&c=\sin\eta=\displaystyle\frac{w_1}{4\rho}=\frac{1}{\rho}\exp(-i\beta\epsilon(5))~,
\end{array}$$
here $\sigma^i~(i=1,~2,~3)$ are the Pauli matrices, $\sigma^0={\bf 1}$,
$w_1=w_2$,
$\eta$ is a free parameter and $\rho$ is an overall irrelevant factor.\\
Notice that for $u = 0$,

\begin{equation}
{\cal R}(0) = \rho \sin \eta\left[\begin{array}{cccc}
                               1 & 0 & 0 & 0\\
                               0 & 0 & 1 & 0\\
                               0 & 1 & 0 & 0\\
                               0 & 0 & 0 & 1 \end{array}
                        \right]= \rho \sin \eta\cdot P~,
\end{equation}
where $P$ is the transposition operator in $V\otimes V$,

\begin{equation}
\begin{array}{rcl}
P:~~V_1\otimes V_2&\to& V_2\otimes V_1~,\\
P:~~a\otimes b&\to& b\otimes a~,
\end{array}
\end{equation}
and that for $u$ and $\eta$ small, $R\approx u${\bf 1}$ + \eta P$.
The introduction  of the spectral parameter $u$ in Eq.(\ref{3.I8}) makes  the
transfer
matrix ${\cal V}^{(n)}$ $u$ dependent.
The spectral parameter
dependent transfer matrix ${\cal V}^{(n)}(u)$ may be regarded as the generating
function of conserved quantities.  The space $V^{\otimes m}$ on which
${\cal V}^{(n)}(u)$ acts, is viewed as the Hilbert space of a quantum
one-dimensional
system.  Upon introducing the operators

\begin{equation}
{\cal H}_i = \frac{\partial^i}{\partial u^i} \log {\cal V}^{(n)}(u)|_{u=0}~,
\end{equation}
the commutation of ${\cal V}^{(n)}(u)$ and ${\cal V}^{(n)}(v)$ implies the
commutation of the infinite set of ${\cal H}_{i}$

\begin{equation}
\left[{\cal H}_i, {\cal H}_j\right] =0~.
\end{equation}
In particular, if ${\cal H}_{1}$ is regarded as the Hamiltonian
of the quantum system, there is an infinite number of conserved
quantities, commuting with ${\cal H}_1$.

Let $d\times d$ matrices $t_{\beta\beta'}$ be

\begin{equation}
(t_{\beta\beta'})_{\alpha'\alpha}=w(\beta,\alpha,\beta',\alpha')~.
\end{equation}
In terms of the matrices $t_{\beta\beta'}$, for the six-vertex model

\begin{equation}
\frac{d}{du}{\cal V}^{(n)}(u)|_{u=0}=\sum_{i=1}^nt_{\beta_1\beta_2}(0)\otimes
 t_{\beta_2\beta_3}(0)\otimes\cdots\otimes \dot{t}_{\beta_i\beta_{i+1}}(0)
 \otimes\cdots\otimes t_{\beta_n\beta_{1}}(0)~.
 \end{equation}
Thus, the Hamiltonian of the $6$-vertex model is

\begin{equation}
{\cal H}_{1} =\left({\cal V}^{(n)}(0)\right)^{-1}\dot{\cal V}^{(n)}(0)=
\sum^{n}_{i=0} 1 \otimes \cdots
\otimes h_{i,i+1} \otimes \cdots \otimes 1~,
\end{equation}
where $h_{i,i+1}$ acts on the $i$ and $i+1$-th vertical
variables, i.e., in the space $V_i \bigotimes V_{i+1}$

\begin{equation}
\begin{array}{rcl}
h_{i,i+1} &=& \displaystyle\frac{1}{\rho\sin\eta}P\dot{\cal
R}(0)=\frac{1}{\sin\eta}
\left[\begin{array}{cccc}
       \cos \eta &   &   & \\
                  & 0 & 1 & \\
                  & 1 & 0 & \\
                  &   &   & \cos \eta \end{array}
\right]\\[8mm]
& =&\displaystyle \frac{1}{2 \sin \eta}\left(\sigma^1 \otimes \sigma^1 +
\sigma^2 \otimes\sigma^2+(1 \otimes 1 + \sigma^3 \otimes
\sigma^3)\cos\eta\right)~.
\end{array}
\end{equation}
Thus, up to an irrelevant constant,

\begin{equation}\label{3.0038}
{\cal H}_1 = \frac{1}{2 \sin \eta} \sum^{n}_{i=1}\left(\sigma^1_i\otimes
\sigma^1_{i+1} + \sigma^2_i \otimes \sigma^2_{i+1} +\cos\eta
\sigma^3_i \otimes\sigma^3_{i+1}\right)~,
\end{equation}
where we have used the notation
$$\sigma^i_k=1_{(1)}\otimes 1_{(2)}\otimes\cdots\otimes 1_{(k-1)}\otimes
  \sigma^i_{(k)}\otimes 1_{(k+1)}\otimes\cdots\otimes 1_{(n)}~.$$
This is the Hamiltonian of a (periodic) chain of $\displaystyle
\frac{1}{2}$-spin,
$s_i=\displaystyle\frac{1}{2}\sigma_i$ interacting with an anisotropic``XXZ"-
interaction.  Thus, we see that the transfer matrix ${\cal V}^{(n)}(u)$ of
the $6$-vertex model commutes with the XXZ
spin-$\displaystyle\frac{1}{2}$ Hamiltonian.
Alternatively we may say that the latter is generated by the former in a
``very anisotropic limit'' where the lattice spacing in the vertical direction
is
let to zero as well as the spectral parameter,

\begin{equation}
{\cal V}^{(n)}(u)=1+u{\cal H}_1~.
\end{equation}

\setlength{\unitlength}{4pt}
\thicklines
\begin{picture}(50,40)(-20,15)
\multiput(25,20)(-5,5){2}{\line(1,1){25}}
\multiput(25,20)(25,25){2}{\line(-1,1){5}}

\multiput(35,20)(-15,15){2}{\line(1,1){15}}
\multiput(35,20)(15,15){2}{\line(-1,1){15}}

\multiput(45,20)(5,5){2}{\line(-1,1){25}}
\multiput(45,20)(-25,25){2}{\line(1,1){5}}

\multiput(25,50)(10,0){3}{\line(1,1){3}}
\multiput(25,50)(10,0){3}{\line(-1,1){3}}

\put(5,25){time}
\put(9,28){\vector(0,1){15}}
\end{picture}

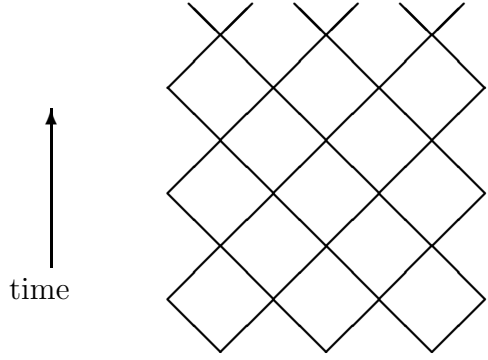
\hspace{7.5em}Fig.15. Diagonal-to-diagonal transfer matrix

\hspace{10.7em}for the vertex model.

Now we consider the six-vertex model, but with a transfer matrix
propagating in the diagonal direction (Fig.15). We restrict ourselves solely
to free boundary condition (depending on the parity of $n$).
The diagonal-to-diagonal transfer matrix ${\cal V}^{(n)}(u)$ acts on
${\cal H} =(V(1))^{\otimes n}$ and related with a gauge transformed $R$
matrix. It is well-known that there are some symmetries of the $R$ matrix.
The symmetries enable us to modify
the $R_{1221}$ and $R_{2112}$ entries of the matrix, retaining their
product unchanged, and preserving the YBE \cite{Akutsu87}.  This is
indeed true, and

\begin{equation}\label{3.I11}
{\cal R}(u) = \left[\begin{array}{cccc}
                   a &          &         & \\
                     &     b    & ce^{xu} & \\
                     & ce^{-xu} &   b     & \\
                     &          &         & a
                    \end{array}\right]
\end{equation}
also satisfies the YBE. The merit of this form is that it
leads to a non trivial limit as $u \rightarrow \pm i\infty$, for
$x = -i$,

\begin{equation}\label{3.666}
{\cal R}(-i\infty) \sim \left[\begin{array}{cccc}
                            1 &            &       &\\
                              &  q^{-1}    &   0   &    \\
                              & 1 - q^{-2} & q^{-1}& \\
                              &            &       & 1 \end{array}
                      \right]~,~~~~
{\cal R}(+i \infty) \sim \left[\begin{array}{cccc}
                        1 &   &         & \\
                          & q & 1-q^{2} & \\
                          & 0 & q       & \\
                          &   &         &   1 \end{array}
                  \right]~,
\end{equation}
where $q = e^{i\eta}$.  There are other more general type of such
"gauge transformations" of the $R$-matrix.  Corresponding to the generalized
form
of $R$ matrix (\ref{3.I11}), the six-vertex model is generalized nicely
to a $d$-state model.  With the gauge transformed $R$ matrix

\begin{equation}
{\cal R}(u) = \left[\begin{array}{cccc}
\sin(\eta+u) &                   &                   & \\
               & \sin u            & \sin\eta ~e^{iu} & \\
               & \sin\eta~ e^{-iu} & \sin u            & \\
               &                   &                   &\sin(\eta+u)
               \end{array}\right]~,
\end{equation}
the Boltzmann weights read

\begin{equation}
a=\sin(\eta+u)~,~~~~b=\sin u~,~~~~c_1=\sin\eta e^{iu}~,~~~~c_2=\sin\eta
e^{-iu}~.
\end{equation}
In terms of the above gauge transformed weights the diagonal-to-diagonal
transfer matrix is

\begin{equation}
{\cal V}^{(n)}(u)=\prod_{i=0}^{\{\frac{n-1}{2}\}}\left(
                  \sin(\eta+u)1-\sin u~ e_{2i+1}\right)
                  \prod_{i=1}^{\{\frac{n}{2}\}}
                  \left(\sin(\eta+u)1-\sin u~ e_{2i}\right)~,
\end{equation}
where
\begin{equation}\label{3.I212}
\begin{array}{rcl}
e_i&=&1_{(1)}\otimes 1_{(2)}\otimes\cdots\otimes 1_{(i-1)}\otimes e_{(i,i+1)}
   \otimes 1_{(i+2)}\otimes\cdots\otimes 1_{(n)}~,\\[7mm]
e&=&\left[\begin{array}{cccc}
           0 &0     &0  &0\\
           0 &q^{-1}&-1 &0\\
           0 &-1    &q  &0\\
           0 &0     &0  &0\end{array}\right]\\[6mm]
 &=&-\displaystyle\frac{1}{2}\left(\sigma^1\otimes\sigma^1+\sigma^2\otimes
    \sigma^2+\cos\eta\sigma^3\otimes\sigma^3\right)+\frac{1}{2}\cos\eta 1-
    \frac{i}{2}\sin\eta \left(\sigma^3\otimes 1-1\otimes\sigma^3\right)~.
\end{array}
\end{equation}
The $e_i$ given in Eq.(\ref{3.I212}) is the quantum analogue of $1 -
P_{i(i+1)}$,
where $P_{i(i+1)}$ is the transposition of the $i$-th and $(i+1)$-th spaces.
It is easy to verify that $e_i$ satisfy the Temperley-Lieb algebra $A_n$
\cite{Lieb},

\begin{equation}
\begin{array}{rcl}
& &e^2_i=\left[2 \right] e_i~,\\
& &e_i e_{i+1} e_i=e_i~,\\
& &e_i e_j=e_j e_i~, ~~~~{\rm if}~ |i - j| \geq 2~.
\end{array}
\end{equation}
The very anisotropic limit of the diagonal-to-diagonal transfer matrix yields

\begin{equation}
{\cal V}^{(n)}(u)=1-\frac{u}{\sin\eta}\sum_{i=1}^{n}e_i~.
\end{equation}
Thus, we obtain the Hamiltonian of the XXZ spin-1/2 chain as

\begin{equation}\label{3.0039}
\begin{array}{rcl}
{\cal H}_1
&=&-\displaystyle\frac{1}{\sin\eta}\displaystyle\sum_{i=1}^{n}e_i\\[3mm]
           &=&\displaystyle\frac{1}{2\sin\eta}\sum^n_{i=1}\left(\sigma^1_i
              \otimes\sigma^1_{i+1} + \sigma^2_i \otimes \sigma^2_{i+1}+
              \cos\eta\sigma^3_i \otimes\sigma^3_{i+1}+i\sin\eta(\sigma^3_i
              -\sigma^3_{i+1})-\cos\eta\right)~.
\end{array}
\end{equation}
This Hamiltonian is different from the standard XXZ Hamiltonian (\ref{3.0038})
by
the boundary terms.

The quantum group $SU_q(2)$ is known as

\begin{equation}
\begin{array}{l}
[S^3,S^\pm]=\pm S^\pm~,\\[1mm]
[S^+,S^-]=[2S^3]~,\\[1mm]
\Delta(S^3)=S^3\otimes 1+1\otimes S^3~,\\[1mm]
\Delta(S^\pm)=S^\pm\otimes q^{-S^3}+q^{S^3}\otimes S^\pm~,\\
\epsilon(S^3)=0=\epsilon(S^\pm)~,\\
S(S^3)=-S^3~,~~~~S(S^\pm)=-q^{\pm 1}S^\pm~.
\end{array}
\end{equation}
For $j=\displaystyle\frac{1}{2}$ the generators of $SU_q(2)$ coincide with
their $q=1$ limit, i.e., the $s^i=\displaystyle\frac{1}{2}\sigma^i$ matrices.
Thus, in $\left(V(1)\right)^{\otimes 2}$ we have

\begin{equation}\label{3.a0}
\begin{array}{l}
S^3=s^3_{(1)}\otimes 1_{(2)}+1_{(1)}\otimes s^3_{(2)}~,\\
S^\pm=s^\pm_{(1)}\otimes q^{-s^3_{(2)}}+q^{s^3_{(1)}}\otimes s^\pm_{(2)}~.
\end{array}
\end{equation}
As well known, there is another comultiplication $\Delta'$. In
$\left(V(1)\right)^{\otimes 2}$, $\Delta'$ is

\begin{equation}\label{3.a1}
\Delta'=P\circ\Delta\circ P~.
\end{equation}
$\Delta'$ and $\Delta$ are related by the universal $R$ matrix of $SU_q(2)$

\begin{equation}\label{3.a2}
\Delta'{\cal R}={\cal R}\Delta~.
\end{equation}
It should be noticed that the explicit form of the universal $R$ matrix in
$\left(V(1)\right)^{\otimes 2}$

\begin{equation}
{\cal R}^{\frac{1}{2}\frac{1}{2}}=q^{\frac{1}{2}}
                 \left[\begin{array}{cccc}
                            1 &            &       &\\
                              &  q^{-1}    &   0   &    \\
                              & 1 - q^{-2} & q^{-1}& \\
                              &            &       & 1 \end{array}
                      \right]
\end{equation}
is the same as the first matrix in Eq.(\ref{3.666}), i.e.,

\begin{equation}
\lim_{u\to -i\infty}{\cal R}(u)={\cal R}^{\frac{1}{2}\frac{1}{2}}~.
\end{equation}
{}From Eqs.(\ref{3.a0})---(\ref{3.a2}), we come to know that the operator
$\breve{\cal R}=P\circ {\cal R}$ commutes with the generators of the quantum
group $SU_q(2)$

\begin{equation}
[\breve{\cal R},SU_q(2)]=0~.
\end{equation}
In $\left(V(1)\right)^{\otimes 2}$, $\breve{\cal R}$ is

\begin{equation}
\breve{\cal R}^{\frac{1}{2}\frac{1}{2}}=q^{-\frac{1}{2}}\left(q-e\right)~,
\end{equation}
and the operator $e$ commutes with the generators of the quantum group
$SU_q(2)$.

Define the generators of the quantum group $SU_q(2)$ in $\left(V(1)\right)
^{\otimes n}$ as

\begin{equation}
\begin{array}{rcl}
& &S^\pm=\bigtriangleup^{(n)} \left(s^\pm \right)=\displaystyle\sum^{n}_{i=1}
   q^{s^3_{(1)}}\otimes q^{s^3_{(2)}}\otimes \cdots \otimes q^{s^3_{(i-1)}}
   \otimes s^\pm_{(i)}\otimes q^{-s^3_{(i+1)}}\otimes\cdots \otimes
   q^{-s^3_{(n)}}~,\\ [3mm]
& &S^3=\bigtriangleup^{(n)} \left(s^3\right)=\displaystyle\sum^{n}_{i=1}
   1_{(1)} \otimes 1_{(2)}\otimes \cdots \otimes 1_{(i-1)} \otimes s^3_{(i)}
   \otimes\cdots \otimes 1_{(n)}~,
\end{array}
\end{equation}
and $s^\pm_{(i)},~s_{(i)}$ are located in the $i$-th position, acting on the
$i$-th spin space. Thus, the Hamiltonian (\ref{3.0039}) is obviously
$SU_q(2)$-invariant, i.e.,

\begin{equation}\label{3.I415}
[{\cal H}_1,SU_q(2)]=0~,
\end{equation}
and the vertex model has an $SU_q(2)$ symmetry besides the
obvious $U(1)$ symmetry due to spin conservation.  It is important
to note that Eq.(\ref{3.I415}) would not hold for the standard weights
(\ref{3.I8}).
The effect of the gauge transformation can be put in boundary
terms only, but these are crucial as far as symmetries and critical
properties are concerned.

\subsection{S.O.S model}

Another family of integrable model is the S.O.S model. The degrees
of freedom of S.O.S model are located on the sites of a square lattice and
interact
through ``interaction-round-face'' around each plaquette. The Boltzmann weight
is
assigned to each unit face depending on the state variable configuration round
the
face (Fig.12). We denote energy of a face by the  state variable configuration
$(a_i,a_j,a_k,a_l)$ as $\epsilon(a_i,a_j,a_k,a_l)$.
The S.O.S model is very general: most of exactly solvable models
can be expressed in the form of S.O.S model. Total energy of the entire lattice
is

\begin{equation}
E=\sum_{\rm all~faces}\epsilon(a_i,a_j,a_k,a_l)~.
\end{equation}
In terms of the Boltzmann weights $w(a_i,a_j,a_k,a_l)$,
the partition function $Z_N$ and the free energy per site $f$ are given by

\begin{equation}
\begin{array}{rcl}
& &Z_N=\displaystyle\sum_{a_1}\cdots\sum_{a_N}\sum_{(i,j,k,l)}
   w(a_i,a_j,a_k,a_l)~,\\[3mm]
& &f=k_BT\displaystyle\lim_{N\to\infty}N^{-1}\log Z_N~.
\end{array}
\end{equation}
Notice that the Boltzmann weight are defined up to a gauge transformation that
has no effect on the partition function in the thermodynamic limit

\begin{equation}
w(a_i,a_j,a_k,a_l)\longrightarrow\frac{f(a_i,a_j)g(a_l,a_i)}{f(a_k,a_l)
g(a_j,a_k)}w(a_i,a_j,a_k,a_l)~.
\end{equation}
The constraint  on the weights expresses that no strongly fluctuating
configuration is allowed. More precisely, if the $l$'s take integers
values (on a finite or infinite range), we have a nonzero Boltzmann weight iff
any
two neighboring heights around the face differ by $\pm 1$.
The row-to-row transfer matrix ${\cal V}^{(n)}$ for the S.O.S model has matrix
elements (Fig.16)

\begin{equation}
{\cal V}^{(n)}_{a,a'}=\prod_{i=1}^nw(a_i,a_{i+1},a_{i+1}',a_i')~,
\end{equation}
where $a=\{a_1,~a_2,~\cdots,~a_n\},~
      a'=\{a'_1,~a'_2,~\cdots,~a'_n\},~a_{n+1}=a_1$,
       and $a'_{n+1}=a'_1$.
It generates the time evolution of the system.
Very similar to the vertex model, we get the
YBE as the consistency conditions of the row-to-row transfer matrix
${\cal V}^{(n)}$ (Fig.17),\\ \\
$\displaystyle\sum_{b''}w(a_i,a_{i+1},b'',a_i''|u)w(a_i'',b'',a_{i+1}',a_i'|u+v)
w(b'',a_{i+1},a_{i+1}'',a_{i+1}'|v) $
\begin{equation}\label{3.abc1}
=\sum_{b''}w(a_i'',a_i,b'',a_i'|v)w(a_i,a_{i+1},a_{i+1}'',b''|u+v)
w(b'',a_{i+1}'',a_{i+1}',a_i'|u)~.
\end{equation}

\setlength{\unitlength}{7pt}
\thicklines
\begin{picture}(50,12)(0,7)
\multiput(10,10)(5,0){8}{\line(0,1){5}}
\multiput(10,10)(0,5){2}{\line(1,0){35}}

\put(10,8.5){$a_1$}
\put(15,8.5){$a_2$}
\put(20,8.5){$a_3$}
\put(10,15.9){$a'_1$}
\put(15,15.9){$a'_2$}
\put(20,15.9){$a'_3$}
\put(40,8.5){$a_n$}
\put(45,8.5){$a_{n+1}=a_1$}
\put(40,15.9){$a'_n$}
\put(45,15.9){$a'_{n+1}=a'_1$}

\end{picture}

\hspace{4em}Fig.16. Row-to-row transfer matrix for the S.O.S model.

\setlength{\unitlength}{7pt}
\thicklines
\begin{picture}(50,21)(0,5)
\multiput(10,10)(0,10){2}{\line(1,0){10}}
\multiput(10,10)(10,0){2}{\line(-1,2){2.5}}
\multiput(7.5,15)(10,0){2}{\line(1,2){2.5}}
\put(7.5,15){\line(1,0){10}}
\put(20,10){\line(1,2){2.5}}
\put(22.5,15){\line(-1,2){2.5}}

\put(27,14.5){$=$}

\multiput(32.5,15)(2.5,5){2}{\line(1,-2){2.5}}
\multiput(32.5,15)(2.5,-5){2}{\line(1,2){2.5}}
\multiput(35,10)(0,10){2}{\line(1,0){10}}
\put(37.5,15){\line(1,0){10}}
\put(45,10){\line(1,2){2.5}}
\put(47.5,15){\line(-1,2){2.5}}

\multiput(17.2,14.7)(20,0){2}{$\bullet$}

\put(11.5,17){$u+v$}
\put(14,11.5){$u$}
\put(19.5,14.5){$v$}

\put(34.5,14.5){$v$}
\put(39.3,16.5){$u$}
\put(39,12){$u+v$}

\multiput(5,14.5)(25,0){2}{$a_i''$}
\multiput(23,14.5)(25,0){2}{$a_{i+1}''$}
\multiput(9.5,8.5)(25,0){2}{$a_{i}$}
\multiput(19.5,8.5)(25,0){2}{$a_{i+1}$}
\multiput(9.5,20.5)(25,0){2}{$a'_{i}$}
\multiput(19.5,20.5)(25,0){2}{$a'_{i+1}$}
\put(16,13.5){$b''$}
\put(37.5,13.5){$b''$}
\end{picture}

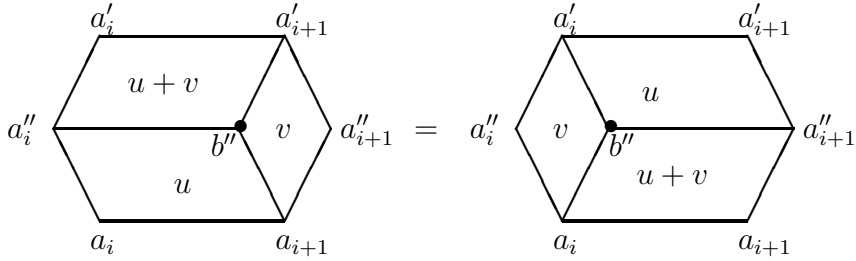
\hspace{6em}Fig.17. YBE for the the S.O.S model.

Defining the operators $X_i(u)$ for the S.O.S model by

\begin{equation}
X_i(u)_{a,a'}=\prod_{j\not=i}\delta_{a_ja'_j}
  w(a_{i-1},a_i,a_{i+1},a_i'|u)~,
\end{equation}
and making use of Eq.(\ref{3.abc1}) yields

\begin{equation}\label{3.a3}
X_i(u)X_{i+1}(u+v)X_i(v)=X_{i+1}(v)X_{i}(u+v)X_{i+1}(u)~.
\end{equation}
The role played by the operator $X_i(u)$ is to evaluate a configuration line by
changing
only one height at site $i$.  The solution of the YBE (\ref{3.a3}) is

\begin{equation}
X_i(u)=\sin(\eta+u)1-\sin u ~e_i~.
\end{equation}
In terms of the operators $X_i(u)$, the diagonal-to-diagonal transfer matrix
${\cal V}^{(n)}(u)$ of the S.O.S model (Fig.18) can be written as

\begin{equation}
{\cal V}^{(n)}(u)=\prod_{i=0}^{\{\frac{n-1}{2}\}}X_{2i+1}
\prod_{i=1}^{\{\frac{n}{2}\}}X_{2i}~.
\end{equation}

\setlength{\unitlength}{4pt}
\thicklines
\begin{picture}(50,19)(0,4.5)
\multiput(10,10)(10,0){8}{\line(1,1){5}}
\multiput(15,15)(10,0){8}{\line(1,-1){5}}
\multiput(45,15)(1,-1){6}{\line(1,1){5}}
\multiput(45,15)(1,1){6}{\line(1,-1){5}}

\put(9.5,8.5){$a_1$}
\put(14.5,15.5){$a_2$}
\put(19.5,8.5){$a_3$}
\put(49.5,8.5){$a_i$}
\put(89.5,8.5){$a_n$}
\put(40,15.5){$a_{i-1}$}
\put(49.5,20.5){$a_i'$}
\put(55.5,15.5){$a_{i+1}$}
\end{picture}

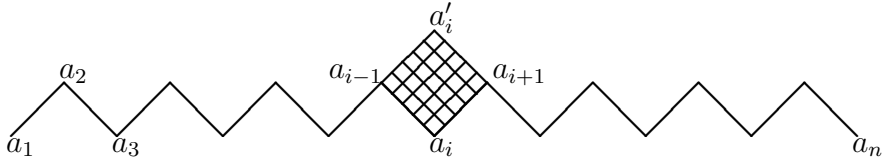
\hspace{3em}Fig.18. Diagonal-to-diagonal transfer matrix for the S.O.S model.

Then, in the very anisotropic limit, we obtain the Hamiltonian of the XXZ
model

\begin{equation}\label{3.0040}
{\cal H}_1=-\frac{1}{\sin\eta}\sum_{i=1}^ne_i~.
\end{equation}
Therefore, the S.O.S model also possesses an $SU_q(2)$ quantum symmetry as the
vertex model does.

\subsection{Configuration space}

In the previous sub-sections, we have shown that both the vertex model and the
S.O.S model
are mapped onto spin-$1/2$ quantum XXZ chains. The generators of the quantum
group
$SU_q(2)$, $S^\pm$ and $S^3$, act on the diagonal-to-diagonal lines of a
lattice,
interpreted as a direct product of spin-$1/2$ representations of
quantum group.
According to the Clebsch-Gordan rule for generic $q$, the configuration space
${\cal H}=(V(1))^{\otimes n}$ can be split into a direct sum of
irreducible highest weight representations $V(2j)$, which can be labeled by the
value of the Casimir operator $C$,

\begin{equation}
{\cal H}=\oplus_jw_j\otimes V(2j)~,
\end{equation}
where $w_j$ is a multiplicity space of dimension
$$\Gamma_j^{(n)}=\left(\begin{array}{c}
                       n\\
                       \displaystyle\frac{1}{2}n-j\end{array}\right)-
                  \left(\begin{array}{c}
                       n\\
                       \displaystyle\frac{1}{2}n-j-1\end{array}\right)~.$$
Eigenvectors of the diagonal-to-diagonal transfer matrix ${\cal V}^{(n)}$ fill
in representations $V(2j)$ of quantum group $SU_q(2)$, and we denote their
eigenvalues by
$\lambda_j^{(\alpha)},~\alpha=1,~2,~\cdots,~\Gamma_j^{(n)}$. The $SU_q(2)$
symmetry manifests itself through the degeneracies of these
eigenvalues of order $2j+1$. Since $\left({\cal V}^{(n)}\right)^\dagger$ is
equivalent to ${\cal V}^{(n)}$ after spin relabelling, eigenvalues are
real or complex conjugate by pairs.

For $q$ being a root of unity ($q^p=\pm 1$),  $\left(V(1)\right)^{\otimes n}$
contains in its decomposition reducible but not fully reducible
representations, which pair up representations that would be distinct
irreducible ones for $q$ generic adding up their $q$-dimensions to
zero. We get pairs of representations that mix in larger structures (type I)
$(V(2{j_k}),V({2j_{k-1}}))$; $(V(2{j_{k-2}}),~V({2j_{k-3}}))$;
$\cdots$.
There

\begin{equation}
\{j_k>j_{k-1}>\cdots>j_1,~~~~0\leq j_1<\frac{1}{2}(p-1)\}~,
\end{equation}
which are related by $j_i=j_1{\rm mod}p$ (if $i$ is odd), and $j_i=p-1-j_1{\rm
mod}p$
(if $i$ is even). Depending on the number of $V's$ we end up with a certain
number of $V({2j_1})$ that cannot mixing and are still irreducible
highest weight representations (type II).
Type-II representations are described by their highest weight vector
$\mid a_j\rangle$ such that $k\mid a_j\rangle=q^j\mid a_j\rangle$ ($0\leq
j<\displaystyle\frac{1}{2}p-1)$.  The number of type-II representations reads

\begin{equation}
\Omega_j^{(n)}=\Gamma_j^{(n)}-\Gamma_{p-1-j}^{(n)}+\Gamma_{j+p}^{(n)}
               -\Gamma_{p-1-j+p}^{(n)}+\cdots~.
\end{equation}
A special situation occurs if
$j_1=\displaystyle\frac{1}{2}(p-1)$ since $p-1-j_1=j_1$. In this case
the representations $V(2{j_k})$ are still irreducible and do not pair.
Thus the configuration space of the integrable lattice models splits into:
type-I representations which have $q$-dimension zero, and are either
mixed or of the kind $V({(np-1)})$,
and type-II representations which have a nonzero $q$-dimension, and are
still isomorphic to $U\left(su(2)\right)$ ones.

\subsection{Solutions}
Now we are in the position to discuss the solutions of the integrable lattice
models.
Although the XXZ Hamiltonian (\ref{3.0039}) and (\ref{3.0040}) is not
Hermitian, the eigenenergies
are real, because it is invariant under complex conjugation and reflection
symmetry
(relabelling sites from right to left). Because the Hamiltonian has $SU_q(2)$
quantum symmetry, each eigenspace corresponds to an irreducible representation
denoted by a Young pattern. Denote a state with $m$ down spins by

$$\mid x_1,x_2,\cdots,x_m\rangle~,$$
where $x_i$'s are the locations of the down spins on the chain. It is
well-known
that if the total  spin of
the chain is $S=\displaystyle\frac{1}{2}(n+1)-m$, the state $\mid
x_1,x_2,\cdots,
x_m\rangle$ are highest weight states. The number of the highest weight
states are given by

\begin{equation}
M_n=\left(\begin{array}{c}
          n\\
          m\end{array}\right)-\left(\begin{array}{c}
                                    n\\
                                    m-2\end{array}\right)~. \end{equation}
Then, the eigenstates of the Hamiltonian are

\begin{equation}
\mid m,h\rangle=\sum_{(x)}f_h(x_1,x_2,\cdots,x_m)\mid
x_1,x_2,\cdots,x_m\rangle~.
\end{equation}
The highest weight state $\mid m,h\rangle$ of the XXZ spin chain with $m$
down spins can be constructed by the Young operators $Y_m$.   For
quantum group $SU_q(2)$ only two-row Young pattern $(n+1-m, m)$ is relevant.
For
each Young pattern we only need one explicit form of the quantum Young
operator corresponding to a Young tableau. To construct the Young operators
explicitly, we first introduce the operators $Z_m^{2m+l}$
\cite{Levy}--\cite{Hou91} as

\begin{equation}
\begin{array}{lcl}
       Z^{2m}_m &=& 1~,\\
       Z^{2m+1}_m&=&g_{2m+1}(1)~,\\
       Z^{2m+2}_m&=&Z^{2m+1}_m g_{2m+2}(2)Z^{2m+1}_m~,\\
       \vdots   &\vdots&  \vdots~,\\
       Z^{2m+l}_m&=&Z^{2m+l-1}_m g_{2m+l}(l)Z^{2m+l-1}_m~,\\
       \vdots   &\vdots&  \vdots~,\\
       Z^n_m &=& Z^{n-1}_m g_n (n-2m)Z^{n}_m~,
\end{array}
\end{equation}
where we have used the notation $g_i(u)=[1+u]-[u] e_i$. It is easy to see that
the operators $Z^{2m+l}_m$ are constructed by $\{e_i|2m<i\leq 2m+l\}$. It
is also not difficult to prove inductively that the following formulas
for the operators $Z^{2m+l}_m$ are satisfied,

\begin{equation}
e_jZ^{2m+l}_m=Z^{2m+l}_me_j=0~,~~~~\forall 2m<j\leq 2m+l~.
\end{equation}
Let

\begin{equation}
S_m=e_1e_3\cdots e_{2m-1}~,
\end{equation}
then $S_m$ and $Z^{2m+l}_m$ are commutative. Thus we may define the quantum
Young
operators as

\begin{equation}
Y_m=S_mZ_m^n=Z_m^nS_m~.
\end{equation}
It is straightforward to prove that

\begin{equation}
\begin{array}{rcl}
& &Z^n_m Z^n_m\propto Z^n_m~,\\
& &S_m w S_m\propto S_m~,~~~ \forall w \in A_{2m}~,\\
& &Y_m w Y_m\propto Y_m~,~~~ \forall w \in A_n~,\\
& &Y_m Y_m\not=0~.
\end{array}
\end{equation}
Therefore, $Y_m$ is a primitive idempotent, and $wY_m~ (w \in A_n)$
is a primitive left ideal. The different Young patterns describe the
inequivalent
irreducible representations. As a result, the primitive left ideal is
$M_n$ dimensional. Let

\begin{equation}
         C^{i_2}_{i_1} = e_{i_2} e_{i_2 -1} \cdots e_{i_1}~,~~~~
         \left(i_2 \geq i_1 \right)~,
\end{equation}
then the $M_n$ basis vectors of the primitive ideals are given by

\begin{equation}
\begin{array}{rcl}
& &{^mC_{(i)}}Z^n_m~,\\
& &{^mC_{(i)}}\equiv {^mC_{i_1i_2\cdots i_m}} \equiv C^{i_1}_1 C^{i_2}_3
   \cdots C^{i_m}_{2m-1}~,\\
& &1 \leq i_1 <i_2<\cdots<i_m\leq n~,~~ i_k \geq 2k-1~.
\end{array}
\end{equation}
{}From the algebra relations of the Temperley-Lieb algebra, we get the
actions of
$e_i$'s on the basis vectors, ${^mC_{(i)}}Z^n_m$, of the primitive left ideal
as
\begin{itemize}
\item For $t=i_k>i_{k-1}+1$~,
      \begin{equation}\label{3.dell1}
      e_t\cdot{^mC_{(i)}}Z^n_m=[2]~{^mC_{(i)}}Z^n_m~.
      \end{equation}
\item For $t=i_{k+1}=i_k+1,~i_{k-\alpha}\geq t-2\alpha~(1\leq \alpha\leq l),~
      i_{k-l-1}<t-2l-2$,
      \begin{equation}\label{3.dell2}
      e_t\cdot{^mC_{(i)}}Z^n_m={^mC_{(i')}}Z^n_m~,
      \end{equation}
      where $$\begin{array}{rcl}
              & &i_p'=i_p~,~~~~\forall p>k~{\rm or}~p<k-l~,\\
              & &i'_{k-\alpha+1}=i_{k-\alpha}~,~~~~1\leq\alpha \leq l~,\\
              & &i'_{k-l}=t-2l-2~.
              \end{array}$$
\item For $t=i_k+1<i_{k+1}$,
      \begin{equation}\label{3.dell3}
      e_t\cdot{^mC_{(i)}}Z^n_m={^mC_{(i')}}Z^n_m~,
      \end{equation}
      where $$\begin{array}{rcl}
              & &i_p'=i_p~,~~~~\forall p\not= k~,\\
              & &i_k'=i_k+1~.
              \end{array}$$
\item For $i_m+1<t\leq n$,
      \begin{equation}\label{3.dell4}
      e_t\cdot{^mC_{(i)}}Z^n_m=0~.
      \end{equation}
\item For  $i_{k-1}+1<t<i_k-1,~i_{k+\alpha-1}\geq t+2\alpha~(1\leq\alpha\leq
l),~
      i_{k+l}=t+2l+1$,
      \begin{equation}\label{3.dell5}
      e_t\cdot{^mC_{(i)}}Z^n_m={^mC_{(i')}}Z^n_m~,
      \end{equation}
      where $$\begin{array}{rcl}
              & &i_p'=i_p~,~~~~\forall p<k,~{\rm or}~p>k+l~,\\
              & &i_{k+\alpha}'=i_{k+\alpha-1}~,~~~~1\leq\alpha\leq l~,\\
              & &i_k'=t~.
              \end{array}$$
\item For $t=i_k-1>i_{k-1}+1$,
      \begin{equation}\label{3.dell6}
      e_t\cdot{^mC_{(i)}}Z^n_m={^mC_{(i')}}Z^n_m~,
      \end{equation}
      where $$\begin{array}{rcl}
              & &i_p'=i_p~,~~~~\forall p\not=k~,\\
              & &i_k'=i_k-1~.
              \end{array}$$
\item For $i_{k-1}+1<t<i_k-1,~i_{k+\alpha-1}\geq t+2\alpha~~(1\leq\alpha
      \leq m-k+1)$,
      \begin{equation}\label{3.dell7}
      e_t\cdot{^mC_{(i)}}Z^n_m=0~.
      \end{equation} \end{itemize}
It is obvious from
Eq.(\ref{3.I212}) that the number of down spins of a state which are not
annihilated by the Young operator ${^mC_{(i)}}Z^n_m$ must be not less than
$m$, and a state of the highest weight of the representation has $m$ down
spins located at the first $2m$ positions. We construct the eigenspace of
the XXZ Hamiltonian by acting the basis of the primitive left ideal,
${^mC_{(i)}} Z_m^n$, on the state $\mid x_1,x_2,\cdots,x_m\rangle$. From
the definition relation of the elements of Temperley-Lieb algebra, we know
that if both $i$-th and $i+1$-th spins are down, the action of $e_i$
returns zero. The Young operators $Y_m$ contain $S_m~(e_1e_3\cdots
e_{2m-1})$, and the direction of any spin among the first $2m$ spins is
different from its neighbor's; the left $n-2m$ spins are all up. Thus the
action of $Z^n_m$ is a constant. For definiteness we define the state with
the highest weight as follows

\begin{equation}
{^m\Psi}_{(i)s}\equiv {^mC}_{(i)}\mid 1,3,5,\cdots,(2m-1)\rangle~,~~~
s=\frac{1}{2}(n+1-2m)~,
\end{equation}
where the constant factor obtained by applying $Z^n_m$ on the state
$\mid 1,2,3,\cdots,(2m-1)\rangle$ has been neglected, the subscript $s$ is
the highest weight eigenvalue of $\Delta^{(n)} \left(s_z \right)$.
Since the different Young patterns describe the inequivalent irreducible
representations, the corresponding energies are generally
different.  However, for the same Young pattern $(n-m,~ m)$
there are $M_n$ Young operators ${^mC}_{(i)}Z^n_m$ and $M_n$ spaces with
the same irreducible representation.
The linear combinations of ${^m\Psi}_{(i)s}$ with the same $m$ form the
eigenstates of the XXZ Hamiltonian

\begin{equation}\label{3.dell8}
\begin{array}{rcl}
& &H\cdot{^m\Phi}_{hs}={^mE}_h ~{^m\Phi}_{hs}~, \\
& &{^m\Phi}_{hs}=\displaystyle\sum_{(i)}a^h_{(i)}~ {^m\Psi}_{(i)s}~,
   ~~~h=1,~2,~\cdots,~M_n.
\end{array}
\end{equation}
Using the lowering operator $\Delta^{(n)}(s^-)$, we get easily the partners
${^m\Phi}_{h\alpha}$ in the representation

\begin{equation}
{^m\Phi}_{h\alpha}=[s-\alpha]^{-1}\Delta^{(n)}(s^-)~ {^m\Phi}_{h(\alpha+1)}~.
\end{equation}
Making use of Eqs.(\ref{3.dell1})---(\ref{3.dell7}), we transform
Eq.(\ref{3.dell8}) to be
a set of coupled linear algebraic equations with respect to
$a^h_{(i)}$, which can be solved by the standard method.

We now discuss a simple example in detail. For $m=0$, it is obvious that there
is only one state $\mid~\rangle_0$ with all spins up and zero eigenenergy.
For $m=1$, the simplest nontrivial case, the equations of motion of
the XXZ system are

\begin{equation}
\begin{array}{rcl}
& &{^1\Phi}_{h\frac{n-1}{2}}=\displaystyle\sum_ia_i^h~{^1C}_i\mid1\rangle~,\\
& &\left([2]-{^1E}_h\right)a_i^h+a_{i+1}^h+a_{i-1}^h=0~,\\
& &a_0^h=a_{n+1}^h=0~.
\end{array}
\end{equation}
By induction, we can easily prove that

\begin{equation}
\frac{a_i^h}{a_1^h}=(-1)^{i-1}\frac{\sin(ih\pi/(n+1))}{\sin(h\pi/(n+1))}~.
\end{equation}
For simplicity, we put $a_1^h=1$, and the eigenenergy is calculated from
$a_{n+1}^h=0$~,

\begin{equation}
{^1E}_h=[2]-2\cos\left(\frac{h\pi}{n+1}\right)~,~~~~h=1,~2,~\cdots,~n.
\end{equation}

\section{Conformal field theory}
During the past ten years, two-dimensional conformal invariant quantum
field theory has become of importance in statistical physics
and string theory \cite{Belavin84}--\cite{Kaku},\cite{Cardy87}.
Conformal field theory
(CFT) turned out due to be relevant in statistical models
exhibiting a second order phase transition.  At such a phase
transition the typical configurations have fluctuations on all length scales,
so  that the field theory describing the model at a critical point is
expected to be invariant under scale transformations.  Unlike in three
and higher dimensions the conformal group is infinite-dimensional in two
dimensions.  This fact imposes significant constraints on two-dimensional
CFT's.  Ultimately, we may hope that these
constraints are strong enough to obtain a complete classification of
two-dimensional critical systems.  CFT's are also essential
to string theory, which assume that the
elementary particles are not point-like but rather
behave as one dimensional objects, called strings. In perturbation theory,
the amplitude of a string can be expressed as sums over all possible
two-dimensional world-sheets of various topology, swept out by a string in
space-time.  The terms in perturbation series only depend on the conformal
equivalence class of the two-dimensional metric on the Riemann surface.

    A CFT is characterized by its scale invariance \cite{Belavin84,Friedan84}.
As a local field
theory, scale invariance implies full conformal invariance.  Scale invariance
is equivalent to the vanishing trace of the energy-momentum tensor.
In complex coordinates this means that $T_{z\bar{z}} = 0$.  Thus,
there are only two independent components of the energy-momentum tensor
$T_{zz},~T_{\bar{z}\bar{z}}$ and their conservation law implies
that
 $T_{zz} \equiv T(z)$ $(T_{\bar{z} \bar{z}} \equiv
T(\bar{z}))$ is a holomophic (antiholomorphic) function of
$z$ $(\bar{z})$. The generators of infinitesimal conformal transformations
are

\begin{equation}\label{4.2.1}
        L_n = \oint_C \frac{dz}{2 \pi i} z^{n+1} T(z)~,~~~~(n\in Z)~,
\end{equation}
and similarly for $\overline{L}_n$.  The contour circles the origin only once.
Its
shape do not matter as a consequence of the conservation law and Cauchy's
theorem.  By making use of Eq.(\ref{4.2.1}), we can write

\begin{equation}
   T(z) = \sum_{n \in Z} L_n z^{-n-2}~,~~~~
   \overline{T} (\bar{z}) = \sum_{n \in Z} \overline{L}_n\bar{z}^{-n-2}~.
\end{equation}
Out of all fields in CFT,  primary fields behave as
$(h,\overline{h})$ tensors, i.e.

\begin{equation}
\Phi_{h,\bar{h}} (z,\bar{z}) d z^h d \bar{z}^{\bar{h}}
\end{equation}
is invariant under conformal transformations. $(h,\bar{h})$ are called the
conformal dimensions of the primary fields.  By analytical properties of
$T(z)$ the correlation function

\begin{equation}
        \langle T (z) \Phi_1 (z_1) \cdots \Phi_N (z_N) \rangle~,
\end{equation}
can be evaluated using the operator product expansion (OPE)

\begin{equation}\label{4.2.5}
 T (z) \Phi (w) = \frac{h}{(z-w)^2} \Phi(w) + \frac{1}{z-w}
                     \partial_w \Phi(w) + \cdots~.
\end{equation}
As a function of $z,~ T(z)$ is a meromorphic quadratic differential
$(h = 2)$ on the sphere.  The only singularities appear at points
$z_1,~ z_2,~ \cdots,~ z_N$ and they are determined by the OPE (\ref{4.2.5}),

\begin{equation}\label{4.2.6}
   \langle T (z) \Phi_1 (z_1) \cdots \Phi_N (z_N)\rangle
           = \sum^N_{j=1} \left( \frac{h_i}{(z-z_j)^2}
             + \frac{1}{z-z_j} \frac{\partial}{\partial z_j} \right)
           \langle\Phi_1(z_1) \cdots \Phi_N (z_N) \rangle~.
\end{equation}
The OPE between two $T$'s is given by

\begin{equation}
T(z) T(w) = \frac{{c}/{2}}{(z-w)^4}+
       \frac{2}{(z-w)^2} T(w)+\frac{1}{z-w}
        \partial_w T(w) +\cdots~.
\end{equation}
This implies that the generators of infinitesimal conformal
transformations, $L_{n}$'s satisfy the following commutators

\begin{equation}
   [L_n,L_m] = (n - m) L_{n+m}+ \frac{c}{12} (n^3 - n) \delta_{n+m,0}~,
\end{equation}
and we obtain the Virasoro algebra.  The symmetry of a CFT under the
Virasoro algebra Vir$\bigotimes\overline{\rm Vir}$ means that the fields of the
theory fall into different conformal families, each generated by a
primary field.  The states generated by the action of $T(z)$ on the
primary field, known as descendent fields, are determined by those of
the primary fields through the use of the conformal Ward identity
(\ref{4.2.6}).

\subsection{Minimal model}
The minimal model in a sense, is the simplest CFT. The Hilbert space of
the theory is a finite sum of irreducible highest weight representation
space of the Virasoro algebra. There are finitely many primary fields, and
correlation functions of conformal fields are expressed in terms of a
finite sum of
holomorphically factorized terms. In principle, conformal invariance allows
one to compute all matrix elements of conformal fields in terms of
finitely many structure constants of the theory. And  the choice
of structure
constants are restrained by the requirement of locality. Imposing this
requirement, one can in principle compute the possible values of the
structure constants.

The Coulomb gas version of the minimal model is defined in terms of
a massless scalar field $\phi$ in the presence of a background charge
$2 \alpha_0$ located at infinity \cite{Dotsenko84,Dotsenko85}.  The
corresponding energy-momentum tensor
defines representations of the Virasoro algebra on the Fock modules ${\cal
F}_\alpha$,

\begin{equation}
\begin{array}{rcl}
& & T(z) = -\frac{1}{4} \partial \phi(z) \partial
                          \phi(z) + i \alpha_0 \partial^2 \phi(z)~,\\
& &\langle\phi(z) \phi(w)\rangle= -2\ln (z - w)~,\\
& &c = 1 - 24\alpha^2_0~.
\end{array}
\end{equation}
The highest weight vector of ${\cal F}_\alpha$ is given by the vertex operator

\begin{equation}
V_\alpha(z)=:e^{i\alpha\phi(z)}:~
\end{equation}
with $U(1)$ charge $\alpha$.

$T(z)$ follows from variation of the action for a free scalar field
with some background charge: a term proportional to $R \phi$ in the
Lagrangian where $R$ is the two-dimensional scalar curvature.  Thus the
$U(1)$ current $\partial \phi(z)$ is anomalous

\begin{equation}
   T(z) \partial \phi(w) = \frac{1}{(z-w)^2} \partial \phi(w) +
                           \frac{1}{z-w} \partial^2 \phi(w) +
                           \frac{4i \alpha_0}{(z-w)^3}+\cdots~.
\end{equation}
The last term represents the anomalous.  The background charge changes the
conformal dimension for the case $\alpha_0 = 0$

\begin{equation}
     T(z):e^{i\alpha\phi(w)}:=\frac{\alpha(\alpha-2\alpha_0)}{(z-w)^2}:
          e^{i\alpha\phi(w)}:+\frac{1}{z-w}\partial_w:
          e^{i\alpha\phi(w)}:+\cdots~.
\end{equation}
The conformal dimension of the vertex operator,
$\Delta_d\left(V_{\alpha}(z)\right)=\alpha(\alpha-2\alpha_0)$ and the
correlation
functions of these vertex operators will vanish unless the background
charge is screened

\begin{equation}
\langle V_{\alpha_1}(z_1)V_{\alpha_2}(z_2)\cdots V_{\alpha_N}(z_N)\rangle
                           =\left\{\begin{array}{l}
                                 \displaystyle\prod_{i<j}(z_i-z_j)^{2\alpha_i
                                  \alpha_j}~,~~~{\rm if}~
                                  \sum_{i=1}^N\alpha_i=2\alpha_0~,\\
                            0~,~~~ {\rm otherwise}~.\end{array}
                             \right.
\end{equation}
The braid relation satisfied by the Coulomb gas vertex operators is

\begin{equation}\label{4.2.14}
      V_{\alpha_1}(z_1)V_{\alpha_2}(z_2)
                  = e^{2\pi i\alpha_1\alpha_2} V_{\alpha_2}
                    (z_2) V_{\alpha_1}(z_1)~.
\end{equation}
For $\alpha=\alpha_\pm$, where $2\alpha_0=\alpha_++\alpha_-$ and
$\alpha_+\alpha_- = -1$, we obtain the screening operators $J(z)=V_
{\alpha_\pm}(z)$ with conformal dimension  equal to one.  It is
important to note that both $V_{\alpha}(z)$ and its dual
$\tilde{V}_\alpha(z)=V_{2\alpha_0-\alpha}
(z)$ have the same conformal dimension.  In particular, we can write
the identity in two ways, $1$ and $:\exp\left(2i\alpha_0\phi\right):$.
The existence
of currents of conformal dimension one allows us to introduce screening
operator $Q$,

\begin{equation}
                   Q = \int_\Gamma J(z)dz~,
\end{equation}
where the contour $\Gamma$ is chosen so that $Q$ acts as an intertwiner,
i.e.,  without
affecting the conformal properties of the correlation function.  For
example if $z_1,~z_2$ are the insertion points of vertex operators, the
screening charge

\begin{equation}
\int^{z_2}_{z_1} dzJ_i(z)
\end{equation}
is an intertwiner.  Inserting $Q$ in correlation functions   does not change
the conformal properties.  If we write down the $4$-point function of an
operator $V_\alpha(z)$, $\langle V_\alpha V_\alpha V_\alpha
\tilde{V}_\alpha\rangle$.
The total charge is $3\alpha +2\alpha_0-\alpha=2\alpha+2\alpha_0$, hence
will vanish generally unless

\begin{equation}
                  2\alpha = -n\alpha_+-m\alpha_-~.
\end{equation}
In the case we can introduce $nQ_+$ and $mQ_-$ screening operators
to obtain a nonvanishing amplitude.  Therefore, the spectrum of vertex
operators with nonvanishing 4-point functions is given by

\begin{equation}\label{4.2.18}
    \alpha_{m,n}= \frac{1-n}{2}\alpha_+ +\frac{1-m}{2}\alpha_-~,~~~~
    n,~m\geq1~.
\end{equation}
A $4$-point block hence  takes the form

\begin{equation}
\begin{array}{rcl}
& &\displaystyle\int_{\Gamma_1}dt_1\int_{\Gamma_2}dt_2\cdots
\int_{\Gamma_n}dt_n \int_{\Gamma'_1} dt'_1 \int_{\Gamma'_2} dt'_2
\cdots \int_{\Gamma'_m} dt'_m\times\\
& &\times\langle V_{\alpha_1}(z_1)V_{\alpha_2}(z_2)V_{\alpha_3}(z_3)
V_{\alpha_4}(z_4)J_+(t_1) J_+(t_2)\cdots J_+(t_n)J_-(t'_1) J_-(t'_2)
\cdots J_-(t'_m) \rangle~.
\end{array}
\end{equation}
The general correlation function in the Coulomb gas representation is\\ \\
$\langle V_{\alpha_1}(z_1) V_{\alpha_2}(z_2)\cdots V_{\alpha_N}(z_N)\rangle$
\begin{equation}
\sim\langle V_{\alpha_1}(z_1)Q_i(z_a,~z_b)\cdots  Q_j(z_c,~z_d)V_{\alpha_2}
   (z_2) \cdots Q_k(z_e,~z_f) V_{\alpha_N}(z_N) \rangle~,
\end{equation}
where $z_a,~z_b,~\cdots,~z_e,~z_f \in \{z_1,~z_2,~\cdots,~z_N\}$,
and total charge equal to zero.

We define the screening operators $X^-_\pm$ \cite{Gomez90,Gomez91} by

\begin{equation}
  X^-_{\pm}V_\alpha(z)=\frac{1}{1-q^{-1}_\pm} \int_G dtJ_\pm V_\alpha(z) \sim
  Q(\infty,~z) V_\alpha(z)~,
\end{equation}
where $q_\pm=e^{2\pi i\alpha^2_\pm}$, and the contour $G$ surrounds the cut
from $z$ to infinity (Fig.19).  Owing to the $SL(2,C)$ invariance of the
correlation function, we can write the $N$-point conformal block as

\begin{equation}
\lim_{z_1\to\infty}\langle \tilde{V}_{\alpha_1}(z_1)X^-_+ \cdots
X^-_-V_{\alpha_2} \cdots X^-_+
X^-_-\cdots V_{\alpha_N}(z_N) \rangle~,
\end{equation}
with vanishing total Coulomb charge. The  screening operator
$X^-_\pm$'s are understood to act only on the first vertex operator
to their right.\\

\setlength{\unitlength}{5pt}
\thicklines
\begin{picture}(50,23)(0,4)
\put(35,10){\oval(8,5)[b]}
\put(31,10){\line(0,1){15}}
\put(39,10){\line(0,1){15}}
\put(31,25){\vector(0,-1){7.5}}
\put(39,10){\vector(0,1){7.5}}
\multiput(35,10.5)(0,1.5){10}{\line(0,1){1}}
\put(34.5,10){$\bullet$}
\put(36,9.4){$z$}
\put(40,17){$G$}
\end{picture}

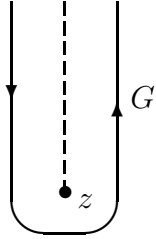
\hspace{7.5em}Fig.19. The defining contour of generators

\hspace{10.7em}of quantum group.

Introduce the operator $k_\pm$ by

\begin{equation}
k_\pm V_\alpha(z) =\exp\left(-2i\pi\alpha_\pm \oint_c \partial\phi \right)
V_\alpha(z)~.
\end{equation}
The Borel subalgebra, generated by the operators $X^-_\pm$ and
$k_\pm$, of the quantum group underlying the minimal model is

\begin{equation}
\begin{array}{rcl}
[X^-_+,X^-_-] &=& 0,~~~~~[k_+,~k_-] =0~,\\
k_\pm X^-_\pm k^{-1}_\pm &=& q^{-1}_\pm X^-_\pm,~~~
k_\pm X^-_\mp k_\pm^{-1}=-X^-_\mp~.
\end{array}
\end{equation}
To define the dual operators of $X^-_\pm$, $X^+_\pm$,  we shall make use of the
transformation law of the screened vertex operator under the Virasoro
algebra.  Explicitly, we define $X^+_\pm$ by\\ \\
$\delta_\xi \left((X^-_+)^n (X^-_-)^mV_\alpha(z)\right) - (X^-_+)^n(X^-_-)^m
\left(\delta_\xi V_\alpha(z)\right)$
\begin{equation}\label{4.2.29}
\begin{array}{rcl}
&=&-(1-q_+^{-1})\xi(\infty)J_+(\infty)X^+_+(X^-_+)^n(X^-_-)^mV_\alpha(z)\\[2mm]
& &-(1-q_-^{-1})\xi(\infty)J_-(\infty)X^+_-(X^-_+)^n(X^-_-)^mV_\alpha(z)~,
\end{array}
\end{equation}
where $\xi(z)$ is the vector field that generates the conformal
transformation.  Making use of the commutator between the Virasoro
generator $L_n$ and the screening operator $J_\pm$,

\begin{equation}
         [L_n,J_\pm] = \frac{d}{dz}\left( z^{n+1}J_\pm(z)\right)~,
\end{equation}
we get\\ \\
$L_n(X^-_+)^n(X^-_-)^mV_\alpha(z) $
\begin{equation}\label{4.2.31}
\begin{array}{rcl}
      &=&(X^-_+)^n(X^-_-)^m\left(L_nV_\alpha(z)\right)\\[3mm]
      & &-\displaystyle\lim_{t\to\infty}
          t^{n+1}J_+(t)|[n]_{q_+^{-1}}(1-e^{4\pi
i\alpha\alpha_+}q^{n-1}_+)(X^-_+)
          ^{n-1}(X^-_-)^mV_\alpha(z)\\[3mm]
      & &-\displaystyle\lim_{t\to\infty}
         t^{n+1}J_-(t)|[m]_{q_-^{-1}}(1-e^{4\pi i
         \alpha\alpha_-}q^{m-1}_-)(X^-_+)^n(X^-_-)^{m-1}V_\alpha(z)~,
\end{array}
\end{equation}
where $\displaystyle|[x]_q=\frac{1-q^{-x}}{1-q^{-1}}$.\\
{}From Eqs.(\ref{4.2.29}) and (\ref{4.2.31}) we obtain that

\begin{equation}\label{4.2.32}
\begin{array}{rcl}
X^+_+(X^-_+)^n(X^-_-)^mV_\alpha(z) &=&
|[n]_{q_+^{-1}}\displaystyle\frac{1-e^{4\pi i \alpha\alpha_+}
    q^{n-1}_+}{1-q^{-1}_+}(X^-_+)^{n-1}(X^-_-)^mV_\alpha(z)~,\\[3mm]
X^+_-(X^-_+)^n(X^-_-)^mV_\alpha(z) &=&
|[m]_{q_-^{-1}}\displaystyle\frac{1-e^{4\pi i \alpha\alpha_-}
           q^{m-1}_-}{1-q^{-1}_-}(X^-_+)^n(X^-_-)^{m-1}V_\alpha(z)~.
\end{array}
\end{equation}
{}From Eqs.(\ref{4.2.29}) and (\ref{4.2.32}) one can verify that
$\delta_\xi$ satisfies
the Virasoro algebra, $X^+_+$ commutes with $X^+_-$, and $X^+_\pm$ commute
with the Virasoro algebra.  Now we can write down the other Borel
subalgebra of the quantum group underlying the minimal model as

\begin{equation}
\begin{array}{rcl}
& &[X^+_+,X^+_-]=0~,\\
& &k_\pm X^+_\pm k_\pm^{-1}=q_\pm X^+_\pm~,~~~~
k_\pm X^+_\mp k_\pm^{-1}=-X^+_\mp~,\\
& &X^+_\pm X^-_\pm - q_\pm X^-_\pm X^+_\pm=\displaystyle\frac{1-k^{-2}_\pm}
   {1-q^{-1}_\pm}~,\\
& &X^+_\pm X^-_\mp-q_\pm X^-_\mp X^+_\pm=0~.
\end{array}
\end{equation}

Let us define the action of the operators $X^-_\pm$ on the space for
the ordinary operator product $V_\alpha(z_1)V_\beta
(z_2)$ corresponding to the comultiplication operation
of Hopf algebra

\begin{equation}
\Delta\left( X^-_\pm\right) \left( V_\alpha(z_1)V_\beta(z_2) \right) =
     \int_{\Delta G}dtJ_\pm(t)V_\alpha(z_1)V_\beta(z_2)~,
\end{equation}
where the contour $\Delta G$  surrounds the operator $V_\alpha(z_1)
V_\beta(z_2)$ (Fig.20).  Deforming the contour $\Delta G$ into the union
$G_1\cup G_2$ with $G_i$ ($i=1,~2$) the  contour surrounding the vertex
at point $z_i$,  we obtain

\begin{equation}\label{4.2.35}
\Delta \left(X^-_\pm\right)\left(V_\alpha(z_1)V_\beta(z_2)\right) =
X^-_\pm\left(V_\alpha(z_1)\right)
V_\beta(z_2) +
k^{-1}_\pm\left(V_\alpha(z_1)\right)X^-_\pm\left(V_\beta(z_2)\right)
{}~.
\end{equation}

\setlength{\unitlength}{5pt}
\thicklines
\begin{picture}(50,23)(0,4)
\put(10,10){\oval(15,5)[b]}
\put(2.5,10){\line(0,1){15}}
\put(17.5,10){\line(0,1){15}}
\put(2.5,25){\vector(0,-1){9.5}}
\put(17.5,10){\vector(0,1){7.5}}
\multiput(7.5,10.5)(0,1.5){10}{\line(0,1){1}}
\multiput(12.5,10.5)(0,1.5){10}{\line(0,1){1}}
\put(7,10){$\bullet$}
\put(8.5,9.4){$z_1$}
\put(12,10){$\bullet$}
\put(13.5,9.4){$z_2$}
\put(8,5){$\Delta G$}

\put(21.5,15){$=$}

\put(35,10){\oval(8,5)[b]}
\put(31,10){\line(0,1){15}}
\put(39,10){\line(0,1){15}}
\put(31,25){\vector(0,-1){9.5}}
\put(39,10){\vector(0,1){7.5}}
\multiput(35,10.5)(0,1.5){10}{\line(0,1){1}}
\multiput(44,10.5)(0,1.5){10}{\line(0,1){1}}
\put(34.5,10){$\bullet$}
\put(36,9.4){$z_1$}
\put(43.5,10){$\bullet$}
\put(45,9.4){$z_2$}
\put(34,5){$G_1$}

\put(50,15){$+e^{2\pi i\alpha_\pm\alpha}$}

\put(69,10){\oval(8,5)[b]}
\put(65,10){\line(0,1){15}}
\put(73,10){\line(0,1){15}}
\put(65,25){\vector(0,-1){9.5}}
\put(73,10){\vector(0,1){7.5}}
\multiput(69,10.5)(0,1.5){10}{\line(0,1){1}}
\multiput(60,10.5)(0,1.5){10}{\line(0,1){1}}
\put(59.5,10){$\bullet$}
\put(61,9.4){$z_1$}
\put(68.5,10){$\bullet$}
\put(70,9.4){$z_2$}
\put(69,5){$G_2$}

\end{picture}

\hspace{7.5em}Fig.20. Comultiplication operation of $X_\pm^-$.

The diagonal operators $k_\pm$ in Eq.(\ref{4.2.35}) arises
from the braiding between the screening operator $J_\pm$ and the vertex
operator $V_\alpha(z)$.  The comultiplication operation for $X^-_\pm$
follows easily from Eq.(\ref{4.2.35}), taking form

\begin{equation}
     \Delta( X^-_\pm) =X^-_\pm \otimes 1 + k^{-1}_\pm \otimes X^-_\pm~.
\end{equation}
Similarly the comultiplication operation for $k_\pm$ is given by

\begin{equation}
        \Delta (k_\pm) = k_\pm \otimes k_\pm~.
\end{equation}
The comultiplication operation for $X^+_\pm$ follows from that for the
Virasoro operators, \\ \\
$\delta_\xi
\left((X^-_+)^n(X^-_-)^mV_\alpha(z_1)(X^-_+)^{n'}(X^-_-)^{m'}V_\beta(z_2)\right)$
\begin{equation}
\begin{array}{rcl}
     &=& \left(\delta_\xi\left( (X^-_+)^n(X^-_-)^mV_\alpha(z_1)\right)\right)
         (X^-_+)^{n'}(X^-_-)^{m'}V_\beta(z_2)\\[3mm]
     & & +(X^-_+)^{n}(X^-_-)^{m}V_\alpha(z_1) \left(\delta_\xi\left(
         (X^-_+)^{n'}(X^-_-)^{m'}V_\beta(z_2)\right)\right)\\[3mm]
     &=& (X^-_+)^{n}(X^-_-)^{m}\left(\delta_\xi V_\alpha(z_1)\right)
         (X^-_+)^{n'}(X^-_-)^{m'}V_\beta
         (z_2)\\[3mm]
     & & + (X^-_+)^{n}(X^-_-)^{m}V_\alpha(z_1)
         (X^-_+)^{n'}(X^-_-)^{m'}\left(\delta_\xi V_\beta(z_2)\right)\\[3mm]
     & & -\left( (1-q^{-1}_+)\xi(\infty)J_+(\infty)\Delta (X^+_+) +
         (1-q^{-1}_-)\xi(\infty)J_-(\infty)\Delta(X^+_-) \right)\times\\[3mm]
     & & ~~\times (X^-_+)^{n}(X^-_-)^{m}V_\alpha(z_1)(X^-_+)^{n}(X^-_-)^{m}
         V_\beta(z_2)~,
\end{array}
\end{equation}
where we have used Eq.(\ref{4.2.29}) and

\begin{equation}
\Delta (X^+_\pm) = X^+_\pm \otimes 1 + k^{-1}_\pm \otimes X^+_\pm~.
\end{equation}
Besides the comultiplication $\Delta$,  the other Hopf operations such as
antipode $S$ and counit $\epsilon$ are introduced by the following
formulas

\begin{equation}
\begin{array}{rcl}
    \epsilon (k_\pm) &=& 1~,~~~~ \epsilon( X^-_\pm) = 0 = \epsilon X^-_\pm~,\\
    S(k_\pm) &=& k^{-1}_\pm~,~~~~S(X^-_\pm)=-k_\pm X^-_\pm~,~~~~S(X^+_\pm)
    =-k_\pm X^+_\pm~.
  \end{array}
\end{equation}
The antipode $S$ is essentially a path reversing operator, while
the counit is a contour killing mapping (i.e. $\epsilon\left(\displaystyle\oint
_CX\right)=0$ for any $X$).  The quantum group generated by
$k_\pm,~X^-_\pm$ and $X^+_\pm$ summarizes the quantum symmetries of
the minimal  model.  For the thermal subalgebra
$\alpha_{n,1}$ (or $\alpha_{1,m})$, the operators $k_+,~X^-_+$
and $X^+_+$ (or $k_-,~X^-_-$ and $X^+_-)$ generate the quantum group
$SU_q(2)$.

For each vertex operator $V_\alpha(z)$ is associated with a representation
space
${\cal V}^\alpha$ generated by the screened vertex operator $e^\alpha_{
n_+,n_-}$,

\begin{equation}\label{4.2.41}
e^\alpha_{n_+,n_-} = (X^-_+)^{n_+}(X^-_-)^{n_-}V_\alpha(z)~.
\end{equation}

The screened vertex operators $e_{n_+,n_-}^\alpha$ give the representations of
the quantum group on the space ${\cal V}^\alpha$

\begin{equation}\label{4.2.42}
\begin{array}{rcl}
& & k_\pm e^\alpha_{n+,n-}=\exp\left(-2\pi i\alpha_\pm(\alpha+n_+\alpha_++n_-
    \alpha_-)\right)e^\alpha_{n_+,n_-}~,\\[3mm]
& & X^-_+e^\alpha_{n_+,n_-}=e^\alpha_{n_++1,n_-}~,~~~~
    X^-_-e^\alpha_{n_+,n_-} = e^\alpha_{n_+,n_- +1}~,\\[3mm]
& & X^+_+e^\alpha_{n_+,n_-}=|[n_+]_{q_+^{-1}}\displaystyle\frac{1-e^{4
    \pi i\alpha\alpha_+}q^{n_+-1}_+}{1-q^{-1}_+}
    e^\alpha_{n_+ -1,n_-}~,\\[3mm]
& & X^+_-e^\alpha_{n_+,n_-}=|[n_-]_{q_-^{-1}}\displaystyle\frac{1-e^{4\pi
    i \alpha\alpha_-}q^{n_- ^{-1}}}{1-q^{-1}_-}
    e^\alpha_{n_+,n_- -1}~.
\end{array}
\end{equation}
The space ${\cal V}^\alpha$ will be finite-dimensional
 provided there
exist positive integers $n^+_\alpha$ and $n^-_\alpha$ such that
$e^\alpha_{n^+_\alpha,n^-_\alpha}=0$.  In this case it is easy
to see from Eq.(\ref{4.2.41}) that the charge $\alpha$ is given
by the Kac's formula (\ref{4.2.18}), with $n=n^+_\alpha$ and $m=n^-_\alpha$.
The dimension of
${\cal V}^\alpha$ is therefore equals to $n^+_\alpha n^-_\alpha = nm$.

In a rational theory, where $\alpha^2_+=p'/p$, the range of the
positive integers $n$ and $m$ that define a finite-dimensional space
${\cal V}^{\alpha_{n,m}}$, is restricted to the intervals $1\leq n\leq p$
and $1\leq m\leq p'$.  This is due to the path ordering of the
screening operator $J_+$ and $J_-$ in Eq.(\ref{4.2.41}).

In order to see
this phenomenon in more detail we will write the screened vertex
operators $e^\alpha_{n_+,n_-}$ alternatively

\begin{equation} \label{4.2.40}
      e^\alpha_{n_+,n_-}(z) = \prod^{n_+-1}_{r=0}\left(1-e^{4\pi i
                              \alpha\alpha_+}q^r_+ \right)
                              \prod^{n_--1}_{r'=0} \left(1-e^{4\pi i
                              \alpha\alpha_-}q^{r'}_- \right)
                              \tilde{e}^\alpha_{n_+,n_-}(z)~,
\end{equation}
where
\begin{equation}
\begin{array}{rcl}
\tilde{e}^\alpha_{n_+,n_-}(z) &=& (P_+)^{n_+}(P_-)^{n_-}V_\alpha(z)\\
                              &=&\displaystyle\int^z_\infty dt_1J_+(t_1) \cdots
                                  \int^z_\infty dt_{n_+}J_+(t_{n_+})
                                  \int^z_\infty dt'_1J_-(t'_1)\cdots
                                  \int^z_\infty dt'_{n_-}J_-(t'_{n_-})
                                                V_\alpha(z),
\end{array}
\end{equation}
here we have used the notation

\begin{equation}
        P_\pm = \int^z_\infty J_\pm(z)dt~.
\end{equation}
It is convenient to path order the integrals entering the definition
of $\tilde{e}^\alpha_{n_+,n_-}(z)$.  Denoting the``path-ordering" operator
by the symbol $T$ (i.e. $|t_1|>|t_2|> \cdots>|z|)$ we obtain

\begin{equation}\label{4.2.43}
   \tilde{e}^\alpha_{n_+,n_-}(z) = |[n_+]_{q_+^{-1}}!|[n_-]_{q_-^{-1}}!
   T(\tilde{e}^\alpha_{m_+,n_-}(z))~.
\end{equation}
Eq.(\ref{4.2.43}) is the $q$-analogue of the path ordering of an integral
operator.
{}From Eqs.(\ref{4.2.40}) and (\ref{4.2.43}), we see that
$e^\alpha_{n_+,n_-}(z)$ is
non-zero, if $1\leq n_+ \leq n^\pm_\alpha-1$, with $1\leq
n=n^+_\alpha
\leq p$ and $1\leq m=n^-_\alpha\leq p'$.

\subsection{WZNW model}

In the previous subsection we have considered the CFT with symmetry algebra
Vir$\otimes\overline {\rm Vir}$. In general, CFT  has
a symmetry algebra ${\cal U}$, the chiral algebra ${\cal U}={\cal U}_L\otimes
{\cal U}_R$ with left- and right-handed components.
The chiral algebras are characteristic of ${\cal U}_L$ (${\cal
U}_R$) always
containing the identity operator and the Virasoro ($\overline{\rm Vir}$)
algebra.

Making use of a set of complex fermions $b^i(z)$ and $c_i(z)$
($i=1,~2,~\cdots,~d$), for an algebra $g$ and a real representation
$(T^a)_{ij}$, ($i,~j=1,~2,~\cdots,~d$), we can construct currents

\begin{equation}
J^a(z)=b^i(T^a)_i^jc_j=bT^ac=\sum_{n\in Z}J_n^az^{-n-1}~.
\end{equation}
The nontrivial OPEs are

\begin{equation}
\begin{array}{rcl}
& &b^i(z)c_j(w)=\displaystyle\frac{\delta^i_{j}}{z-w}+\cdots~,\\[2mm]
& &c_i(z)b^j(w)=\displaystyle\frac{\delta_i^{j}}{z-w}+\cdots~,
\end{array}
\end{equation}
and the commutation relations of $J_n^a$ follow from the OPE

\begin{equation}
J^a(z)J^b(w)=\frac{{\rm Tr}T^aT^b}{(z-w)^2}+if^{abc}\frac{J^c(w)}{z-w}+\cdots~.
\end{equation}
Choosing an appropriate basis for the Lie algebra $g$,

\begin{equation}
{\rm Tr}T^aT^b=k\delta^{ab}~,
\end{equation}
we can always write down the commutation relations of the currents algebra as

\begin{equation}
[J_n^a,J_m^b]=if^{abc}J_{n+m}^c+n\delta^{ab}k\delta_{n+m,0}~.
\end{equation}
The operators $J_n^a$'s are associated with the loop algebra of $g$,

\begin{equation}
\epsilon(z)=\sum_{a,n}\epsilon_n^aT^az^n~,
\end{equation}
where $\epsilon_n^a$'s are the generators of the loop algebra, their
commutation relations always generate a Kac-Moody algebra. If the ground state
of a CFT is invariant under the Kac-Moody algebra, the fields of the
theory  are catalogued into families, each of which provides an irreducible
representation of the Kac-Moody algebra.

A primary field has the properties

\begin{equation}\label{4.2.50}
J^a(z)\Phi_j(w)=-(T^a)_{jk}\frac{\Phi_k(w)}{z-w}+\cdots~.
\end{equation}
And the primary fields also have the OPEs

\begin{equation}\label{4.2.51}
\begin{array}{rcl}
& & T(z)\Phi_i(w)=\displaystyle\frac{h}{(z-w)^2}\Phi_i(w)+\frac{1}{z-w}\partial
_w\Phi_i(w)+\cdots~,\\[3mm]
& & T(z)J^a(w)=\displaystyle\frac{1}{(z-w)^2}J^a(w)+\frac{1}{z-w}\partial
_wJ^a(w)+\cdots~.
\end{array}
\end{equation}
And then

\begin{equation}
J^a(z)J^b(w)=\frac{k\delta^{ab}}{(z-w)^2}+if^{abc}\frac{J^c(w)}{z-w}
   +\cdots~.
\end{equation}
Eqs.(\ref{4.2.50}) and (\ref{4.2.51}) allow us to determine the
correlation functions explicitly, provided that all the fields involved are
primary ones,\\
$\langle T(z)\Phi_{i_1}(z_1)\Phi_{i_2}(z_2) \cdots\Phi_{i_N}(z_N) \rangle$
\begin{equation}
\begin{array}{c}
=\displaystyle

\sum_{j=1}^N\left(\frac{h_j}{(z-z_j)^2}+\frac{1}{z-z_j}\frac{\partial}{\partial
z_j}
   \right)\langle\Phi_{i_1}(z_1)\Phi_{i_2}(z_2) \cdots\Phi_{i_N}(z_N)
\rangle~,\\[2mm]
\langle J^a(z)\Phi_{i_1}(z_1)\Phi_{i_2}(z_2) \cdots\Phi_{i_N}(z_N) \rangle
   =-\displaystyle\sum_{j=1}^N\frac{(T^a)_{i_jk_j}}{(z-z_j)}
   \langle\Phi_{k_1}(z_1)\Phi_{k_2}(z_2) \cdots\Phi_{k_N}(z_N) \rangle~,
\end{array}
\end{equation}
which are the ward identities corresponding to the conformal and gauge
symmetries.

One field theory possessing the above features is
the Wess-Zumino-Novikov-Witten (WZNW) model \cite{Novikov82,Feigin90,Gepner86}.

If dim($g$)=$d$, rank$(g)=l$, the Coulomb gas version of the
corresponding WZNW model is depicted by $(d-l)/2$
free  $\beta$-$\gamma$ pairs  and $l$ free real scalars $\vec{\phi}$ with a
boundary term (charge at infinity) \cite{Wakimoto,Feigin90,Gerasimov}.
The energy-momentum tensor is

\begin{equation}\label{4.2.55}
   T(z) = - \sum_{\vec{\alpha}\in\Delta_+} :\beta_{\vec{\alpha}} \partial
          \gamma_{\vec{\alpha}}:
          - \frac{1}{2}(\partial\vec{\phi})^2 - \frac{i}{\nu}\vec{\rho} \cdot
            \partial^2\vec{\phi}~,
 \end{equation}
where $\nu=\sqrt{k+h^*}$, $h^*=\vec{\theta} \cdot (\vec{ \theta}+2\vec{\rho})/
\vec{\theta}^2$ is the dual Coxeter number, $\vec{\theta}$
is the highest root ($\vec{\theta}^2=2$) and
$2\vec{\rho}=\displaystyle\sum_{\vec{\alpha}\in\Delta_+}\vec{\alpha}$. \\
Thus, the OPEs are

\begin{equation}
\begin{array}{rcl}
& &\gamma_{\vec{\alpha}}(z)\beta_{\vec{\alpha}'}(w)=-\delta_{\vec{\alpha}\vec{
   \alpha}'}\displaystyle\frac{1}{z-w}+ \cdots~,\\
& &\vec{ a}\cdot\vec{ \phi}(z)\vec{ b}\cdot\vec{ \phi}(w) =-\vec{ a}\cdot\vec{
b}
  \log(z-w) + \cdots~,
\end{array}
\end{equation}
where $\vec{a}$ and $\vec{b}$ are arbitrary vectors in the roots space.\\
The central charge of this system indeed is

\begin{equation}
      c=(d-l)+(l-\frac{12}{\nu^2}\vec{ \rho}^2)=\frac{kd}{k+h^*}~,
\end{equation}
where we have used the Freudenthal-de Vries formula $h^*=\displaystyle
\frac{12}{d}\vec{ \rho}^2$. The
vertex operators of primary fields are

\begin{equation}
V_{\vec{\lambda}}(z)=\exp\left((i/\nu)\vec{ \lambda}\cdot
   \vec{\phi}(z)\right)
\end{equation}
with conformal weight
$$\Delta_d(\lambda)=\vec{ \lambda}\cdot(\vec{ \lambda}+2\vec{\rho})/(2\nu^2).$$
Their descendants with respect to the Kac-Moody algebra are

\begin{equation}
V_{\vec{\lambda},\vec{m}}(z)= P_{\vec{
m}}\left(\gamma(z)\right)V_{\vec{\lambda}}(z)~, \end{equation}
where $\vec{ m}$ is an element of the weight lattice.  Unitary
modules correspond to background charges leading to rational central
terms, satisfying $k \in I -\{0\}$ and $\lambda_i \in I$.  The Coulomb
gas duality shows up in that
$\Delta_d(V_{\vec{\lambda}}(z))=\Delta_d(V_{-2\vec{\rho}-\vec{ \lambda}}(z))$.
The screening currents associated with the simple roots acquire a compact
form

\begin{equation}
       J_i(z)=R_i\left(\beta(z),~\gamma(z)\right)K_i(z)~,
\end{equation}
with $R_i$ a polynomial of degree one in the $\beta$'s (hence
$\Delta_d(R_i)=\Delta_d(\beta_i)=1$, since $\Delta_d(\gamma_i)=0$),

\begin{equation}
K_i(z)=\exp\left(-(i/\nu)\vec{\alpha}_i\cdot\vec{\phi}(z)\right)~,
\end{equation}
whose conformal weight $\Delta(K_i)_d=\vec{\alpha}_i\cdot
(\vec{ \alpha}_i-2\vec{\rho})/2\nu^2=0$.
The Kac-Moody currents can be classified into

$$\{x^+_i,~x^-_i,~k_i\}~,~~~~1\leq i\leq d~.$$
For the vertex operators $V_{\vec{\lambda},\vec{ m}}(z)$ and the screening
currents $J_i(z)$ we have the following braid relations

\begin{equation}\label{4.2.61}
\begin{array}{rcl}
& & V_{\vec{\lambda},\vec{ m}}(z) V_{\vec{\lambda}',\vec{ m}'}(w)=q^
    {\vec{\lambda}\cdot\vec{\lambda}'/2}V_{\vec{\alpha}',\vec{
m}'}(w)V_{\vec{\alpha}
    ,\vec{ m}}(z)~,\\[2mm]
& & J_i(z)V_{\vec{ \lambda},\vec{ m}}(w)=q_i^{-\lambda_i/2}V_{\vec{\lambda}
    ,\vec{ m}}(w)J_i(z)~,\\[2mm]
& & J_i(z)J_j(w)=q_i^{a_{ij}/2}J_j(w)J_i(z)~,
\end{array}
\end{equation}
where $\lambda_i=2\vec{\lambda}\cdot\vec{\alpha}_i/\vec{\alpha}^2_i,~q=
e^{2\pi i/(k+h^*)}$, $q_i=q^{\vec{\alpha}^2_i/2}$ and
$a_{ij}=2\vec{\alpha}_i\cdot\vec{\alpha}_j/\vec{alpha}_i^2$. Since $h^*$ is
always an integer for the unitary representation
($k$ an integer) $q$ is a root of unity.  More generally, $q$ is a
root of unity whenever the level $k$ is rational.

Let us define the screening charges by

\begin{equation}
                    Q_i = \int_{z_a}^{z_b} dzJ_i(z)~,
\end{equation}
so as to balance the total Coulomb charge of the correlation function for
maintaining transformation invariance, where $z_a$ and $z_b$ are arbitrary
insertion points $\{z_1,~z_2,~\cdots,~
z_N\}$. The most general setting for the correlation function in the
background charge formalism will be given by\\ \\
$\langle V_{\vec{\lambda}_1,\vec{ m}_1}(z_1) V_{\vec{\lambda}_2,\vec{
m}_2}(z_2)
\cdots V_{\vec{\lambda}_N,\vec{ m}_N}(z_N)\rangle$
\begin{equation}
\sim\langle V_{\vec{\lambda}_1,\vec{ m}_1}(z_1)Q_i(z_a,z_b)\cdots Q_j(z_c,z_d)
                     V_{\vec{ \lambda}_2,\vec{ m}_2}(z_2) \cdots Q_k(z_e,z_f)
                     V_{\vec{\lambda}_N,\vec{ m}_N}\rangle~,
\end{equation}
where $\displaystyle\sum^N_{i=1} \vec{m}_i = 0$, and the total Coulomb
charge on the r.h.s. plus the background charge vanishes.
Define the  screening operators $X^-_i$ as

\begin{equation}
   X^-_i V_{\vec{\lambda},\vec{ m}}(z) = \frac{1}{1-q^{-1}_i}\int_{G} dtJ_i(t)
                         V_{\vec{\lambda},\vec{ m}}(z)~\sim~Q_i(\infty,z)
                         V_{\vec{\lambda},\vec{ m}}(z),
\end{equation}
where the contour $G$ is the same as that we used in the minimal model case.

Then, the $N$-point conformal block are

\begin{equation}
\lim_{z_1\to\infty}\langle \tilde{V}_{\vec{\lambda}_1,\vec{m}_1}(z_1)
X^-_i\cdots
   X^-_j V_{\vec{\lambda}_2,\vec{m}_2}(z_2)
   \cdots X^-_k \cdots X^-_l V_{\vec{\lambda}_N,\vec{m}_N}(z_N)\rangle~,
\end{equation}
with $\displaystyle\sum^N_{i=1} \vec{m}_i=0$, and vanishing total Coulomb
charge.
Notice that in the above equation we have used the $SL(2,C)$ invariance.
This allows us to adopt the convention that one vertex will be taken in its
dual form with the definition for the dual of vertex operator
$$\tilde{V}_{\vec{\lambda},\vec{ m}}(z)=
V_{-\vec{\lambda}-2\vec{\rho},-\vec{ m}}(z)~.$$

We define the operators $k_i$ by

\begin{equation}
k_iV_{\vec{\lambda},\vec{m}}(z)=\exp\left(i\frac{\nu}{\vec{\alpha}^2_i}\oint
dz\partial \vec{\phi}_i(z)-1\right)V_{\vec{\lambda},\vec{m}}(z)~.
\end{equation}
$k_i$ and $X^-_i$ generate the Borel subalgebra of the quantum
group underlying the WZNW model, which turn out to be $U_q(g)$, where
$g$ is the zero mode of the Kac-Moody algebra $\hat{g}$

\begin{equation}
[k_i,k_j] = 0,~~~~~k_iX^-_jk_i^{-1}=q_i^{-1/2}X^-_j~.
\end{equation}
Define the adjoint action of the quantum group ad$X^-_i$ by

\begin{equation}
\left({\rm ad}X^-_i\right)X^-_j=\frac{1}{1-q_i^{-1}}\int_{G}dzX^-_iJ_j(z)~,
\end{equation}
yielding

\begin{equation}
({\rm ad}X^-_i)X^-_j = X^-_iX^-_j - q_i^{a_{ij}/2}X^-_jX^-_i~.
\end{equation}
The consistency of the Borel subalgebra generated by $k_i$ and $X^-_i$
is ensured by the Serre relations

\begin{equation}\label{4.2.73}
\left({\rm ad}X^-_i\right)^{1-a_{ij}}X^-_j = 0~,~~~~(i \neq j)~.
\end{equation}
Owing to the product rule

\begin{equation}
\left({\rm ad}X^-_i\right)\left(X^-_j X^-_k\right) = \left({\rm ad}X^-_i\right)
(X^-_j)X^-_k + q_i^{a_{ij}/2}X^-_j\left({\rm ad}X^-_i\right)(X^-_k)~,
\end{equation}
Equation (\ref{4.2.73}) yields the explicit form of the Serre relation

\begin{equation}
\sum^{1-a_{ij}}_{k=0}(-1)^k q_i^{-k(1-a_{ij}-k)}
\left\vert\left[\begin{array}{c}1-a_{ij}\\ k \end{array}\right]\right.
_{q_i^{-1}}(X^-_i)^kX^-_j(X^-_i)^{1-a_{ij}-k}= 0~,
\end{equation}
where $\displaystyle\left\vert\left[\begin{array}{c}
             n\\
             k\end{array}\right]_q\right.=\frac{|[n]_q!}{|[k]_q!|[n-k]_q}$.

The dual to $X^-_i,~X^+_i$, and $k_i$ generated the other Borel subalgebra
of the quantum group.  We define $X^+_i$ by

\begin{equation}\label{4.2.74}
\nu^2K_i(\infty)X^+_iX^-_IV_{\vec{\lambda},\vec{m}}(z)
=(x^-_i)^{(0)}X^-_IV_{\vec{\lambda},\vec{m}}(z)-
 X^-_I(x^-_i)^{(0)}V_{\vec{\lambda},\vec{m}}(z)~,
\end{equation}
where the short-hand $X^-_I=X^-_{i_1i_2\cdots i_n}=X^-_{i_1}
X^-_{i_2}\cdots X^-_{i_n}$ and $(x^-_i)^{(0)}$ is the zero mode of the
Kac-Moody currents
$x^-_i$.  The acting of the zero mode of the Kac-Moody currents $(x^-_i)^{(0)}$
on vertex operators can be computed directly,\\ \\
$(x^-_i)^{(0)}(X^-_j)^nV_{\vec{\lambda},\vec{m}}(z)$
\begin{equation}
\begin{array}{rcl}
&
&=[(x^-_i)^{(0)},(X^-_j)^n]V_{\vec{\lambda},\vec{m}}(z)+(X^-_j)^n(x^-_i)^{(0)}
   V_{\vec{\lambda},\vec{m}}(z)\\[2mm]
& &=\nu^2\delta_{ij}K_i(\infty)\displaystyle\sum^{n-1}_{k=0}
   q^{n-1-k}_i(X^-_i)^{n-1-k}\frac{1-k_i^{-1}}{1-q^{-1}_i}
   (X^-_i)^kV_{\vec{\lambda},\vec{m}}(z)+(X^-_j)^n(x^-_i)^{(0)}V_{\vec{\lambda}
   ,\vec{m}}(z)~.
\end{array}
\end{equation}
Then we have

\begin{equation}
X^+_iV_{\vec{\lambda},\vec{m}}(z)=0~,
\end{equation}
and

\begin{equation}
\begin{array}{rcl}
& &k_iX^+_jk_i^{-1}=q_i^{1/2}X^+_j~,\\[2mm]
&
&X^+_iX^-_j-q^{a_{ij}/2}_iX^-_jX^+_i=\delta_{ij}\displaystyle\frac{1-k_i^{-2}}
                                    {1-q^{-1}_i}~.
\end{array}
\end{equation}
The adjoint action of $X^+_i$ on $X^+_j$ is also given  by

\begin{equation}
\left({\rm ad}X^+_i\right)X^+_j = X^+_iX^+_j - q_i^{a_{ij}/2}X^+_jX^+_i~.
\end{equation}
Consequently, the corresponding Serre relations are

\begin{equation}
\left({\rm ad}X^+_i\right)^{1-a_{ij}}X^+_j=0~.
\end{equation}
The explicit form of the Serre relations are

\begin{equation}
\sum^{1-a_{ij}}_{k=0}(-1)^k q^{-k(1-a_{ij}-k)}_i
\left\vert\left[\begin{array}{c}
       1-a_{ij}\\
       k\end{array}\right]_{q^{-1}_i}\right.
(X^+_i)^kX^+_j(X^+_i)^{1-a_{ij}-k}=0~.
\end{equation}

Simple contour deformations are sufficient for finding the comultiplication,

\begin{equation}
\begin{array}{rcl}
& &\Delta(k_i)=k_i \otimes k_i~,\\[2mm]
& &\Delta(X^\pm_i)=X^-_i \otimes 1 +k_i^{-1} \otimes X^-_i~,\\[2mm]
& &\Delta(X^+_i)=X^+_i \otimes 1 + k_i^{-1}\otimes X^+_i~.
\end{array}
\end{equation}
The last comultiplication, for $X^+_i$, arises from an integration likewise in
Fig.(21),
however,  due to th factor $K_i(\infty)$ in Eq.(\ref{4.2.74}), we have the
braiding phase that makes the comultiplication non-commutative.

\setlength{\unitlength}{5pt}
\thicklines
\begin{picture}(50,13)(0,4)
\put(10,10){\oval(15,5)}
\put(7,10){$\bullet$}
\put(8.5,9.4){$z_1$}
\put(12,10){$\bullet$}
\put(13.5,9.4){$z_2$}
\put(8,5){$\Delta G$}

\put(21.5,9.5){$=$}

\put(35,10){\oval(8,5)}
\put(34.5,10){$\bullet$}
\put(36,9.4){$z_1$}
\put(43.5,10){$\bullet$}
\put(45,9.4){$z_2$}
\put(34,5){$G_1$}
\put(69,10){\oval(8,5)}
\put(53,9.5){$+$}

\put(59.5,10){$\bullet$}

\put(61,9.4){$z_1$}
\put(68.5,10){$\bullet$}
\put(70,9.4){$z_2$}
\put(69,5){$G_2$}

\end{picture}

\hspace{7.5em}Fig.21. Comultiplication operation of $X_i^+$.

In this basis the counit $\epsilon$ and antipode $S$ are,

\begin{equation}
\begin{array}{l}
\epsilon(k_i)=1~,~~~~ \epsilon(X^-_i)=0=\epsilon(X^+_i)~,\\
S(k_i)=k_i^{-1}~,~~~S(X^-_i)=-X^-_ik_i^{-1}~,~~~S(X^+_i) = -k_iX^+_i~.
\end{array}
\end{equation}
Now we discuss the contour representations of the quantum group $U_q(g)$.
Let us first review the simplest case of $SU_q(2)$, from which the general
case follows, mainly because any Lie algebra can be viewed as a
superposition of the $SU(2)$'s associates with the various simple
roots.  The conventional notation $\lambda=2j$ with $j$ the spin
of the representation is adopted.  The braidings among vertex operators
$V_j={\rm
exp} (ij\phi/\nu)$ of spin $j$ and the screening current $J=\beta{\rm exp}
(-i\phi/\nu)$ follow from relation (\ref{4.2.61}),

\begin{equation}
\begin{array}{rcl}
& &V_j(z)V_{j'}(w)=q^{jj'}V_{j}(w)V_j(z)~,\\
& &J(z)V_j(w)=q^{-j}V_j(w)J(z)~,\\
& &J(z)J(w)=qJ(w)J(z)~.
\end{array}
\end{equation}
The screened vertex operators $e_n^j(z)$ give the representations of the
quantum group $SU_q(2)$

\begin{equation}
\begin{array}{l}
e^j_n(z)=\left(X^-\right)^nV_j(z)~,\\
X^-e^j_n(z)=e^j_{n+1}(z)~,\\
X^+e^j_n(z)=|[n]_{q^{-1}}|[2j-n+1]_qe^j_{n-1}(z)~,\\
ke^j_n=\exp(j-n)e^j_n~.
\end{array}
\end{equation}
By making use of path-ordering, these operators can be rewritten
in terms of the actions of integrals
$P=\displaystyle\int_\infty^zdtJ(t)$ on $V_j(z)$,

\begin{equation}
e_n^j = (X^-)^nV_j(z) =|[2j]_q|[2j-1]_q \cdots|[2j-n+1]_q P^nV_j(z)~.
\end{equation}
Clearly, if $n>2j$, $e_n^j=0$, and  $\dim\{(e_n^j)\}=2j+1$.
It is reasonable to interpret $e_n^j$ as a $q$-multiplet of spin
$j$.
The time-ordering of the screenings is

\begin{equation}
P^n=|[n]_{q^{-1}}!T(P^n)=|[n]_{q^{-1}}!\int_\infty^zdt_1\int_\infty^{t_1}dt_2
\cdots\int_\infty^{t_{n-1}}dt_nJ(t_n)\cdots J(t_1)~.
\end{equation}
Similarly, for an arbitrary simple group $X^-_{i_1}\cdots
X^-_{i_n} V_{\vec{\lambda},\vec{m}}(z)$ can be decomposed into products of
integrals
like

\begin{equation}
P_i = \int^z_\infty dtJ_i(t)~,
\end{equation}
by means of the recursion formula \\ \\
$X^-_iP_{i_1} \cdots P_{i_n} V_{\vec{\lambda},\vec{m}}(z)$
\begin{equation}
=\frac{1}{1-q^{-1}_i}\left(P_iP_{i_1} \cdots P_{i_n}
V_{\vec{\lambda},\vec{m}}(z)
-q^{-\lambda_i+(a_{ii_1}+
\cdots +a_{ii_n})/2}_i P_{i_1}\cdots
P_{i_n}P_i\right)V_{\vec{\lambda},\vec{m}}(z)~.
\end{equation}
Furthermore, the products of screening $P_i$ must be time-ordered.
Consider, for example, a product of two $P_i$.  It is simple to
see that

\begin{equation}
P_{i_1}P_{i_2}V_{\vec{\lambda},\vec{m}}(z) = \left(T\left(P_{i_1}P_{i_2}\right)
-q^{a_{i_1i_2}/2}_{i_1}T\left(P_{i_2}P_{i_1}\right)\right)
V_{\vec{\lambda},\vec{m}}(z)~,
\end{equation}
and the following recursion relation in general holds

\begin{equation}
P_iT(P_{i_1}\cdots P_{i_n})V_{\vec{\lambda},\vec{m}}(z)
=\sum^n_{k=0}q_i^{(a_{ii_1}+\cdots+a_{ii_k})/2}
T(P_{i_1}P_{i_2}\cdots P_{i_k}P_iP_{i_{k+1}}\cdots
                       P_{i_n}V_{\vec{\lambda},\vec{m}}(z)~,
\end{equation}
where

\begin{equation}
T(P_{i_1}\cdots P_{i_n}) = \int^z_\infty dt_n \int^{t_n}_\infty
dt_{n-1}\cdots \int^{t_2}_\infty
dt_1J_{i_1}(t_1)J_{i_2}(t_2)\cdots J_{i_n}(t_n)~.
\end{equation}

The resultant contour representation of the quantum group is
consistent, i.e, it fulfill the Serre relation.

\section{Molecular spectrum}

The Schr\"{o}dinger equation for a diatomic molecule is

\begin{equation}
\frac{1}{m}\sum_i\left(\frac{\partial^2\Psi}{\partial x_i^2}
+\frac{\partial^2\Psi}{\partial y_i^2}+\frac{\partial^2\Psi}{\partial z_i^2}
\right)+
\sum_k\frac{1}{M_k}\left(\frac{\partial^2\Psi}{\partial x_k^2}+
\frac{\partial^2\Psi}{\partial y_k^2}+\frac{\partial^2\Psi}{\partial z_k^2}
\right)+\frac{8\pi^2}{h^2}\left(E-V\right)\Psi=0~,
\end{equation}
where $x_i,y_i,z_i$ are coordinates for electrons with identical
masses  $m$, while
the $x_k,y_k,z_k$ are coordinates for the nuclei with masses $M_k$.
According to Born and Oppenheimer approximation \cite{Born},,
the wave function $\Psi$ can be separated into

\begin{equation}\label{6.4}
\Psi=\psi_e\left(\cdots,x_i,y_i,z_i,\cdots
\right)\psi_{\rm vib-rot}\left(\cdots,x_k,y_k,z_k,\cdots\right)~,
\end{equation}
where the $\psi_e,~\psi_{\rm vib-rot}$ are the solutions of the following
equations,

\begin{equation}\label{6.5}
\begin{array}{l}
\displaystyle \sum_i\left(\frac{\partial^2\psi_e}{\partial x_i^2}+
\frac{\partial^2\psi_e}{\partial y_i^2}+\frac{\partial^2\psi_e}{\partial z_i^2}
\right)+\frac{8\pi^2m}{h^2}\left(E_{e}-V_e\right)\psi_e=0~,\\[2mm]
\displaystyle \sum_k \frac{1}{M_k}\left(\frac{\partial^2\psi_{\rm
vib-rot}}{\partial x_k^2}+
\frac{\partial^2\psi_{\rm vib-rot}}{\partial y_k^2}+\frac{\partial^2\psi_{\rm
vib-rot}}
{\partial z_k^2}
\right)+\frac{8\pi^2}{h^2}\left(E-E_{e}-V_n\right)\psi_{\rm vib-rot}=0~.
\end{array}
\end{equation}
The first equation is the Schr\"{o}dinger equation of  electrons
moving in the field of  fixed nuclei, represented by an effective potential
$V_e$.  The second one describes the  motion in the
effective potential $E_{e}+V_n$, with $V_n=\displaystyle\frac{z_1z_2e^2}{r}$,
 the Coulomb potential between the two nuclei of electric charge  $z_1e$
and $z_2e$ in distance $r$. It can be cast into the
following form

\begin{equation}
\left(-\frac{h^2}{8\pi^2 M_1}\bigtriangledown _1^2-
\frac{h^2}{8\pi^2 M_2}\bigtriangledown _2^2+V_n(r)\right)\psi_{\rm vib-rot}
=E_t\psi_{\rm vib-rot}~,
\end{equation}
where $E_t=E-E_e$. In the center-of-mass frame the equation reads

\begin{equation}
\left(-\frac{h^2}{8\pi^2 M}\bigtriangledown^2+V_n(r)\right)
\psi_{\rm vib-rot}=E_t\psi_{\rm vib-rot}~,
\end{equation}
where $\displaystyle M=\frac{M_1M_2}{M_1+M_2}$ is the reduced mass. As the
tangent and radial variables can be separated, we have

\begin{equation}
\psi_{\rm vib-rot}=R(r)\psi_{\rm rot}(\theta,\phi)~, ~~R(r)={1\over r}
\psi_{\rm vib}(r)~,
\end{equation}
and

\begin{equation}\label{6.8}
\Psi=\psi_e\cdot {1\over r}\psi_{\rm vib}\psi_{\rm rot}~.
\end{equation}
The Hamiltonian of the diatomic molecule are composed of three parts
for electronic transition, vibration and rotation, i.e.,

\begin{equation}\label{6.9}
H=H_e+H_{\rm vib}+H_{\rm rot}~.
\end{equation}
The phenomenological description of the vibrational and rotational
energy is given by the following formulas \cite{Herzberg}--\cite{Mizushima}

\begin{equation}\label{6.22}
\begin{array}{l}
E_{\rm vib}=\displaystyle hc\omega_e(v+{1\over 2})-hc\omega_ex_e(v+{1\over
2})^{2}
+hc\omega_ey_e(v+{1\over 2})^3+\cdots~,\\[2mm]
E_{\rm rot}=BhcJ(J+1)-DhcJ^2(J+1)^2+HhcJ^3(J+1)^3+\cdots~.
\end{array}
\end{equation}
When the coefficients are selected appropriately, the spectra given by the
above formulas may fit with  experimental results very well.

The interaction between vibration and rotation can be taken into account by
the explicit $v$ dependence  of the coefficients $B$, $D$, $\cdots$, in the
second formula of Eq.(\ref{6.22}).
To lowest order, the set of vibration-rotational constants can be
written as

\begin{equation}\begin{array}{rcl}
B_v&=&B-\alpha_e(v+\displaystyle\frac{1}{2})+\cdots~,\\[2mm]
D_v&=&D+\beta_e(v+\displaystyle\frac{1}{2})+\cdots~,
\end{array}\end{equation}
where $\alpha_e$ and $\beta_e$ are constants  much smaller than $B$ and $D$
respectively.
Thus, we have the energy levels of the vibrating and rotating diatomic
molecule

\begin{equation}\label{6.6429}
\begin{array}{rcl}
E_{\rm vib-rot}(v,J)&=&\displaystyle hc\omega_e\left(v+{1\over 2}\right)
              -hc\omega_ex_e\left(v+{1\over 2}
\right)^{2}+hc\omega_ey_e\left(v+{1\over 2}\right)^3\\[2mm]
           &&\displaystyle-hc\alpha_e(v+\frac{1}{2})J(J+1)
             -hc\beta_e(v+\frac{1}{2})J^2(J+1)^2\\[2mm]
           &&+hcB_eJ(J+1)-hcD_eJ^2(J+1)^2+\cdots~.
\end{array}\end{equation}
The complete phenomenological description for vibrational and rotational
structure of diatomic molecule is given by the Dunham expansion
\cite{Dunham29,Dunham32}

\begin{equation}
E_{\rm vib-rot}(v,J)=hc\sum_{ij}Y_{ij}\left(v+\frac{1}{2}\right)^i
\left(J(J+1)\right)^j~,
\end{equation}
where $Y_{ij}$ are the coefficients of the Dunham expansion. A successful
theory for vibrating and rotating spectra of diatomic molecules
must recover  leading terms of the Dunham expansion.

\subsection{Vibrating diatomic molecular spectrum}

The Hamiltonian for a quantum  $q$-oscillator system \cite{ZC91a,ZC91b} is

\begin{equation}\label{6.23}
H_{q-{\rm vib}}={1\over 2}\left(a_q^\dagger a_q+
a_q a_q^\dagger \right)hc\nu_{vib}~,
\end{equation}
where  $a_q$, $a_q^\dagger$ are annihilation and creation
operators for the deformed system. These operators are related with
the operators $a,a^\dagger$ of the harmonic oscillator system by

\begin{equation}\label{6.6432}
a_q=\sqrt{\frac{[N +1+b\gamma]_q}{N +1}}a~,~~~~
a_q^\dagger=a^\dagger\sqrt{\frac{[N+1+b\gamma]_q}{N+1}}~,
\end{equation}
where $N =a^\dagger a$. Making use of the basic commutators

\begin{equation}\label{6.26}
[a,a^\dagger]=1,~~~[a,a]=[a^\dagger,a^\dagger]=0,
\end{equation}
we have the following commutation relations \cite{Macfarlane}--\cite{Song90}

\begin{equation}\label{6.37}
\begin{array}{rcl}
\displaystyle  \left[a_q,a_q^\dagger\right]&=&\left[N+1+b\gamma\right]_q
-\left[N+b\gamma\right]_q~,\\
\displaystyle  \left[N, a_q\right]&=&-a_q~,~~~
\displaystyle  \left[N, a_q^\dagger\right]=a_q^\dagger~.
\end{array}\end{equation}
The Hopf algebra structure can be defined by

\begin{equation}\label{6.37a}
\begin{array}{l}
\Delta(N')=N'\otimes 1+1\otimes N'-i\displaystyle\frac{\alpha}{\gamma}
           1\otimes 1~,\\
\Delta(a_q)=\left(a_q\otimes q^{N'/2}+iq^{-N'/2}\otimes
           a_q\right)e^{-i\alpha/2}~,\\
\Delta(a_q^\dagger)=\left(a_q^\dagger\otimes q^{N'/2}+
           iq^{-N'/2}\otimes a_q^\dagger\right)e^{-i\alpha/2}~,\\
\epsilon(N')=\displaystyle i\frac{\alpha}{\gamma}~,~~~~
           \epsilon(a_q)=0=\epsilon(a_q^\dagger)~,\\
S(N')=-N'+i\displaystyle\frac{2\alpha}{\gamma}\cdot 1~,\\
S(a_q)=-q^{-1/2}a_q~,~~~~~S(a_q^\dagger)=-q^{1/2}a_q^\dagger~,
\end{array}
\end{equation}
where $N'=N+\displaystyle\frac{1}{2}+b\gamma,~\alpha=2k\pi+\frac{\pi}{2},
{}~k\in Z$.\\
Eqs.(\ref{6.37}) and (\ref{6.37a}) constitute the quantum Weyl-Heisenberg
group $H_q(4)$.

Use is made of Eq.({\ref{6.6432}) to cast the Hamiltonian for the
$q$-oscillator system into

\begin{equation}\label{6.e316}
H_{q-{\rm vib}}={1\over 2}\left([N+b\gamma ]_q+[N+1+b\gamma
]_q\right)hc\nu_{vib}~.
\end{equation}
The representations of quantum group $H_q(4)$ are constructed by

\begin{equation}\label{6.29}
\mid{v}\rangle\rangle=\left([v+b\gamma ]_q!\right)^{-1/2}\left(a_q^\dagger
\right)^v\vert 0\rangle~.
\end{equation}
The actions of the operators $a_q^\dagger$, $a_q$ on the Fock states yield

\begin{equation}\label{6.30}
\begin{array}{l}
\displaystyle
a_q^\dagger\vert {v}\rangle\rangle=\sqrt{[v+1+b\gamma ]_q}\vert v+1\rangle
\rangle~,\\
\displaystyle a_q\vert {v}\rangle\rangle=\sqrt{[v+b\gamma ]_q}\vert v-1
\rangle\rangle~,\\
\displaystyle a_q\vert {0}\rangle=0~,
\end{array}\end{equation}
and the energy levels of the system are

\begin{equation}\label{6.31}
\begin{array}{rcl}
E_{q-{\rm vib}}(v)&=&\displaystyle{1\over 2}\left([v+1+b\gamma ]_q+[v+b\gamma
]_q\right)h
                     c\nu_{vib}\\[2mm]
                  &=&\displaystyle\frac{hc\nu_{\rm
vib}}{2\sinh\frac{\gamma}{2}}
                     \sinh\left(\gamma(v+\frac{1}{2}+b\gamma)\right)\\[2mm]
                  &=&\displaystyle hc\nu_{\rm vib}\left(\frac{\sinh(\gamma c)}
                     {2\sinh\left(\frac{\gamma}{2}\right)}+\cosh(\gamma
c)\left(
                     v+\frac{1}{2}\right)\right.\\[2mm]
                  & &~~~~~\displaystyle\left.+\frac{\gamma\sinh(\gamma c)}{2}
                     \left(v+\frac{1}{2}\right)^2+\frac{\gamma^2\cosh(\gamma
c)}{6}
                     \left(v+\frac{1}{2}\right)^3+\cdots\right)~,
\end{array}
\end{equation}
where $c\equiv b\gamma$.

It is easy to see that the above equation gives the coefficients
$\omega_e,~\omega_ex_e,~\omega_ey_e,~\cdots$ of Eq.({\ref{6.22}).

For generic $q$, the representation for the quantum group
$H_q(4)$ is isomorphic to that for the Lie group $H(4)$

\begin{equation}\label{6.32}
\begin{array}{rcl}
\vert v\rangle\rangle&=&\left([v+b\gamma ]_q !\right)^{-1/2}
                        \left(a_q^\dagger\right)^v\vert 0\rangle\\
                     &=&\left(v!\right)^{-1/2}\left(a^\dagger\right)^v
                        \vert 0\rangle\\
                     &=&\vert v\rangle~.
\end{array}\end{equation}
where $\vert v\rangle$ are the Fock states for the harmonic oscillator system.
Therefore, the representations of $H_q(4)$
in space coordinates  can be exactly expressed by Hermite polynomials, i.e.,

\begin{equation}\label{6.33}
\tilde{\psi}_v(x)= N_v H_v(X)e^{-X^2/2}~,
\end{equation}
where $X=\beta x$, $x$ is the change of
internuclear distance away from equilibrium position,
$\beta=\left((2\pi)^2mc\nu_{\rm vib}/h\right)^{1/2}$ and $N_v=\left(\sqrt{\pi}
2^vv!\right)^{-1/2}$.

If there is a dipole moment for the nuclei in equilibrium position, as
it should be for the molecule of unlike atoms, this dipole
moment  changes upon  varying the internuclear distance. As a first
order  approximation, the dipole moment is assumed to vary linearly
 with the internuclear distance, i.e., $M=M_0+M_1x$,
where $M_0$ is the dipole moment at equilibrium, $M_1$ is the rate of
change of the dipole moment with the
internuclear distance. Therefore, the transition matrix elements are

\begin{equation}\label{6.35}
R^{v^\prime v^{\prime\prime}}_q=\displaystyle
\int\tilde{\psi}^\ast_{v^\prime} M
\tilde{\psi}_{v^{\prime\prime}} dx =
\displaystyle M_0\delta_{v^\prime v^{\prime\prime}}+M_1\frac{N_{v^\prime}
N_{v^{\prime\prime}}}{\beta}\int X H_{v^\prime}(X) H_{v^{\prime\prime}}
(X) e^{-X^2}dx~.
\end{equation}
By the recursion relation of the Hermite polynomial, the second term
in the above equation vanishes unless
$v^\prime=v^{\prime\prime}\pm 1$. The selection rule
for the $q$-oscillator system is $\Delta v=\pm 1$. Therefore the infrared
spectrum  of the $q$-oscillator system is

\begin{equation}
\label{6.39}
\nu=\frac{1}{2}\left([v+2+b\gamma ]_q-[v+b\gamma ]_q\right)\nu_{vib}~.
\end{equation}

An external electric field ${\bf F}$ induces
a dipole moment ${\bf p}$ in the diatomic system. Its magnitude
is  proportional to that of the field,

\begin{equation}\label{6.40}
\vert{\bf p}\vert=\alpha\vert {\bf F}\vert~,
\end{equation}
where $\alpha$ is the polarizability.
The transition matrix elements corresponding to the induced dipole moment
are

\begin{equation}\label{6.41}
\left|{\bf p}\right|^{v^\prime v^{\prime\prime}}=\int \tilde{\Psi}_{v^\prime}
^\ast {\bf p}\tilde{\Psi}_{v^{\prime\prime}}dx~,
\end{equation}
where $\tilde{\Psi}_{v^\prime}$ and $\tilde{\Psi}_{v^{\prime\prime}}$
are the (time dependent)
wave functions of the $q$-oscillator system at states $v^\prime$ and
$v^{\prime\prime}$ \ respectively. The evolution factors for
 $\tilde{\Psi}_{v^\prime}^\ast$, $\tilde{\Psi}_{v^{\prime\prime}}$ and
${\bf p}$ should be $e^{2\pi i(E_{q-{\rm vib}}(v^\prime)/h)t}$,
$e^{-2\pi i(E_{q-{\rm vib}}(v^{\prime\prime})/h)t}$ and $e^{2\pi ic
\nu_{\rm ext}t}$,
where the $c\nu_{\rm ext}$ is the frequency of the external electric field.
$\left|{\bf p}\right|^{v^\prime v^{\prime\prime}}$ evolves by
frequency $c\nu_{\rm ext}+(E_{q-{\rm vib}}(v^\prime)-E_{q-{\rm vib}}(
v^{\prime\prime}))/h$ with the amplitude

\begin{equation}\label{6.42}
\left|{\bf p}\right|^{v^\prime v^{\prime\prime}}=
\vert {\bf F}\vert \int \tilde{\psi}_{v^\prime}^\ast \alpha
\tilde{\psi}_{v^{\prime\prime}}dx~.
\end{equation}
To the lowest order assuming a linear variation of $\alpha$
with the displacement $x$ from the equilibrium position, i.e.,

\begin{equation}\label{6.43}
\alpha=\alpha_{0v}+\alpha^1_vx~,
\end{equation}
we have

\begin{equation}\label{6.e23}
\left|{\bf p}\right|^{v^\prime v^{\prime\prime}}=
\vert {\bf F}\vert\alpha_{0v}\int
\tilde{\psi}^\ast_{v^\prime}\tilde{\psi}_{v^{\prime\prime}}dx+
\vert {\bf F}\vert\alpha_{v}^1\int x
\tilde{\psi}^\ast_{v^\prime}\tilde{\psi}_{v^{\prime\prime}}dx~.
\end{equation}
Because of the orthogonality of the wave functions of the $q$-oscillator
system, the first term in the above equation is zero unless
$v^\prime=v^{\prime\prime}$, which gives the Reighley scattering. The
integration in the second
term vanishes unless $v^\prime=v^{\prime\prime}\pm 1$. The selection
rule for vibrational Raman spectrum is $\Delta v=\pm 1$.
We then obtain the vibrational Raman
spectrum from the energy levels given in Eq.(\ref{6.e316}),

\begin{equation}\label{6.46}
\begin{array}{rcl}
\nu&=&{\nu^\prime}\pm \left(E_{q-{\rm vib}}(v^\prime+1)-E_{q-{\rm vib}}
(v)\right)\\[3mm]
&=&\displaystyle{\nu^\prime}\pm \frac{1}{2}\left([v+2+b\gamma ]_q-[v+b\gamma
]_q\right)\nu_{vib}~,
\end{array}
\end{equation}
where $\nu^\prime$ is the wave number of the incident photon.

The electronic transitions are involved in the
visible and ultraviolet spectral regions. Denote the vibrational spectrum of
the
electronic state by $E_{\rm e-vib}$, leading to

\begin{equation}\label{6.56x}
E_{\rm e-vib}=E_0+E_{\rm q-vib}~.
\end{equation}
Then,  the vibrational spectrum of electronic transitions is

\begin{equation}\label{6.47}
\begin{array}{rcl}
\nu&=&\left(E^\prime_{\rm e-vib}-E^{\prime\prime}_{\rm e-vib}\right)/hc\\
   &=&(E_0^\prime-E_0^{\prime\prime})/hc
+(G^\prime_{q^\prime}-G^{\prime\prime}_{q^{\prime\prime}})
=\nu_e+\nu_{v}~,
\end{array}
\end{equation}
where $\nu_e$ is a certain  constant depending on the transition of
electronic states, $G_q\equiv E_{q-vib}/hc$. By Eq.(\ref{6.e316}) we rewrite
Eq.(\ref{6.47}) into the following:

\begin{equation}\label{6.48}
\begin{array}{rcl}
\nu&=&\displaystyle\nu_e+\frac{1}{2}\left([v^\prime+1+b^\prime\gamma^\prime
]_{q^\prime}+[v^\prime+b^\prime\gamma^\prime ]_{q^\prime}\right)
\nu_{vib}^\prime\\[2mm]
& &-\displaystyle\frac{1}{2}\left(
[v^{\prime\prime}+1+b^{\prime\prime}\gamma^{\prime\prime} ]_{q^{\prime\prime}}
+[v^{\prime\prime}+b^{\prime\prime}\gamma^{\prime\prime}
]_{q^{\prime\prime}}\right)\nu_{vib}^{\prime\prime}~.
\end{array}
\end{equation}
An investigation of the selection rules shows that for electronic transitions
there is no strict selection rule for the vibrational quantum number $v$.
In principle, each vibrational state of the upper electronic state can be
combined with each vibrational state of lower electronic
state, i.e., there is no restriction to the quantum numbers $v^\prime$ and
$v^{\prime\prime}$ in Eq.(\ref{6.48}), and therefore Eq.(\ref{6.48}) gives very
complicated
spectral structures.

If the quantum number $v^\prime$ in Eq.(\ref{6.48}) is fixed, then the
$v^{\prime\prime}$ progression
 is formed. In the $v^{\prime\prime}$ progression
the upper vibrational state is fixed while the lower vibrational state
is different. Then Eq.(\ref{6.48}) is rewritten in the following form

\begin{equation}\label{6.49}
\nu={\nu_0^{\prime\prime}}-{1\over 2}\left([v^{\prime\prime}+1+b^{\prime\prime}
\gamma^{\prime\prime} ]_{q^{\prime\prime}}
+[v^{\prime\prime}+b^{\prime\prime}\gamma^{\prime\prime}
]_{q^{\prime\prime}}\right){\nu_{vib}^{\prime\prime}}~,
\end{equation}
where the quantum number ${\nu_0^{\prime\prime}}$ is $\nu_e$
 plus the fixed vibrational
spectrum in the upper electronic state, and therefore is a constant.

If the quantum number $v^{\prime\prime}$ is chosen to be a constant, the
$v^\prime$ progression is formed in which different vibrational  state
in an upper electronic state combine with the vibrational state of
lower electronic state. It is expressed by the following formula

\begin{equation}
\label{6.50}
\nu={\nu_0^{\prime}}+{1\over 2}\left([v^{\prime}+1+b^{\prime}\gamma^{\prime}
 ]_{q^{\prime}}
+[v^{\prime}+b^{\prime}\gamma^{\prime}
]_{q^{\prime}}\right){\nu_{vib}^\prime}~,
\end{equation}
where $\nu_0^\prime$ is $\nu_e$ minus the fixed vibrational spectrum of
the lower electronic state, and therefore is a constant.

\subsection{Rotating diatomic molecular spectrum}

The Hamiltonian for the $q$-rotator system \cite{ZC91b},\cite{ZC91bb}--
\cite{Bonatsos90} is

\begin{equation}\label{6.1a}
H_{q-{\rm rot}}=\frac{h^2}{8\pi^2I}C_{I,q}~,
\end{equation}
where
\begin{equation}\label{6.2a}
C_{I,q}=J^\prime_- J^\prime_++[J_3^\prime]_q[J_3^\prime+1]_q~,
\end{equation}
is the Casimir operator for the quantum group $SU_q(2)$

\begin{equation}
\begin{array}{l}
[J_3',J_\pm']=\pm J_\pm'~,\\[1mm]
[J_+',J_-']=[2J_3']~,\\[1mm]
\Delta(J_3')=J_3'\otimes 1+1\otimes J_3'~,\\
\Delta(J_\pm')=J_\pm'\otimes q^{-J_3'}+q^{J_3'}\otimes J_\pm'~,\\
\epsilon(J_3')=0=\epsilon(J_\pm')~,\\
S(J_3')=-J_3'~,~~~~~S(J_\pm')=-q^{\pm 1}J_\pm'~.
\end{array}
\end{equation}
It is well-known  that the generators of the quantum group $SU_q(2)$ are
related with  the generators of the Lie group $SU(2)$ by

\begin{equation}\label{6.8a}
\begin{array}{l}
J_+^\prime=\displaystyle
\sqrt{\frac{[J_3+J_0]_q[J_3-1-J_0]_q}{(J_3+J_0)
(J_3-1-J_0)}}J_+~,\\[3mm]
J_-^\prime=\displaystyle J_-\sqrt{\frac{[J_3+J_0]_q[J_3-1-J_0]_q}{(J_3+J_0)
(J_3-1-J_0)}}~,\\[3mm]
J_3^\prime=J_3~.
\end{array}
\end{equation}
For generic $q$, the representations (in coordinates space) of
$SU_q(2)$ can be chosen to be the spherical harmonics , i.e.,

\begin{equation}\label{6.9a}
\tilde{\psi}_{JM}({\bf x})=Y_{JM}(\theta, \phi)~.
\end{equation}
The actions of the generators of $SU_q(2)$ yield

\begin{equation}\label{6.10a}
\begin{array}{l}
 J^\prime_\pm\tilde{\psi}_{JM}({\bf x})=\sqrt{[J\mp M]_q
[J\pm M+1]_q}\tilde{\psi}_{J,M\pm 1}({\bf x})~,\\[2mm]
J^\prime_3\tilde{\psi}_{JM}({\bf x})=M\tilde{\psi}_{JM}({\bf x})~.
\end{array}
\end{equation}
The action of the Casimir operator $C_{I,q}$ yields

\begin{equation}\label{6.11a} C_{I,q}\tilde{\psi}_{JM}({\bf
x})=[J+1]_q[J]_q\tilde{\psi}_{JM}({\bf x})~.
\end{equation}
Therefore, the
eigenvalues of the Hamiltonian for the $q$-rotator system are

\begin{equation}\label{6.ppa}
\begin{array}{rcl}
E_{q-{\rm rot}}&=&\displaystyle\frac{h^2}{8\pi^2I}[J]_q[J+1]_q\\[2mm]
               &=&\displaystyle\frac{h^2}{8\pi^2I}\left(\left(1-\frac{1}{6}
                  \gamma^2+\frac{7}{360}\gamma^4\right)J(J+1)+
                  \gamma^2\left(\frac{1}{3}-\frac{7}{90}\gamma^2\right)\left(
                  J(J+1)\right)^2\right.\\[2mm]
               & &\displaystyle\left. +\frac{2\gamma^4}{45}\left(J(J+1)
                  \right)^3+\cdots\right)~.
\end{array}
\end{equation}
It is easy to see that Eq.(\ref{6.ppa}) gives the coefficients
$B_0,~D_0,~H_0,~\cdots$ of Eq.(\ref{6.22}).

There is internal dipole moment $M_0$ in the diatomic molecules
system, as it is always for molecules consisting of unlike atoms
with  spatial components,

\begin{equation}\label{6.12a}
\begin{array}{l}
M_{0x}=M_0\sin\theta\cos\phi~,\\[2mm]
M_{0y}=M_0\sin\theta\sin\phi~,\\[2mm]
M_{0z}=M_0\cos\theta~.
\end{array}\end{equation}
The dipole transition matrix elements are

\begin{equation}\label{6.13a}
\begin{array}{l}
R_x^{J^\prime M^\prime J^{\prime\prime} M^{\prime\prime}}
  =\displaystyle M_0\int \tilde{\psi}_{J^\prime M^{\prime}}^\ast
  \sin\theta\cos\phi\tilde{\psi}_{J^{\prime\prime}M^{\prime\prime}}
  d\tau~,\\[2mm]
R_y^{J^\prime M^\prime J^{\prime\prime} M^{\prime\prime}}
  =\displaystyle M_0\int \tilde{\psi}_{J^\prime M^{\prime}}^\ast
  \sin\theta\sin\phi\tilde{\psi}_{J^{\prime\prime}
  M^{\prime\prime}}d\tau~,\\[2mm]
R_z^{J^\prime M^\prime J^{\prime\prime} M^{\prime\prime}}
  =\displaystyle M_0\int \tilde{\psi}_{J^\prime M^{\prime}}^\ast
  \cos\theta\tilde{\psi}_{J^{\prime\prime}M^{\prime\prime}}d\tau~,
\end{array}
\end{equation}
where $d\tau=\sin\theta d\theta d\phi$.
Applying the recursion relation of the spherical harmonics,
we can cast $R_z^{J^\prime M^\prime J^{\prime\prime} M^{\prime\prime}}$ into

\begin{equation}
\begin{array}{rcl}
R_z^{J^\prime J^\prime J^{\prime\prime} M^{\prime\prime}}
 &=&M_0\left(a_{J^{\prime\prime},M^{\prime\prime}}\displaystyle
    \int Y_{J^\prime M^\prime}^\ast Y_{J^{\prime\prime}+1, M^{\prime\prime}}
    \sin\theta d\theta d\phi\right.\\[4mm]
 & &\left.+a_{J^{\prime\prime}-1,M^{\prime\prime}}\displaystyle
    \int Y_{J^\prime, M^\prime}^\ast
    Y_{J^{\prime\prime}-1, M^{\prime\prime}}\sin\theta d\theta d\phi\right)~,
\end{array}\end{equation}
where

\begin{equation}\label{6.15a}
a_{J,M}=\sqrt{\frac{(J+1)^2-M^2}{(2J+1)(2J+3)}}~.
\end{equation}
By the orthogonality of the spherical harmonics,
the above matrix elements vanish unless $J^{\prime\prime}=J^\prime\pm 1$.
The similar result  is valid for
$R_x^{J^\prime M^\prime J^{\prime\prime} M^{\prime\prime}}$ and
$R_y^{J^\prime M^\prime J^{\prime\prime} M^{\prime\prime}}$. As a result the
selection rule of the emission (absorption) of the $q$-rotator model
is $\Delta J=\pm 1$. The emission (absorption) spectrum is

\begin{equation}\label{6.y1}\begin{array}{rcl}
\nu&=&\frac{\displaystyle E_{q-rot}(J+1)-E_{q-rot}(J)}{\displaystyle
hc},\\[2mm]
&=&B\left([J+1]_q[J+2]_q-[J+1]_q[J]_q\right),
\end{array}\end{equation}
where $B=\displaystyle\frac{h}{8\pi^2 Ic}$.

With an external electric field ${\bf F}$ an induced dipole moment
is formed.
Suppose that the external field is along  $z$-axis,
and the induced dipole moment  along the $z$-axis is

\begin{equation}\label{6.17a}
{\bf p}_z=\alpha_{zz}{\bf F}_z~,
\end{equation}
where $\alpha_{zz}$ is a component of the polarizability tensor
in the fixed frame.  In terms of
the polarizability measured in the frame rotating with the molecules, it
is expressed by

\begin{equation}\label{6.18a}
\alpha_{zz}=\alpha_{x_mx_m}+
\left(\alpha_{z_mz_m}-\alpha_{x_mx_m}\right)\cos^2\theta~,
\end{equation}
where $\alpha_{x_{m}x_{m}}$ and $\alpha_{z_mz_m}$ are the components of the
polarizability tensor measured in the frame fixed on the rotating molecule.
The  corresponding matrix elements are

\begin{equation}
\begin{array}{rcl}
\displaystyle\int \alpha_{zz}{\psi}^\ast_{J^\prime M^\prime}
{\psi}_{J^{\prime\prime}M^{\prime\prime}}d\tau
&=&\alpha_{x_mx_m}\displaystyle\int Y^\ast_{J^\prime M^\prime}
   Y_{J^{\prime\prime}M^{\prime\prime}}d\tau\\[3mm]
& &+\left(\alpha_{z_mz_m}-\alpha_{x_mx_m}\right)\displaystyle
   \int\cos^2\theta Y^\ast_{J^\prime M^\prime}
   Y_{J^{\prime\prime} M^{\prime\prime}}d\tau~.
\end{array}
\end{equation}
According to the recursion relations of the spherical harmonics, the above
equation can be written as \\ \\
$\displaystyle\int \alpha_{zz}\tilde{\psi}^\ast_{J^\prime M^\prime}
\tilde{\psi}_{J^{\prime\prime}M^{\prime\prime}}d\tau$
\begin{equation}
\begin{array}{rcl}
&=&\left(\alpha_{x_mx_m}+\left(\alpha_{z_mz_m}-\alpha_{x_mx_m}
   \right)\left(\left(a_{J^{\prime\prime},M^{\prime\prime}}\right)^2
   +\left(a_{J^{\prime\prime}-1,M^{\prime\prime}}\right)^2\right)\right)
   \displaystyle\int Y^\ast_{J^\prime M^\prime}
   Y_{J^{\prime\prime}M^{\prime\prime}}d\tau\\[3mm]
& &+\left(\alpha_{z_mz_m}-\alpha_{x_mx_m}\right)
   a_{J^{\prime\prime},M^{\prime\prime}}a_{J^{\prime\prime}+1,M^{\prime\prime}}
   \displaystyle\int Y^\ast_{J^\prime M^\prime}
   Y_{J^{\prime\prime}+2, M^{\prime\prime}}d\tau\\[3mm]
& &+\left(\alpha_{z_mz_m}-\alpha_{x_mx_m}\right)

a_{J^{\prime\prime}-1,M^{\prime\prime}}a_{J^{\prime\prime}-2,M^{\prime\prime}}
   \displaystyle\int Y^\ast_{J^\prime M^\prime}
   Y_{J^{\prime\prime}-2, M^{\prime\prime}}d\tau~.
\end{array}
\end{equation}
It is obvious that the first term in the above equation vanishes
unless $J^\prime=J^{\prime\prime}$,
i.e., it gives the lines without shifting; the second and third terms
vanish unless
$J^\prime=J^{\prime\prime}\pm 2$, i.e., it gives the shifted lines.
The same results can be obtained for the other two components.
Therefore the selection rule for the rotational Raman spectrum is
$\Delta J=\pm 2$. The rotational Raman spectrum can be expressed as

\begin{equation}\label{6.y2}
\begin{array}{rcl}
\nu&=&\nu_o+\frac{\displaystyle\left(E_{q-{\rm rot}}(J+2)-
E_{q-{\rm rot}}(J)\right)}{\displaystyle hc}~,\\[2mm]
&=&\nu_0+B\left([J+3]_{q}
[J+2]_{q}-[J+1]_{q}[J]_{q}\right)~,
\end{array}
\end{equation}
where $\nu_0$ is the wave number of the incident photon.

The rotational spectra of diatomic molecules involve the electronic
transitions and vibrational transitions, obeying the selection rule identical
to  that for rigid rotator, i.e., $\Delta J=0,\pm 1$.
The rotational structures are

\begin{equation}\label{6.x0}
\begin{array}{clclll}
R&~{\rm branch:}&~
\nu&=&\nu_0+\frac{\displaystyle {E^\prime_{q-{\rm rot}}(J+1)-
      E^{\prime\prime}_{q-{\rm rot}}(J)}}
      {\displaystyle hc}~,\\[2mm]
&& &=&\nu_0+B^\prime[J+1]_{q^\prime} [J+2]_{q^\prime}
			-B^{\prime\prime}[J]_{q^{\prime\prime}}
      [J+1]_{q^{\prime\prime}}~;\\[4mm]
Q&~{\rm branch:}&~
\nu&=&\nu_0+\frac{\displaystyle {E^\prime_{q-{\rm rot}}(J)-
      E^{\prime\prime}_{q-{\rm rot}}(J)}}
      {\displaystyle hc}~,\\[2mm]
&& &=&\nu_0+B^\prime [J]_{q^\prime}[J+1]_{q^\prime}
			-B^{\prime\prime}[J]_{q^{\prime\prime}}
      [J+1]_{q^{\prime\prime}}~;\\[4mm]
P&~{\rm branch:}&~
\nu&=&\nu_0+\frac{\displaystyle {E^\prime_{q-{\rm rot}}(J-1)-
      E^{\prime\prime}_{q-{\rm rot}}(J)}}
      {\displaystyle hc}~,\\[2mm]
&& &=&\nu_0+B^\prime[J-1]_{q^\prime}     [J]_{q^\prime}
      -B^{\prime\prime}[J]_{q^{\prime\prime}}
      [J+1]_{q^{\prime\prime}}~, \end{array} \end{equation} where $\nu_0$
is a quantity solely depending on the electronic transitions and
vibrational
structure, and $E^\prime_{q\rm -rot}$, $E^{\prime\prime}_{q\rm -rot}$ are
the eigenvalues of the $q$-rotator system in the upper and lower
electronic states respectively. Since internuclear distances are different
for different electronic states, the moments of inertia $I$'s of the
system are different for different electronic states. $B^\prime$
and $B^{\prime\prime}$ are different,  because they are proportional to
$I^{-1}$.

\subsection{Vibrating-rotating structure}

The Hamiltonian describing the
vibrating-rotating structure of diatomic molecules \cite{ZC91c,ZC91d}
is
\begin{equation}\label{6.e388}
H_{q(J)-\rm vib}={1\over 2}\left(a_{q(J)}^\dagger a_{q(J)}+a_{q(J)}a_{q(J)}
^\dagger\right)hc\nu_{\rm vib}~.
\end{equation}
This Hamiltonian differs from that of  the previous subsections, as
the quantization parameter $q$ is no longer a constant, but takes different
values at different rotational levels. The dependence of $q$ on the
rotational quantum number $J$ twists the effective rotational levels and
provides the necessary interaction of vibration and rotation.

It is obvious that this Hamiltonian commutes with the rotational one,
thus $H_{q(J)-{\rm vib}}$ and $H_{\rm rot}$ have common eigenstates.
The total Hamiltonian of the system reads

\begin{equation}
\begin{array}{rcl}
H_{q-\rm vib-rot}&=&H_{q(J)-{\rm vib}}+H_{\rm rot}\\[3mm]
                 &=&\displaystyle{1\over 2}\left(a_{q(J)}^\dagger
                    a_{q(J)}+a_{q(J)}a_{q(J)}^\dagger\right)hc
                    \nu_{\rm vib}+\displaystyle\frac{h^2}{8\pi^2I}C_I~,
\end{array}
\end{equation}
where $C_I$ is the Casimir operator
of the Lie group $SU(2)$. The dependence of $q$ on $J$ cannot be ignored
unless the interaction between vibration and rotation is negligible.

The quantum group $H_{q(J)}(4)$ is

\begin{equation}\label{6.27}
\begin{array}{l}
\displaystyle  \left[a_{q(J)},a_{q(J)}^\dagger\right]=\left[N+1+b\gamma(J)
\right]_{q(J)}-\left[N+b\gamma(J)\right]_{q(J)}~,\\
\displaystyle  \left[N, a_{q(J)}\right]=-a_{q(J)}~,~~~
\displaystyle  \left[N, a_{q(J)}^\dagger\right]=a_{q(J)}^\dagger~,\\
\Delta(N')=N'\otimes 1+1\otimes N'-i\displaystyle\frac{\alpha}{\gamma(J)}
           1\otimes 1~,\\
\Delta(a_{q(J)})=\left(a_{q(J)}\otimes q(J)^{N'/2}+iq(J)^{-N'/2}\otimes
           a_{q(J)}\right)e^{-i\alpha/2}~,\\
\Delta(a_{q(J)}^\dagger)=\left(a_{q(J)}^\dagger\otimes q(J)^{N'/2}+
           iq(J)^{-N'/2}\otimes a_{q(J)}^\dagger\right)e^{-i\alpha/2}~,\\
\epsilon(N')=\displaystyle i\frac{\alpha}{\gamma(J)}~,~~~~
           \epsilon(a_{q(J)})=0=\epsilon(a_{q(J)}^\dagger)~,\\
S(N')=-N'+i\displaystyle\frac{2\alpha}{\gamma(J)}\cdot 1~,\\
S(a_{q(J)})=-q(J)^{-1/2}a_{q(J)}~,~~~~~S(a_{q(J)}^\dagger)=-q(J)^{1/2}
                                    a_{q(J)}^\dagger~,
\end{array}
\end{equation}
where $N=a^\dagger a$, $N'=N+\displaystyle\frac{1}{2}+b\gamma(J)$.
It is interesting to note that all these structures are  implicitly
$J$-dependent.

The Hamiltonian (\ref{6.e388}) can be cast into

\begin{equation}\label{6.28}
H_{{q(J)}-{\rm vib}}=\frac{1}{2}\left([N+b\gamma(J) ]_{q(J)}+[N+1+b\gamma(J)
]_{q(J)}\right)hc\nu_{\rm vib}~.
\end{equation}
The representation of $H_{q(J)}(4)$ in coordinate space is also
expressed by $H_v(x)$, the Hermite polynomial
\begin{equation}
\tilde{\psi}_{q(J)-\rm vib}(v,x)= N_v H_v(X)e^{-X^2/2}.
\end{equation}
The energy levels of the system are

\begin{equation}
E_{{q(J)}-{\rm vib}}(v,J)
={1\over 2}\left([v+1+b\gamma(J) ]_{q(J)}+[v+b\gamma(J) ]_{q(J)}\right)hc\nu_
{\rm vib}~.
\end{equation}
Thus the vibration-rotational spectrum of diatomic molecules has the form

\begin{equation}
E_{q-\rm vib-rot}(v,J)
={1\over 2}\left([v+1+b\gamma(J)]_{q(J)}+[v+b\gamma(J)]_{q(J)}\right)hc\nu_
{\rm vib}+\frac{h^2}{8\pi^2I}J(J+1)~.
\end{equation}
The vibration-rotational energy levels of a diatomic molecule at a certain
electronic state can be written as

\begin{equation}\begin{array}{rcl}
E&=&E_0+E_{q-\rm vib-rot}\\[2mm]
 &=&E_0+\displaystyle \frac{\nu_{\rm vib}}{2}
    \left(\left[v+b(J)\gamma(J)\right]_{q(J)}
     +\left[v+b(J)\gamma(J)+1\right]_{q(J)}\right)hc\displaystyle
    +\displaystyle\frac{h^2}{8\pi^2I} J(J+1)\\[3mm]
 &=&E_0+\displaystyle {\nu_{\rm vib}}\frac{1}{2\sinh(\gamma(J)/2)}
    \sinh\left(\gamma\left(J\right)\left(v+{1\over 2}+c(J)\right)\right)
    +\displaystyle\frac{h^2}{8\pi^2I}J(J+1),
\end{array}
\end{equation}
where $E_0$ is the pure electron-transition energy and
$c(J)\equiv b(J)\gamma(J)$.

When $J=0$, there is no rotational excitation, and

\begin{equation}
\begin{array}{rcl}
E&=&E_0+E_{q(J)-{\rm vib}}(v)\\
 &=&E_0+\displaystyle
\frac{1}{2\sinh(\gamma_0/2)}\sinh\left(\gamma_0\left(v+{1\over 2}+
c_0\right)\right)hc\nu_{\rm vib}~,
\end{array}
\end{equation}
which is just the vibrational spectrum.

When $v=0$,  there is no vibrational excitation, and

\begin{equation}
\begin{array}{rcl}
E&=&E_0+E_{q(J)-{\rm vib}}(0)\\
 &=&E_0+\displaystyle hc\nu_{\rm vib}\frac{1}{2\sinh(\gamma(J)/2)}
    \sinh\left(\gamma
    \left(J\right)\left({1\over 2}+c(J)\right)\right)\\[2mm]
 & &+\displaystyle \frac{h^2}{8\pi^2I}J(J+1)~,
\end{array}
\end{equation}
which coincides in leading terms with the spectrum of the $q$-rotator
system. It should be noted that when $v=J=0$,

\begin{equation}
E=E_0+\frac{1}{2\sinh(\gamma(0)/2)}
\sinh\left(\gamma(0)\left(\frac{1}{2}+c(0)\right)\right)hc\nu_{\rm vib}
\end{equation}
which is $T_e$, the electronic term.

For simplicity assuming

\begin{equation}\label{6.e407}
\begin{array}{l}
\gamma(J)=\gamma_0+\gamma_1J(J+1)~,\\
c(J)=c_0+c_1J(J+1)~,
\end{array}\end{equation}
we have

\begin{equation}\label{6.e408}
\begin{array}{rcl}
E&=&E_0+\left(\displaystyle 2\sinh
    \left({1\over2}\gamma_0+{1\over2}\gamma_1J(J+1)\right)
    \right)^{-1}\times\\[3mm]
 & &\times \sinh\left(\left(\displaystyle\gamma_0+\gamma_1J(J+1)\right)
    \left(v+{\displaystyle 1\over\displaystyle
    2}+c_0+c_1J(J+1)\right)\right)hc\nu_{\rm vib}\\[4mm]
 & &+\displaystyle\frac{h^2}{8\pi^2I}J(J+1)~.
\end{array}
\end{equation}
This is the general form of the vibration-rotational energy levels of a
diatomic molecule. The second term represents the vibrational spectra in
interaction with the rotational, while the third describes the rigid rotation.
If we expand Eq.(\ref{6.e408}) into Taylor series, the parameters
$T_e$, $\omega_e$, $\omega_ex_e$, $\omega_ey_e$ and $\alpha_e$ introduced
in the conventional phenomenological treatment are reproduced as coefficients
of $\left(v+{1\over2}\right)^i\left(J\left(J+1\right)\right)^j$.

The  total Hamiltonian for this system is

\begin{equation}
H=H_{\rm e}+H_{q(J)-{\rm vib}}+H_{\rm rot}~,
\end{equation}
which has the symmetry of $H_q(4)\otimes{\rm SU}(2)$.

The Hilbert space should be constructed from representations of the
symmetry $H_q(4)\otimes{\rm SU}(2)$, namely

\begin{equation}
\Psi_{q-\rm vib-rot}(v,J,x)=N_vH_v(X)e^{-X^2/2}Y_{JM}(\theta,\phi).
\end{equation}
The selection rule for infrared spectrum resulted from
$H_q(4)\otimes{\rm SU}(2)$ symmetry says that $v$ can change by arbitrary
integer
although $\Delta v=\pm 1$ gives the most intense transitions due to dipole
nature of the interaction, and $J$ can change only by $1$ due to the
observation of the total angular momentum. Of course $\Delta v=0$ is also
allowed, but this does not give  rise to any rotation-vibrational spectrum but
the pure rotational one. If we now consider a particular transition from
$v^\prime$ to $v^{\prime\prime}$, the spectrum (in wavenumber) should be

\begin{equation}
\begin{array}{rcl}
\nu&=&\displaystyle\frac{\nu_{\rm vib}}{2}
      \left[2\left(v^{\prime}+{1\over 2}+c_0+
      c_1J^{\prime}(J^{\prime}+1)\right)
      \right]_{ \left<\left(\gamma_0+\gamma_1J^{\prime}(J^{\prime}+1)
      \right)/2\right>}\\[4mm]
   & &-\displaystyle\frac{\nu_{\rm
vib}}{2}\left[2\left(v^{\prime\prime}+{1\over 2}+c_0
      +c_1J^{\prime\prime}(J^{\prime\prime}+1)\right)
      \right]_{\left<\gamma_0+
      \gamma_1J^{\prime\prime}(J^{\prime\prime}+1)/2\right>}\\[4mm]
   & &+B_e\left(J^\prime(J^{\prime}+1)-J^{\prime\prime}(J^{\prime
      \prime}+1)\right)~,
\end{array}\end{equation}
where the notation $[x]_{\left<\gamma\right>}= [x]_q$ is implied.

{}From the selection rule $\Delta J=1$ or $-1$, we have

\begin{equation}
\begin{array}{rcl}
\nu_R&=&\displaystyle \frac{\nu_{\rm vib}}{2}
        \left[2\left(v^\prime+{1\over 2}+c_0+c_1(J+1)(J+2)\right)\right]
        _{\left<\left(\gamma_0+\gamma_1(J+1)(J+2)\right)/2\right>}\\[4mm]
     & &-\displaystyle \frac{\nu_{\rm vib}}{2}
        \left[2\left(v^{\prime\prime}+{1\over 2}+c_0+c_1J(J+1)\right)\right]
        _{\left<\left(\gamma_0+\gamma_1J(J+1)\right)/2\right>}\\[4mm]
     & &+B_e\left((J+1)(J+2)-J(J+1)\right)~,
\end{array}
\end{equation}
and

\begin{equation}
\begin{array}{rcl}
\nu_P&=&\displaystyle \frac{\nu_{\rm vib}}{2}
        \left[2\left(v^\prime+{1\over 2}+c_0+c_1J(J-1)

\right)\right]_{\left<\left(\gamma_0+\gamma_1J(J-1)\right)/2\right>}\\[4mm]
     & &-\displaystyle \frac{ \nu_{\rm vib}}{2}
        \left[2\left(v^{\prime\prime}+{1\over 2}+c_0+c_1J(J+1)\right)\right]_
        {\left<\left(\gamma_0+\gamma_1J(J+1)\right)/2\right>}\\[4mm]
     & &+B_e\left(J(J-1)-J(J+1)\right)~,
\end{array}
\end{equation}
where $J^{\prime\prime}$ is replaced by $J$. Since $J$ can
take a whole series of values, these two formulae represent two series
of lines, which are called $R$, and $P$ branch respectively.

The selection rule for the Raman spectrum is $\Delta J=0,\pm 1$. Accordingly,
for a given Raman vibrational band,
there are three branches, for which the spectrum is readily obtained from

\begin{equation}
\begin{array}{rcl}
\Delta \nu&=&\displaystyle\frac{\nu_{\rm vib}}{2}\left[2\left(v^\prime
             +{1\over 2}+c_0+c_1J^\prime (J^\prime +1)\right)\right]_
             {\left<\left(\gamma_0+\gamma_1J^\prime (J^\prime +1)
             \right)/2\right>}\\[4mm]
          & &-\displaystyle\frac{\nu_{\rm vib}}{2}\left[2\left
             (v^{\prime\prime}+{1\over 2}+c_0+
             c_1J^{\prime\prime}(J^{\prime\prime}+1)\right)\right]_{\left<
             \left(\gamma_0+\gamma_1J^{\prime\prime}(J^{\prime\prime}+1)
             \right)/2\right>}\\[4mm]
          & &+B_e\left(J^\prime(J^{\prime}+1)-J^{\prime\prime}(J^{\prime
             \prime}+1)\right),
\end{array}
\end{equation}
by substituting $J^\prime=J^{\prime\prime}+2$ for $S$ branch,
$J^\prime=J^{\prime\prime}-2$ for $O$ branch and $J^\prime=J^{\prime\prime}$
for $Q$ branch (and redenoting $J^{\prime\prime}=J$):

\begin{equation}
\begin{array}{rcl}
\left(\Delta \nu\right)_S&=&\displaystyle
     \frac{\nu_{\rm vib}}{2}\left[2\left(v^{\prime}+{1\over 2}+
     c_0+c_1(J+2)(J+3)\right)\right]_{\left<\left(\gamma_0+\gamma_1(J+2)(J+3)
     \right)/2\right>}\\[4mm]
 & &-\displaystyle\frac{\nu_{\rm vib}}{2}\left[2\left(v^{\prime\prime}+{1\over
2}+
     c_0+c_1J(J+1)\right)\right]_{\left<\left(\gamma_0+\gamma_1J(J+1)
     \right)/2\right>}\\[4mm]
 & &+B_e(4J+6)~,
 \end{array}
 \end{equation}
where $J=0,1,\cdots$;

\begin{equation}
\begin{array}{rcl}
\left(\Delta \nu\right)_O&=&\displaystyle\frac{\nu_{\rm vib}}{2}
   \left[2\left(v^{\prime}+{1\over 2}+
   c_0+c_1(J-2)(J-1)\right)\right]_{\left<\left(
   \gamma_0+\gamma_1(J-2)(J-1)\right)/2\right>}\\[4mm]
& &-\displaystyle\frac{\nu_{\rm vib}}{2}\left[2\left(v^{\prime\prime}+
   {1\over 2}+c_0+c_1J(J+1)\right)\right]_{\left<\left(\gamma_0+\gamma_1J(J+1)
   \right)/2\right>}\\[4mm]
& &+B_e(-4J+2)~,
\end{array}
\end{equation}
where $J=2,3,\cdots$;

\begin{equation}
\begin{array}{rcl}
\left(\Delta \nu\right)_Q&=&\displaystyle\frac{\nu_{\rm vib}}{2}
   \left[2\left(v^{\prime}+{1\over 2}+
   c_0+c_1J(J+1)\right)\right]_{\left<\left(\gamma_0+\gamma_1J(J+1)
   \right)/2\right>}\\[4mm]
 & &-\displaystyle\frac{\nu_{\rm vib}}{2}\left[2\left(v^{\prime\prime}+{1\over
2}+
   c_0+c_1J(J+1)\right)\right]_{\left<\left(\gamma_0+\gamma_1J(J+1)
   \right)/2\right>}~,
\end{array}
\end{equation}
where $J=0,1,\cdots$.

Now we examine the vibrational-rotational structure of
elec\-tronic tran\-sitions, for which the wavenumber of the
transition is \begin{equation}
\nu=\left({E_0^\prime-E_0^{\prime\prime}}+{E_{q(J)-{\rm vib}}^\prime-
E_{q(J)-{\rm vib}}^{\prime\prime}}+{E_{rot}^\prime-
E_{\rm rot}^{\prime\prime}}\right)/hc~,
\end{equation}
where $E_0^\prime$, $E_{q(J)-{\rm vib}}^\prime$, $E_{\rm rot}^\prime$ and
$E_0^{\prime\prime}$, $E_{q(J)-{\rm vib}}^{\prime\prime}$, $E_{\rm
rot}^{\prime\prime}$
are the electronic energy and  the vibration-rotational terms of the upper and
lower electronic state, respectively. The difference of the present spectra
from those of infrared and Raman lies in  $E_{q(J)-{\rm vib}}^\prime$,
$E_{\rm rot}^\prime$ and $E_{q(J)-{\rm vib}}^{\prime\prime}$,
$E_{\rm rot}^{\prime\prime}$ belong to different electronic states
and  have generally  different magnitudes.

The selection rule tells us that the upper and lower states may have
different electronic angular
momenta $\Lambda$. If at least one of the two states has nonzero $\Lambda$,
the selection rule is $\Delta J=J^\prime-J^{\prime\prime}=0,\pm1$. However, if
$\Lambda=0$ in both electronic states, (i.e.,
$^1\Sigma\to~^1\Sigma$), the transition of $\Delta J=0$ is forbidden and
only the transitions of $\Delta J=\pm 1$ are allowed, as for most infrared
bands.
Expectedly, there are  three or two series of lines (branches), for which the
wavenumbers are  the following \\
$R$ branch:
\begin{equation}\begin{array}{rcl}
\nu&=&\displaystyle\frac{E_0^\prime-E_0^{\prime\prime}}{hc}+\displaystyle
      \frac{\nu_{\rm vib}^\prime}{2}\left[2\left(v^{\prime}+{1\over 2}+
      c_0+c_1(J+1)(J+2)\right)\right]_{\left<\left(
      \gamma_0+\gamma_1(J+1)(J+2)\right)/2\right>}\\[4mm]
   & &-\displaystyle\frac{\nu_{\rm vib}^{\prime\prime}}
      {2}\left[2\left(v^{\prime\prime}+{1\over 2}+
      c_0+c_1J(J+1)\right)\right]_{\left<\left(\gamma_0+\gamma_1J(J+1)
      \right)/2\right>}\\[4mm]
    & &+\left(B_e^\prime(J+2)(J+1)-B_e^{\prime\prime}J(J+1)\right)~;
\end{array}
\end{equation}
$Q$ branch:
\begin{equation}
\begin{array}{rcl}
\nu&=&\displaystyle\frac{E_0^\prime-E_0^{\prime\prime}}{hc}+\displaystyle
      \frac{\nu_{\rm vib}^\prime}{2}\left[2\left(v^{\prime}+{1\over 2}+
      c_0+c_1J(J+1)\right)\right]_{\left<\left(\gamma_0+\gamma_1J(J+1)
      \right)/2\right>}\\[4mm]
    & &-\displaystyle\frac{\nu_{\rm vib}^{\prime\prime}}{2}\left[2\left(
       v^{\prime\prime}+{1\over 2}+
       c_0+c_1J(J+1)\right)\right]_{\left<\left(\gamma_0+\gamma_1J(J+1)
       \right)/2\right>}\\[3mm]
     & &+\left(B_e^\prime-B_e^{\prime\prime}\right)J(J+1)~;
\end{array}
\end{equation}
$P$ branch:
\begin{equation}
\begin{array}{rcl}
\nu&=&\displaystyle\frac{E_0^\prime-E_0^{\prime\prime}}{hc}+\displaystyle
      \frac{\nu_{\rm vib}^\prime}{2}\left[2\left(v^{\prime}+{1\over 2}+
      c_0+c_1(J-1)J\right)\right]_{\left<\left(\gamma_0+\gamma_1(J-1)J
      \right)/2\right>}\\[4mm]
   & &-\displaystyle\frac{\nu_{\rm vib}^{\prime\prime}}
      {2}\left[2\left(v^{\prime\prime}+{1\over 2}+
      c_0+c_1J(J+1)\right)\right]_{\left<\left(\gamma_0+\gamma_1J(J+1)
      \right)/2\right>}\\[3mm]
   & &+\left(B_e^\prime(J-1)J-B_e^{\prime\prime}J(J+1)\right)~.
\end{array}
\end{equation}
This completes the description for vibrating-rotating
diatomic molecules in the quantum group theoretic approach.

\bigskip\bigskip\bigskip

\centerline{\bf Acknowledgments}
I would like to thank Prof. H. Y. Guo for his encouragement to
prepare this Review. The author is indebted to Prof. Yang-Zhong Zhang for
his reading the manuscript and helps in  rhetoric.

\end{document}